\documentclass[twocolumn,showpacs,prc,superscriptaddress,amsmath,amssymb]{revtex4-1}
\usepackage[english]{babel}
\usepackage[utf8]{inputenc}
\usepackage[colorinlistoftodos, color=green!40, prependcaption]{todonotes}
\usepackage{amsthm}
\usepackage{mathtools}
\usepackage{physics}
\usepackage{xcolor}
\usepackage[left=23mm,right=13mm,top=35mm,columnsep=15pt]{geometry} 
\usepackage{adjustbox}
\usepackage{placeins}
\usepackage{lipsum}
\usepackage{csquotes}
\usepackage[pdftex, pdftitle={Article}, pdfauthor={Author}]{hyperref} 
\usepackage{color}

\usepackage{amssymb}
\usepackage{amsmath}
\usepackage{acronym}
\usepackage[OT2,T1]{fontenc}
\usepackage{gensymb}
\usepackage{wasysym}
\usepackage{graphicx}
\usepackage{multirow} 
 \usepackage{lineno}

\newcommand{\geant}{{\sc{Geant4}}}
\newcommand{\labr}{LaBr$_{3}$:Ce}
\newcommand{\cebr}{CeBr$_{3}$}

\newcommand{\talys}{{\sf{TALYS}}}

\newcommand{\eli}{Extreme Light Infrastructure-Nuclear Physics (ELI-NP)/Horia Hulubei National Institute for Physics and Nuclear Engineering (IFIN-HH), Str. Reactorului 30, M\u{a}gurele 077125, Romania}
\newcommand{\uio}{Department of Physics, University of Oslo, N-0316 Oslo, Norway}
\newcommand{\dfn}{Department of Nuclear Physics (DFN)/Horia Hulubei National Institute for Physics and Nuclear Engineering (IFIN-HH), Str. Reactorului 30, M\u{a}gurele 077125, Romania}
\newcommand{\umi}{Universit\`{a} degli Studi di Milano, Milano, Italy}
\newcommand{\infnmi}{INFN sez. Milano, Milano, Italy}
\newcommand{\ifjpan}{The Henryk Niewodniczanski Institute of Nuclear Physics, Polish Academy of Sciences, ul. Radzikowskiego 152, 31-342 Krakow, Poland}
\newcommand{\istu}{Department of Physics, Faculty of Science, Istanbul University, Vezneciler/Fatih, 34134, Istanbul, Turkey}
\newcommand{\ctupr}{Institute of Experimental and Applied Physics, Czech Technical University in Prague, Husova 240/5, 110 00 Prague 1, Czech Republic}

\begin{document}
\title{Statistical properties and photon strength functions of the ${}^{112,114}$Sn isotopes below the neutron separation threshold}
\author{P.-A.~S{\"o}derstr{\"o}m}\email[Correspondence email address: ]{par.anders@eli-np.ro}\affiliation{\eli}
\author{M.~Markova}\affiliation{\uio}
\author{N.~Tsoneva}\affiliation{\eli}
\author{Y.~Xu}\affiliation{\eli}
\author{A.~Ku\c{s}o\u{g}lu}\affiliation{\eli}\affiliation{\istu}
\author{S.~Aogaki}\affiliation{\eli}
\author{D.~L.~Balabanski}\affiliation{\eli}
\author{S.~R.~Ban}\affiliation{\eli}
\author{R.~Borcea}\affiliation{\dfn}
\author{M.~Brezeanu}\affiliation{\eli}
\author{F.~Camera}\affiliation{\umi}\affiliation{\infnmi}
\author{M.~Ciema\l{}a}\affiliation{\ifjpan}
\author{Gh.~Ciocan}\affiliation{\dfn}
\author{C.~Clisu}\affiliation{\dfn}
\author{C.~Costache}\affiliation{\dfn}
\author{F.~C.~L.~Crespi}\affiliation{\umi}\affiliation{\infnmi}
\author{M.~Cuciuc}\affiliation{\eli}
\author{A.~Dhal}\affiliation{\eli}
\author{I.~Dinescu}\affiliation{\dfn}
\author{N.~M.~Florea}\affiliation{\dfn}
\author{A.~Giaz}\affiliation{\umi}\affiliation{\infnmi}
\author{M.~Kmiecik}\affiliation{\ifjpan}
\author{V.~Lelasseux}\affiliation{\eli}
\author{R.~Lica}\affiliation{\dfn}
\author{N.~M.~M\u{a}rginean}\affiliation{\dfn}
\author{C.~Mihai}\affiliation{\dfn}
\author{R.~E.~Mihai}\affiliation{\dfn}\affiliation{\ctupr}
\author{D.~Nichita}\affiliation{\eli}
\author{H.~Pai}\affiliation{\eli}
\author{I.~P.~P\^{a}rlea}\affiliation{\eli}
\author{T.~Petruse}\affiliation{\eli}
\author{A.~Rotaru}\affiliation{\eli}
\author{C.~O.~Sotty}\affiliation{\dfn}
\author{A.~Sp\u{a}taru}\affiliation{\eli}
\author{L.~Stan}\affiliation{\dfn}
\author{D.~A.~Testov}\affiliation{\eli}
\author{D.~Tofan}\affiliation{\dfn}
\author{S.~Toma}\affiliation{\dfn}
\author{T.~Tozar}\affiliation{\eli}
\author{A.~Turturic\u{a}}\affiliation{\dfn}
\author{G.~V.~Turturic\u{a}}\affiliation{\eli}
\author{S.~Ujeniuc}\affiliation{\dfn}
\author{C.~A.~Ur}\affiliation{\eli}
\author{O.~Wieland}\affiliation{\umi}\affiliation{\infnmi}

\date{\today} 

\begin{abstract}
Here, we report on the measurements of the $\gamma$-ray strength functions and nuclear level densities of ${}^{112,114}$Sn performed for the first time at the 9~MV Tandem accelerator facilities at IFIN-HH using the Oslo method. We extract thermodynamic properties and gross and fine properties of the pygmy dipole resonance for systematic comparison in the chain of Sn isotopes. The results are compared with microscopic models implemented in the \talys\ reaction code and the fully microscopic quasiparticle-phonon model for the underlying nuclear structure of the dipole strength in ${}^{112,114}$Sn. The quasiparticle-phonon model results show the importance of complex configurations to the low-energy dipole response in the pygmy dipole resonance energy region. The experimental data are further included in the cross-section and reaction rate calculations for the $(\mathrm{n},\gamma)$ reaction of the $p$-process nuclei ${}^{112,114}$Sn showing a significant increase in reaction rates at high temperatures compared to existing nuclear databases.
\end{abstract}


\maketitle

\section{Introduction} \label{sec:introduction}

    The concept of \acp{gSF} is one of the critical components both in understanding the collective behaviour of atomic nuclei and calculating reaction rates in nuclear astrophysics \cite{Goriely2019b,Kawano2020}. In the statistical regime, the \ac{gSF} describes the average probability for an internal nuclear decay for a given $\gamma$-ray energy. This function is almost entirely dominated by the \ac{GDR}, commonly visualised as a collective vibration of the bulk protons against the bulk neutrons, exhausting almost the full electric dipole strength, with a width of typically $\Gamma_{\mathrm{GDR}} = 2.5$-$5$~MeV, and a centroid of, typically, $E_{\mathrm{GDR}} = 13$-$20$~MeV \cite{Harakeh2001}. Additional structures are commonly observed on the low-energy tail of the \ac{GDR}. These include, for example, an observed fine structure of the \ac{GDR} \cite{Tamii2011}, the magnetic dipole modes: scissors mode \cite{Heyde2010} and spin-flip giant resonance, the low-energy upbend \cite{Schwengner2013}, and also additional \ac{LEDR} around the neutron separation threshold. This \ac{LEDR} is sometimes referred to, in general, as being a \ac{PDR}, while sometimes the term \ac{PDR} describes a specific interpretation of the \ac{LEDR} as neutron skin oscillations against the bulk protons and neutrons \cite{Paar2007,Savran2013,Bracco2019b,Lanza2023}. The exact nature of the \ac{LEDR}, and its relation to and nature of the \ac{PDR}, is a current topic for debate in the scientific community, where some recent experiments \cite{Spieker2020,Weinert2021} have inspected the single-particle content of the low-lying strength relative to the typical collective motion picture.

    For measuring \acp{gSF} and \acp{NLD} below the neutron separation threshold, the Oslo method \cite{Guttormsen1987,Guttormsen1996,Schiller2000,Larsen2011} has been used extensively over the years. In fact, to our knowledge, it has been the only measurement approach to simultaneously determine \acp{gSF} and \acp{NLD}. In particular, the results reported in this work serve as a continuation of the work in Oslo to measure the Sn chain of isotopes \cite{Agvaanluvsan2009a, Agvaanluvsan2009b, Toft2010, Toft2011, Markova2021, Markova2022, Markova2023, Markova2024}, where ${}^{111-113,116-122,124}$Sn have been presented. While mainly used at the \ac{OCL} in its original form, extensions have been developed like the inverse Oslo method for unstable nuclei \cite{Ingeberg2020, Ingeberg2022} and $\beta$-Oslo method for constraining the statistical neutron-capture cross sections on short-lived nuclei via $\beta$ decay \cite{Liddick2019}. Also, other approaches to the \ac{LEDR} are under development, like the neutron pick-up from $(\mathrm{d},\mathrm{p})$ reactions, tested recently on ${}^{120}$Sn \cite{Weinert2021}, where the single-particle nature can be highlighted. 
    With the projected $\gamma$-ray beams at \ac{ELI-NP} \cite{Filipescu2015, Gales2016, Gales2018, Tanaka2020, Constantin2024}, that is expected to provide users with narrow bandwidth $\gamma$ rays for photoexcitation and decay studies, there will be new opportunities to measure the \acp{gSF} directly, as, for example, has been done at the \ac{HIgS} facility at the \ac{TUNL} using the $(\vec{\gamma},\gamma'\gamma'')$ technique \cite{Isaak2019}, meaning inelastic scattering of incoming polarised photons, $\vec{\gamma}$, with a two-step decay as $\gamma'$ and $\gamma''$. 
    
    In the case of the future $\gamma$-ray beam at \ac{ELI-NP}, detailed studies of the \ac{GDR} and \ac{PDR} will be the focus of the \ac{ELIGANT} family of instruments. In particular, the \ac{ELIGANT-GN} setup \cite{Camera2016, Krzysiek2019a, Soderstrom2022} is optimised for detailed investigations of the nuclear structure of these excitation modes via neutron and $\gamma$-ray decay branchings, while the general cross-section measurements will be performed using the \ac{ELIGANT-TN} moderated neutron counter \cite{Clisu2023}. A primary motivation for these measurements is understanding the production of $p$ nuclei \cite{Filipescu2015, Soderstrom2023b}. However, for cross sections in stellar environments \cite{Mohr2004, Utsunomiya2006, Rauscher2012, Rauscher2013a, Rauscher2014}, complementary measurements of \acp{gSF} and \acp{NLD} will be necessary to understand how the reaction rates impact the $p$ process. For the work reported here, we chose the ${}^{112,114}$Sn nuclei partially for their nature as exclusively $p$-process nuclei. In particular, the ${}^{111}\mathrm{Sn}(\mathrm{n},\gamma){}^{112}\mathrm{Sn}$ reaction has been discussed as a key reaction in Type Ia supernova regions with densities  $0.4-1 \times 10^{9}$~g~cm$^{-3}$ \cite{Nishimura2018}. Another reason is for the recent wealth of information that has been published on the Sn isotopes \cite{Markova2021, Markova2022, Markova2023, Markova2024} making them an ideal case for completing the systematic investigation of the Sn isotopic chain. 

\section{Experimental setup} \label{sec:experiment}

    The experiment reported here was carried out the at the 9~MV Tandem accelerator at the \ac{IFIN-HH} using the special version of the \ac{ROSPHERE} setup \cite{Bucurescu2016} comprising 21 \labr\ and \cebr\ detectors from \ac{ELI-NP}, in particular from \ac{ELIGANT-GN} \cite{Camera2016, Soderstrom2022} and two detectors from the beam diagnostics \cite{Weller2016} setups for the \ac{ELI-NP} $\gamma$-ray beam system. This special version of \ac{ROSPHERE} have been used for detailed studies of light nuclei and high-energy resonant states
    \cite{Kusoglu2024a,Kusoglu2024b,Wieland2024a,Soderstrom2024a}, and
    is detailed in Reference~\cite{Aogaki2023}. However, the geometry in this experiment differed slightly from what was reported in Reference~\cite{Aogaki2023}, with the exact location of the detector types at different positions. The total angular coverage of the scintillators in this configuration was 11.95\% of the solid angle.

    In addition to the $\gamma$-ray detectors, the setup included a $\Delta E - E$ telescope consisting of two annular \acp{DSSSD} in the backward direction. These detectors were of the type Micron S7 with thicknesses of 65~$\mu$m and 1000~$\mu$m, respectively, placed at a distance of 28~mm from the target for the thin detector and 44~mm from the target for the thick detector. This provided us with an angular coverage of 122$^{\circ}$-136$^{\circ}$ where both detectors were in the line-of-sight from the target. Due to the limited feed-throughs in the target chamber, the $\Delta E$ detector was only read out from the 16~sectors. The readout from the $E$ detector consisted of 64 channels with 16~sectors and 48 rings for reconstructing the scattering angle required to obtain the nuclear excitation energy.

    The \ac{DAQ} system employed in this experiment was fully digital with the \labr\ and \cebr\ detectors read out using CAEN V1730 digitisers running \ac{DPP-PSD} firmware, and the silicon detectors as well as the \ac{HPGe} detectors read out by CAEN V1725 digitisers running \ac{DPP-PHA} firmware. The $\gamma$-ray detectors were individually calibrated using a ${}^{137}$Cs source for low energies and calibrated at high energies using the  4.44~MeV and 9~MeV transitions from an isotopic \ac{PuBe} source \cite{Soderstrom2021} and a special sphere for neutron capture in nickel \cite{Soderstrom2023a} for the highest energy. The silicon detectors were aligned using a standard three-$\alpha$ source consisting of ${}^{239}$Pu, ${}^{241}$Am, and ${}^{244}$Cm, and the absolute calibration was obtained from the elastic scattering of the protons using the in-beam data.

    We used thin, self-supporting tin targets in the centre of the array for this experiment. The two targets were made of  ${}^{112}$Sn with an isotopic purity of 99.6\% and a thickness of 1.7~mg/cm$^{2}$, and ${}^{114}$Sn with an isotopic purity of 87.1\% and a thickness of 1.8~mg/cm$^{2}$. The main isotopic contaminants in the ${}^{112}$Sn target were ${}^{114}$Sn with 0.3\% and ${}^{115}$Sn with 0.1\%, while for ${}^{114}$Sn the main contaminants were ${}^{115}$Sn with 6.5\%, ${}^{112}$Sn with 4.6\%, ${}^{120}$Sn with 1.3\%, ${}^{116}$Sn with 1.1\%, and ${}^{119}$Sn with 0.4\%.

The experiment was performed using a proton energy of 12.7~MeV and a typical beam current of 0.5~nA. The data were collected for 68~h for ${}^{112}$Sn and 
    70~h for ${}^{114}$Sn. The trigger rate in the silicon detectors was about 1.1~kHz total rate for each detector, and around 10~kHz on average from the \labr\ and \cebr\ detectors.

\section{Analysis} \label{sec:analysis}

\begin{figure}[h!]
\begin{center}
\includegraphics[width=\columnwidth]{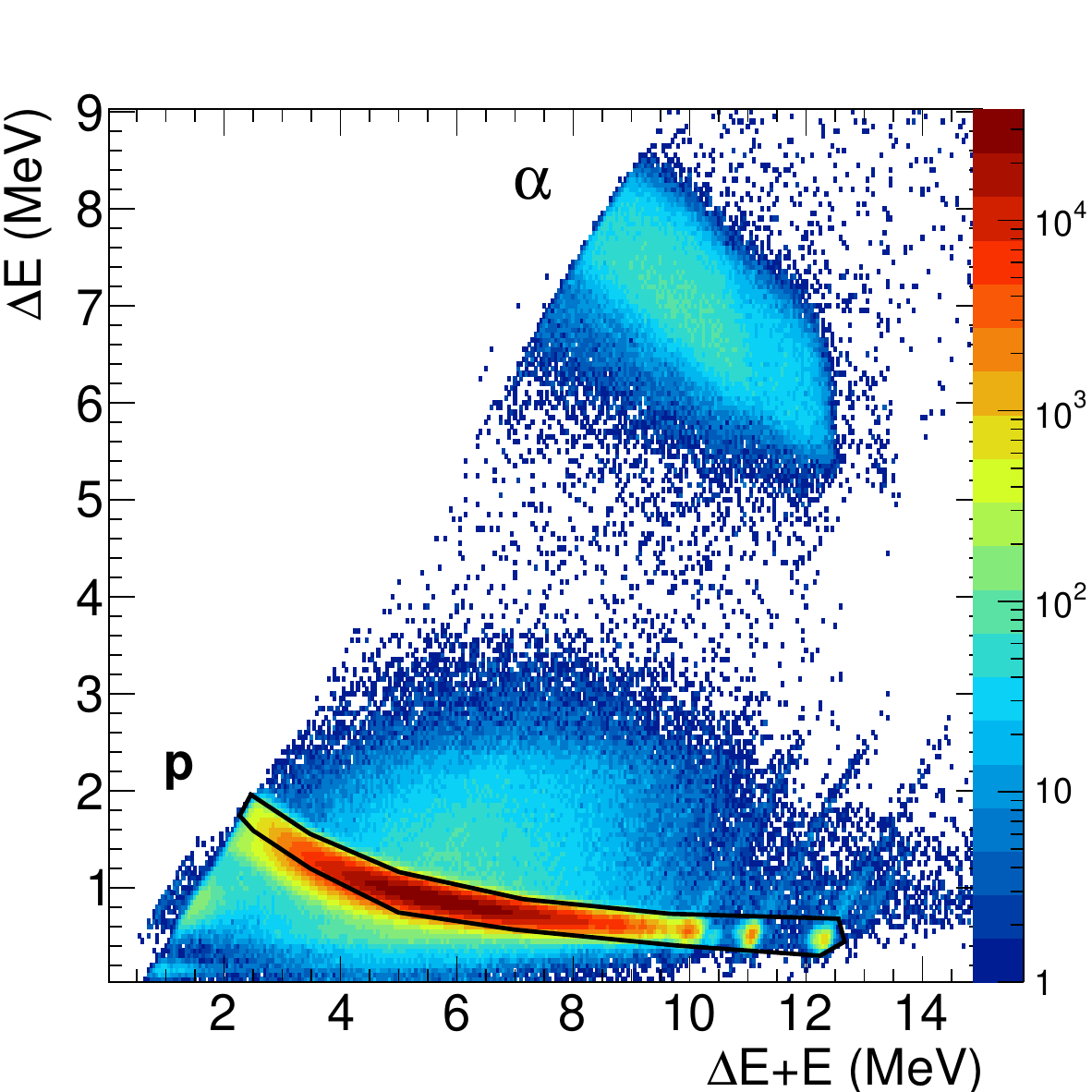}
\end{center}
\caption{Total energy deposition ($\Delta E + E$) versus the energy deposited in the first silicon layer $\Delta E$. The structures corresponding to the protons and $\alpha$ particles have been labelled. The gate used to select protons is shown as a black outline. Note that due to reflections in one of the connectors, the low-energy region exhibits a shadow double of the main proton band that has been excluded from the analysis. Pile-up events can be seen as diagonal structures extending above the proton band.}\label{fig:dee}
\end{figure}

\begin{figure*}[ht!]
\begin{center}
\includegraphics[width=0.32\textwidth]{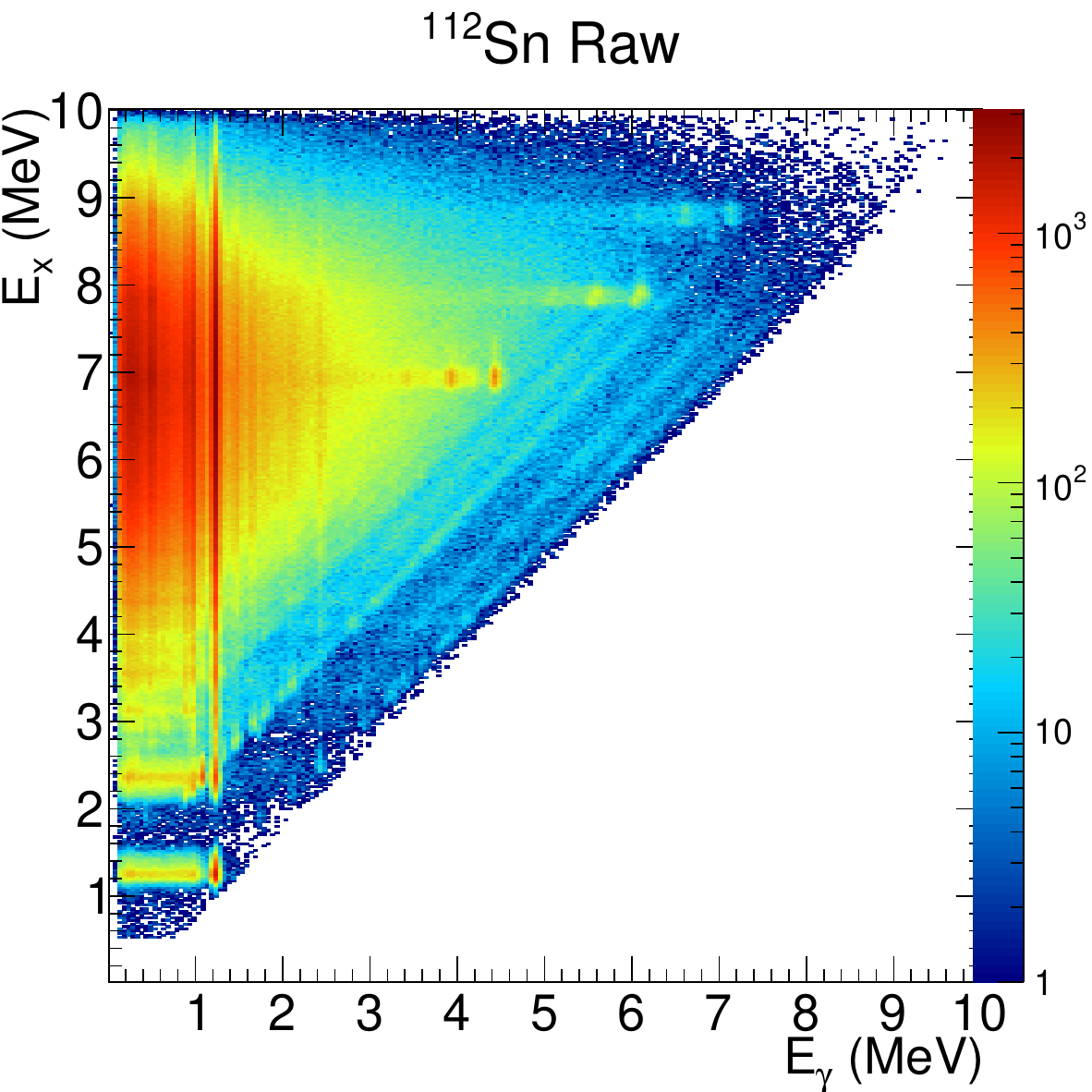}
\includegraphics[width=0.32\textwidth]{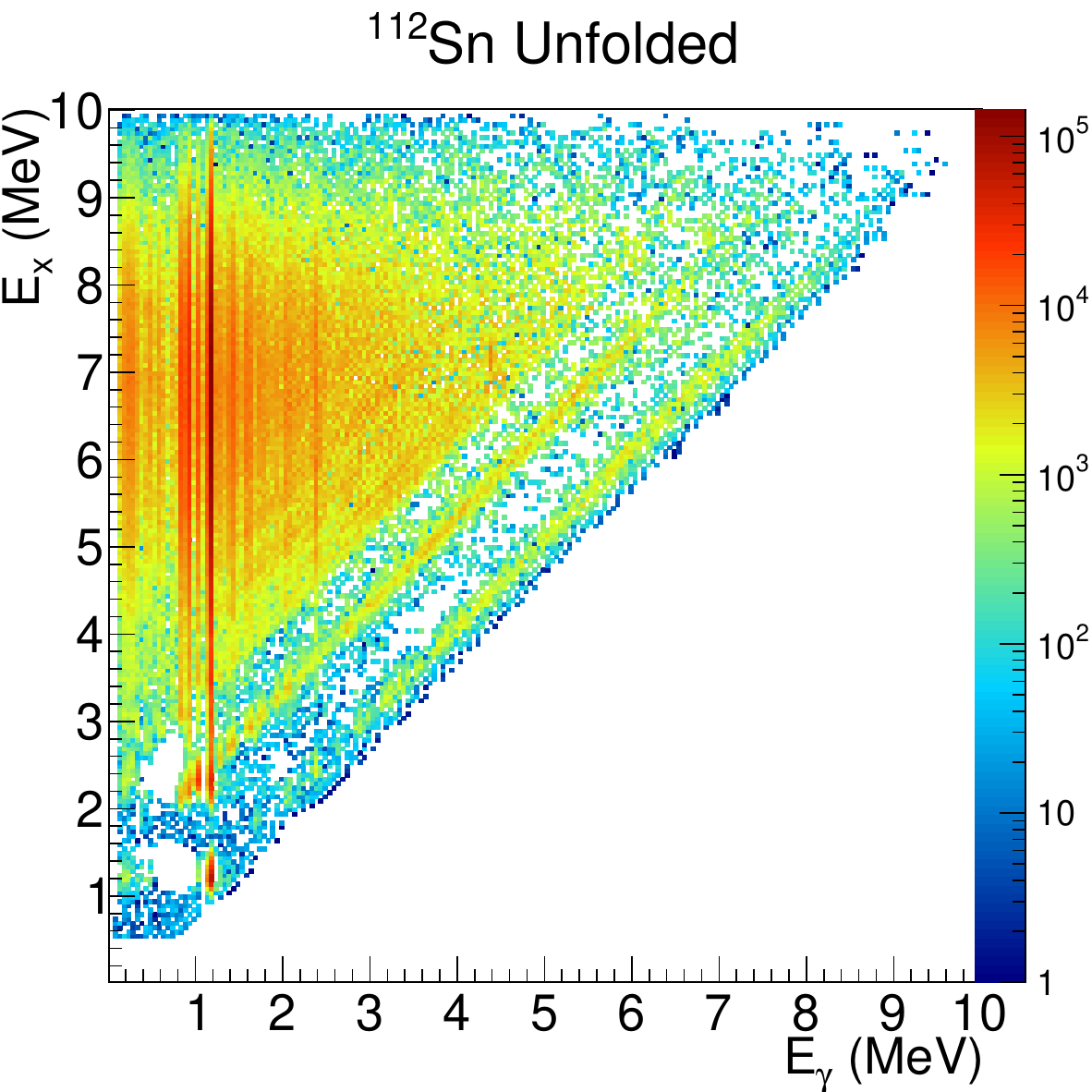}
\includegraphics[width=0.32\textwidth]{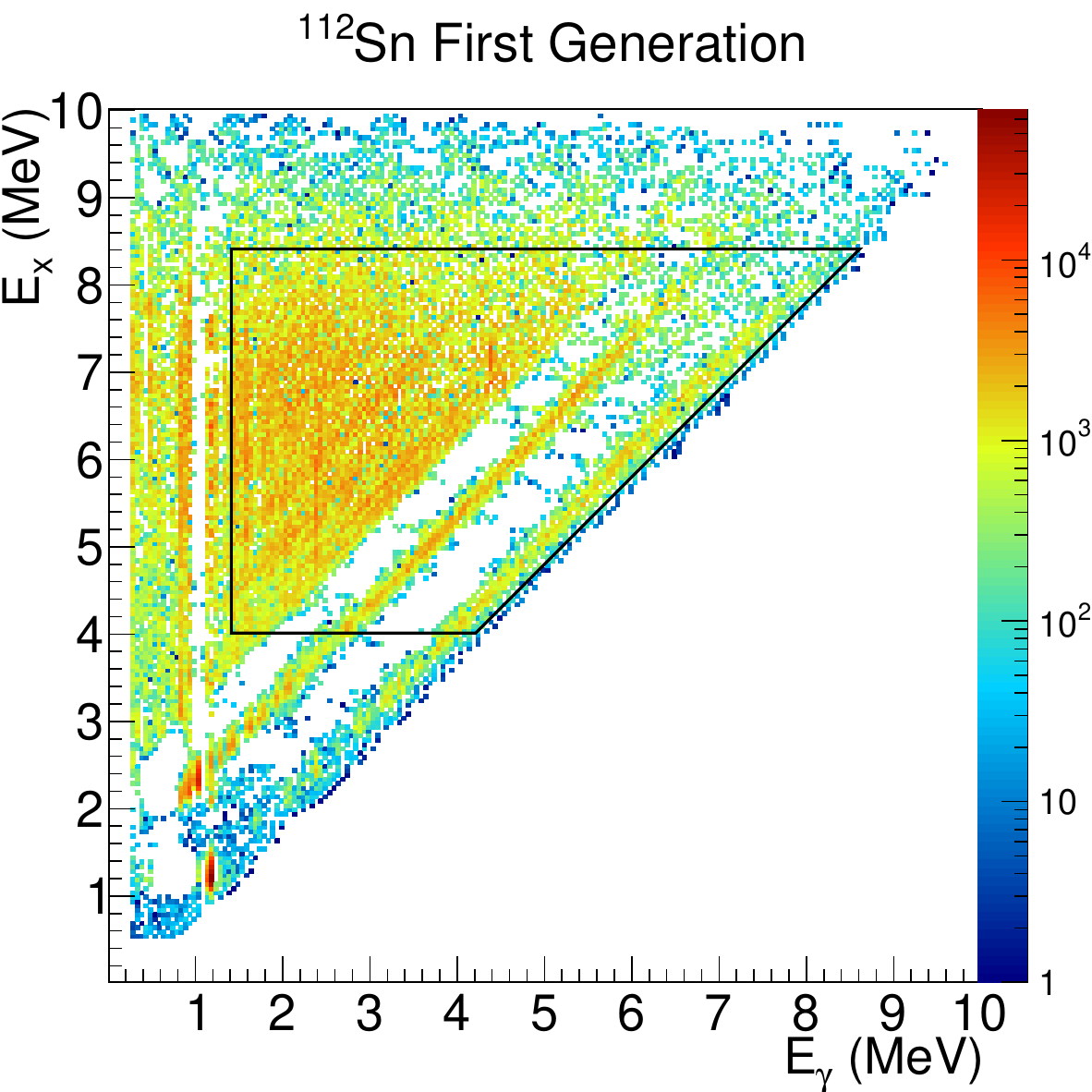}\\
\includegraphics[width=0.32\textwidth]{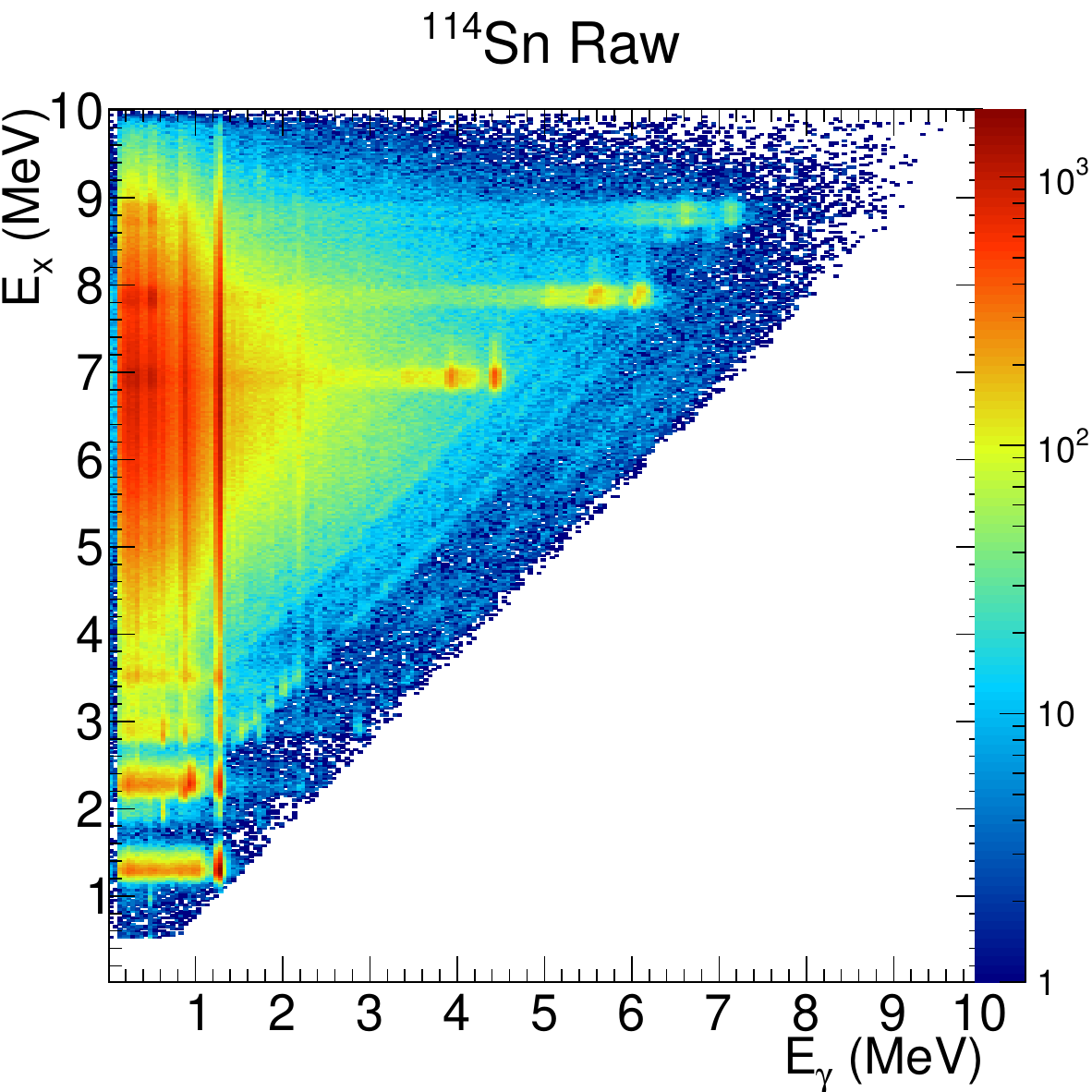}
\includegraphics[width=0.32\textwidth]{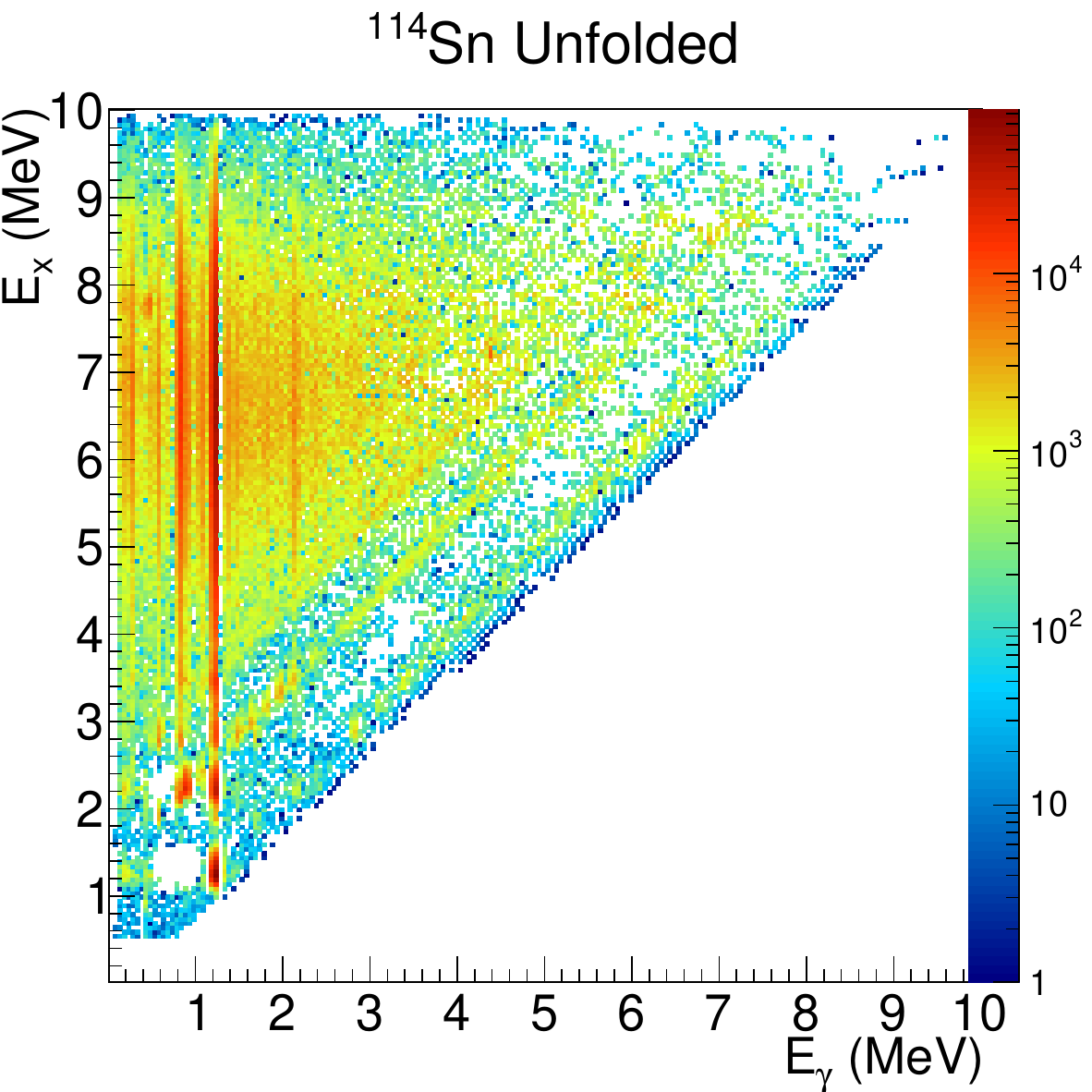}
\includegraphics[width=0.32\textwidth]{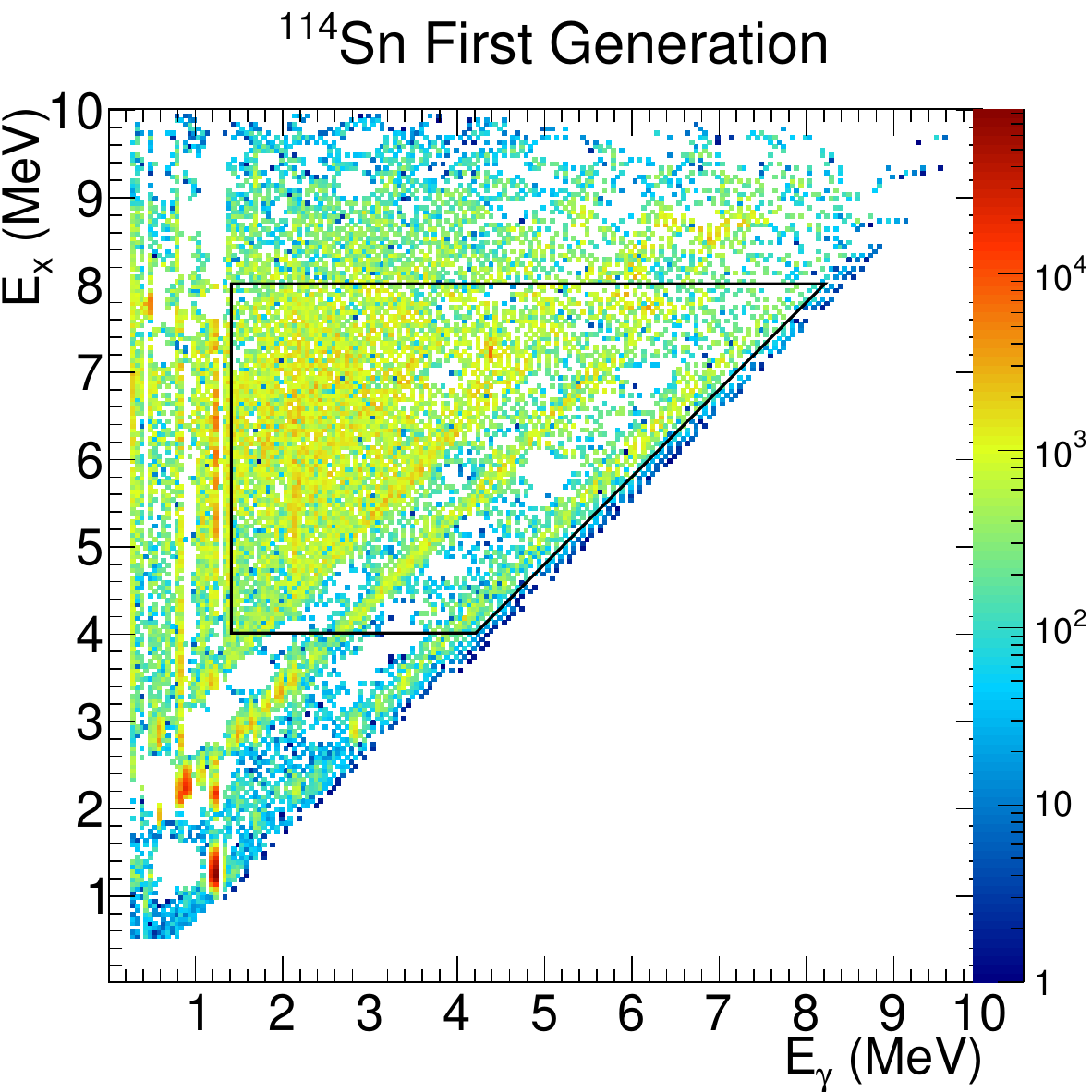}
\end{center}
\caption{Raw (left), unfolded (middle) and first generation (right) matrices for ${}^{112}$Sn (top) and ${}^{114}$Sn (bottom). The area selected for further analysis of the first-generation matrices is shown as a black outline.}\label{fig:matrices}
\end{figure*}

To make sure that only good events, with unambiguous excitation energy of the nucleus, were kept in the analysis, events were selected in the silicon detectors if they had a multiplicity of precisely one in the $\Delta E$ detector, a multiplicity of exactly one in the sectors of the $E$ detectors, and either had a multiplicity of one or two in the rings of the $E$ detector. If the multiplicity was two in the rings, then additional constraints were put in that the two rings had to be neighbours and that the sum of energies in the two rings had to equal the energy recorded in the sector. Furthermore, the $\Delta E$ and $E$ sectors had to have a geometric overlap for the event to be accepted. With these events selected, we can create a two-dimensional histogram based on the standard $\Delta E$-$E$ method, shown in Figure~\ref{fig:dee}, of the energy in the $\Delta E$ detector versus the total energy recorded in both detectors.
Here, the inelastically scattered protons appear clearly as a continuous band, with isolated regions corresponding to the scattering of the lowest excited discrete states and elastically scattered protons in the bottom right edge of the distribution. A significant background above this region, noticeable above the discrete peaks, was identified as originating from a pile-up in the $\Delta E$ detector, which unfortunately did not allow for the identification of deuterons. However, at the highest silicon energies, we can see a clear distribution corresponding to $\alpha$ particles produced in ${}^{112}$Sn$(\mathrm{p},\alpha){}^{109}$In reactions.
For each event, the excitation energy was calculated from the beam energy, the scattering angle, and the energy deposited in the silicon detectors, and the associated $\gamma$-rays, as measured by the \labr\ and \cebr\ detectors, were plotted against the excitation energy in a so-called $E_{x}$-$E_{\gamma}$ matrix, as shown in Figure~\ref{fig:matrices}, where each energy $E_{x}$ has its own associated measured $\gamma$-ray distribution vector $\vec{S}$. Each element in $\vec{S}$ corresponds to the number of counts in the associated energy bin.

\begin{figure}[ht!]
\begin{center}
\includegraphics[width=\columnwidth]{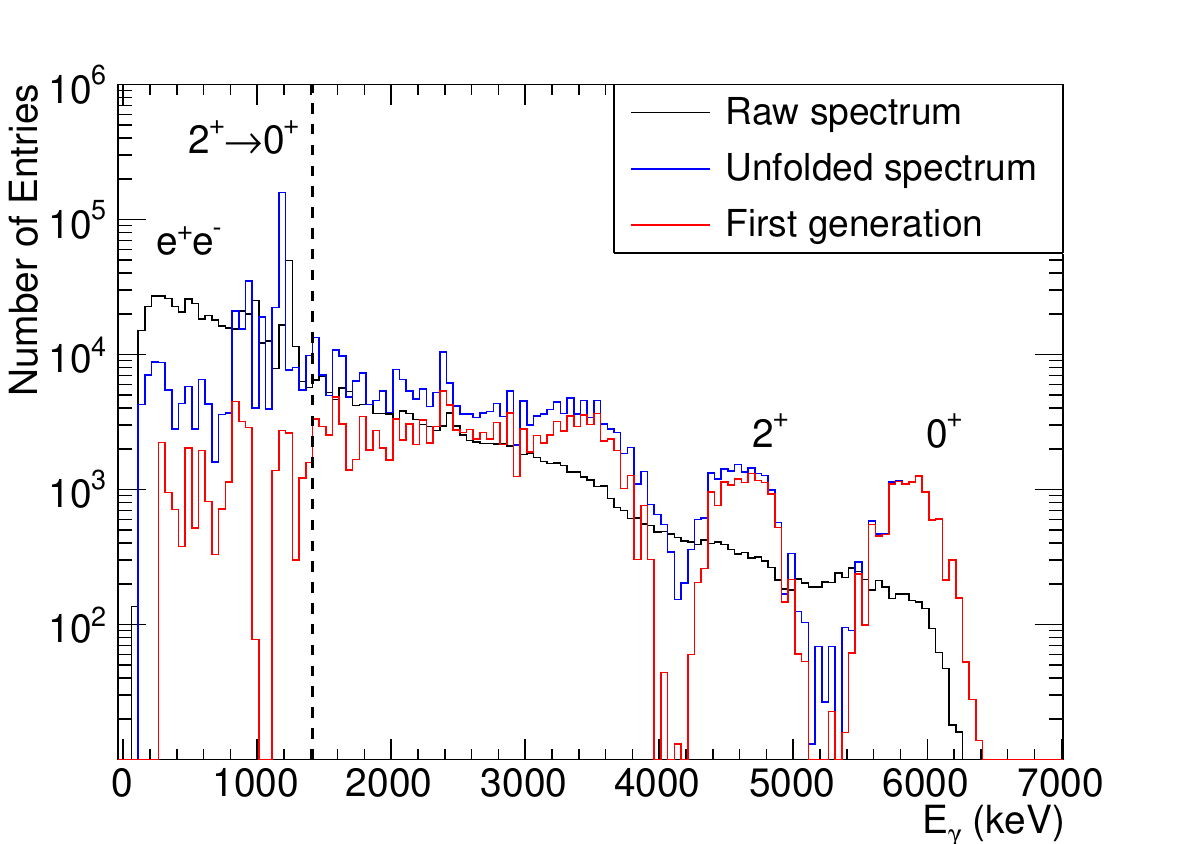}
\end{center}
\caption{Projected raw, unfolded, and first-generation spectra from ${}^{112}$Sn in the excitation energy range 5.7-6.2~MeV, highlighting the impact on the transitions to the $0^{+}$ and $2^{+}$ states. The low-energy cut-off for the analysis is marked with a dashed line, below which discrete transitions and the risk of oversubtraction may impact the result. The peak associated with the $2^{+} \to 0^{+}$ transition and the e$^{+}$e$^{-}$ annihilation peak are labelled to highlight the performance of the unfolding and first-generation procedures.}\label{fig:sn112proj}
\end{figure}

To obtain the actual emitted distribution of all $\gamma$-rays emitted in a reaction, $\vec{A}$, we need to remove the detector response from $\vec{S}$. For this purpose, the detector response was simulated in \geant\ \cite{Agostinelli2003} using an in-house developed software, \ac{GROOT} \cite{Lattuada2017}, providing a detector response matrix, $\mathbf{R}$. For details about the simulations, see Reference~\cite{Aogaki2023}.
The raw matrices were unfolded using the iterative unfolding method using a response matrix $\mathbf{R}$, as discussed in Reference~\cite{Guttormsen1996} and evaluated for these particular detectors in Reference~\cite{Soderstrom2019b}. The main algorithm behind this procedure is the successive folding of a starting guess spectrum, $\vec{A}{'}_{i}$, iterating over $i$ and a refolded spectrum $\vec{S}{'}_{i}$. The starting guess was chosen as the spectrum itself $\vec{A}{'}_{0}=\vec{S}$. This gives the first step of the procedure as
\begin{equation}
 \vec{S}{'}_{0} = \mathbf{R}\vec{A}{'}_{0} = \mathbf{R}\vec{S},
\end{equation} 
where the next starting guess was modified as
\begin{equation}
 \vec{A}{'}_{1} = \vec{A}{'}_{0} + \left(\vec{S}-\vec{S}{'}_{0}\right),
\end{equation} 
and the procedure is repeated for each iteration $i$
\begin{equation}
\begin{split}
 \vec{A}{'}_{i} &= \vec{A}{'}_{i-1} + \left(\vec{S}-\vec{S}{'}_{i-1}\right),\\
 \vec{S}{'}_{i} &= \mathbf{R}\vec{A}{'}_{i},
\end{split} 
\end{equation} 
until convergence. Tens of iterations were typically required to reach a Kolmogorov-Smirnof similarity of $1-\alpha_{\mathrm{KS}}=10^{-6}$. The unfolded matrices are shown in Figure~\ref{fig:matrices}, and a specific example of a projected spectrum is shown in Figure~\ref{fig:sn112proj}.
As the unfolding procedure inevitably introduces a small number of positive-negative oscillations in the output, the spectra are partially smoothed by redistributing positive-value bins into negative-value bins within a Gaussian window defined by the detector resolution.

\begin{figure}[ht!]
\begin{center}
\includegraphics[width=\columnwidth]{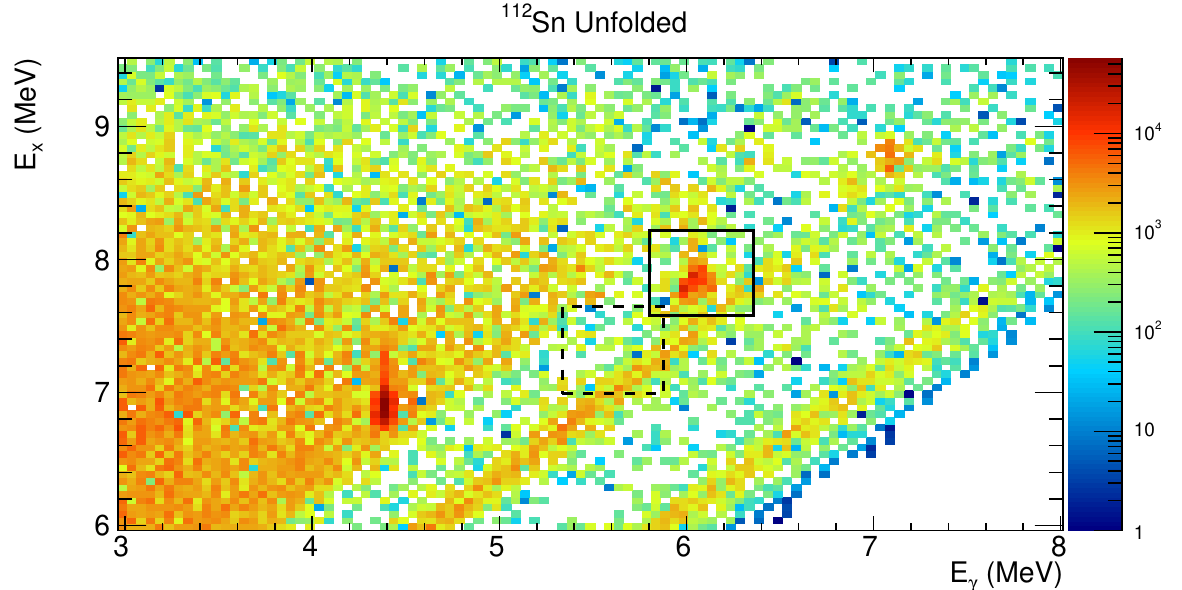}\\
\includegraphics[width=\columnwidth]{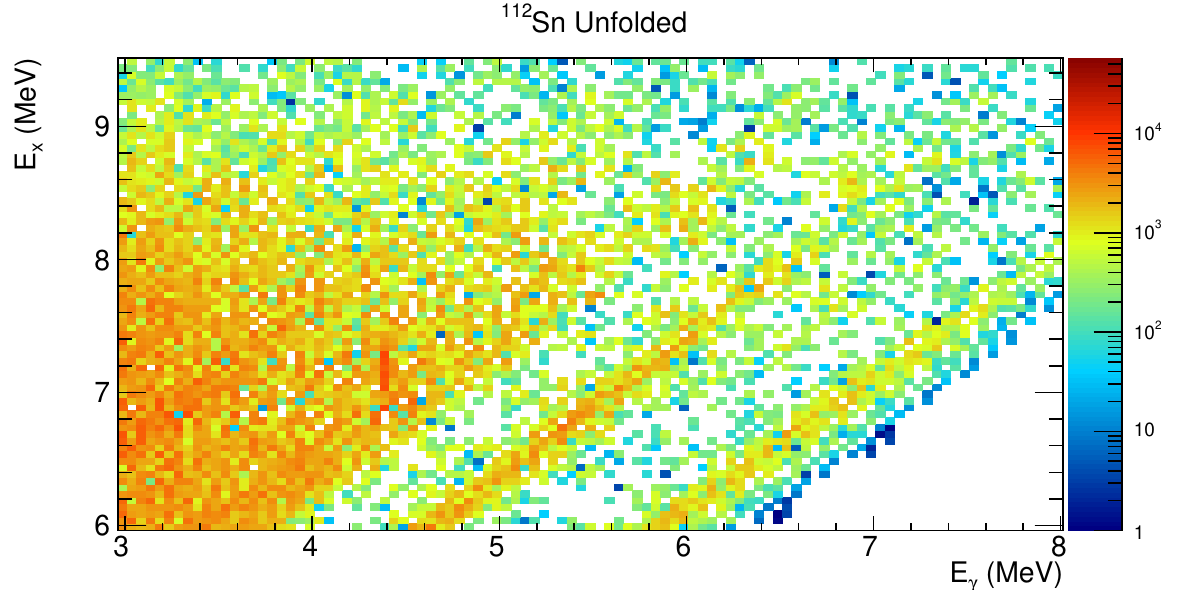}\\
\end{center}
\caption{Part of the unfolded matrix, expanded to highlight the region with contaminants, from ${}^{112}$Sn before (top) and after (bottom) the subtraction of the contaminant peaks. The fitting region for one ${}^{16}$O contaminant is marked with a solid line, and the region used for background estimation is marked with a dashed line.}\label{fig:uf_rd2}
\end{figure}

As seen in the raw matrices, there are three significant contaminants in the measured data that correspond to the presence of trace amounts of ${}^{12}$C and ${}^{16}$O in the target. While these contaminants stretch out over the entire $\gamma$-ray energy range, they become well located at their specific energies in the unfolding procedure. Thus, a simple fit can subtract them from the unfolded matrix. These fits were performed using a Gaussian function along the diagonals of the data, where the underlying background was estimated from the neighbouring region. This procedure is illustrated in Figure~\ref{fig:uf_rd2}.

While the unfolded spectrum contains all the $\gamma$ rays originating from a specific excited state, the strength functions only relate to each cascade's first $\gamma$ ray. These $\gamma$ rays are typically called the first-generation $\gamma$ rays. The spectra need to be separated into the first-generation spectrum, $\vec{F}(E_{\gamma})$, and the spectrum of the later generation of $\gamma$ rays, $\vec{A}(E_{\gamma})-\vec{F}(E_{\gamma})$.

Here, we will outline the procedure to obtain the first generation, closely following the description in Reference~\cite{Larsen2011}, which goes deeper into detail. In general, to obtain $\vec{F}(E_{x})$ from $\vec{A}(E_{x})$, one needs to subtract all the contributions to $\vec{A}(E_{x})$ from all the possible states in the cascade below $E_{x}$, as
\begin{equation}
    \vec{F}(E_{x}) = \vec{A}(E_{x}) - \sum_{i}^{i<x} n_{i}(E_{x}) w_{i}(E_{x})\vec{A}(E_{i}),\label{eq:firstgensub}
\end{equation}
using an appropriate choice of weight vectors $\vec{w}(E_{x})$ and normalisation vectors $\vec{n}(E_{x})$, where each element $n_{i}(E_{x})$ correspond to the relative population cross-section between the state at $E_{x}$ and $E_{i}$ where $(E_{x}-E_{i})=E_{\gamma}$, making sure that the subtraction in Eq.~(\ref{eq:firstgensub}) is always carried out on the same number of cascades. Several methods exist to obtain $\vec{n}$ \cite{Larsen2011}. Here, we have used the multiplicity normalisation. In short, this means that for a given excitation energy, $E_{x}$, the average $\gamma$-ray multiplicity, $\langle M_{\gamma} \rangle$, can be obtained from the average $\gamma$-ray energy, $\langle E_{\gamma} \rangle$, as $\langle M_{\gamma} \rangle = E_{x}/\langle E_{\gamma} \rangle$. Following this, the normalisation between two different excitation energies can be obtained from
\begin{equation}
    n_{i}(E_{x})= \frac{\sum_{j} A_{j}(E_{i})/\langle M_{\gamma} \rangle(E_{i})}{\sum_{j} A_{j}(E_{x})/\langle M_{\gamma} \rangle(E_{x})}.
\end{equation}
This can be easily understood as that the total number of $\gamma$ rays at a given energy divided by the average $\gamma$-ray multiplicity equals the total number of excitations at that energy, and the ratio between the total number of excitations at two different energies correspond to the ratio of the population cross sections.
The weight function $\vec{w}(E_{x})$ is simply the normalised probability to decay from the excited state $E_{x}$ to $E_{i}$,
\begin{equation}
    \vec{w}(E_{x}) = \frac{\vec{F}(E_{x})}{\sum_{j}F_{j}(E_{x})},
\end{equation}
which directly contains the $\vec{F}(E_{x})$ we are looking for. Thus, we can use an appropriate starting guess of $\vec{w}(E_{x})$ and iteratively solve Equation ~(\ref{eq:firstgensub}). The resulting first-generation matrices are shown in Figure~\ref{fig:matrices}, and a specific example of a projected spectrum is shown in Figure~\ref{fig:sn112proj}.

Note that, here, a critical assumption has been made: the decay of the state is independent of the formation mechanism, such that the decay of a set of states with an energy $E_{x}$ is the same if they were populated directly by $(\mathrm{p},\mathrm{p}')$ or by $\gamma$-ray feeding from above. For a discussion on this assumption, see Reference~\cite{Larsen2011}.

The area in the $E_{\gamma}$ and $E_{x}$ matrix selected for further analysis is shown in Figure~\ref{fig:matrices}. The lower and higher limits of $E_{x}$ were selected to contain the quasi-continuum region above the discrete excited states, where we assume the criterion discussed above is expected to hold but below the energy where the neutron-separation threshold starts to reduce the statistics. The lower energy limit of $E_{\gamma}$ was selected such that artefacts from under- and over-subtracting low-energy transitions will be minimal, further illustrated in Figure~\ref{fig:sn112proj}.

\subsection{The Oslo method}

\begin{figure}[ht!]
\begin{center}
\includegraphics[width=\columnwidth]{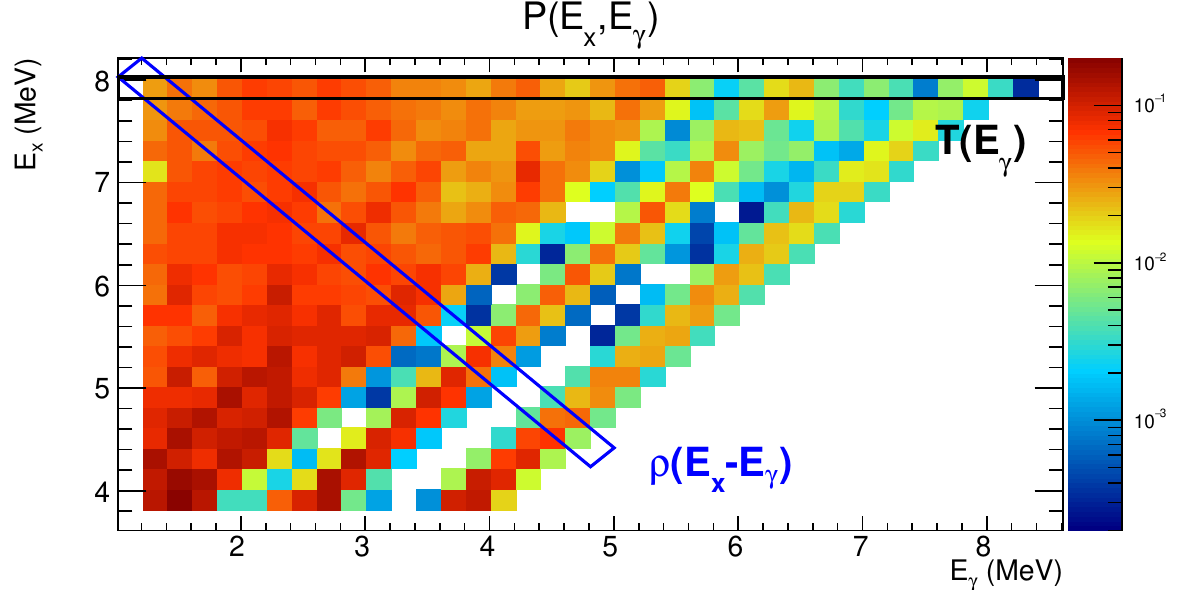}\\
\includegraphics[width=\columnwidth]{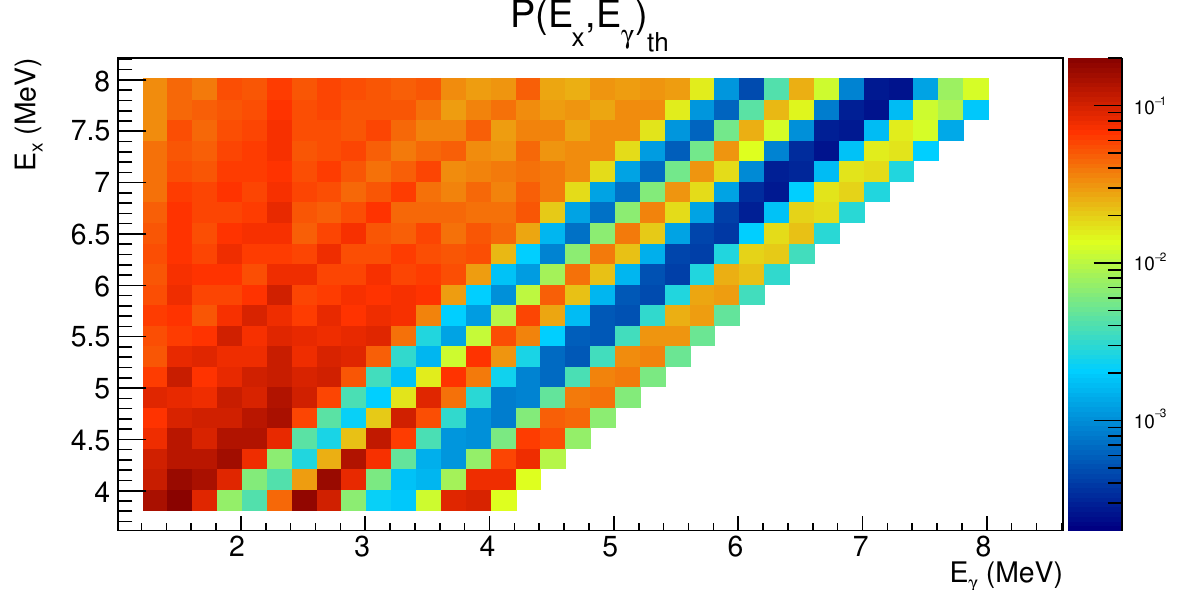}\\
\includegraphics[width=\columnwidth]{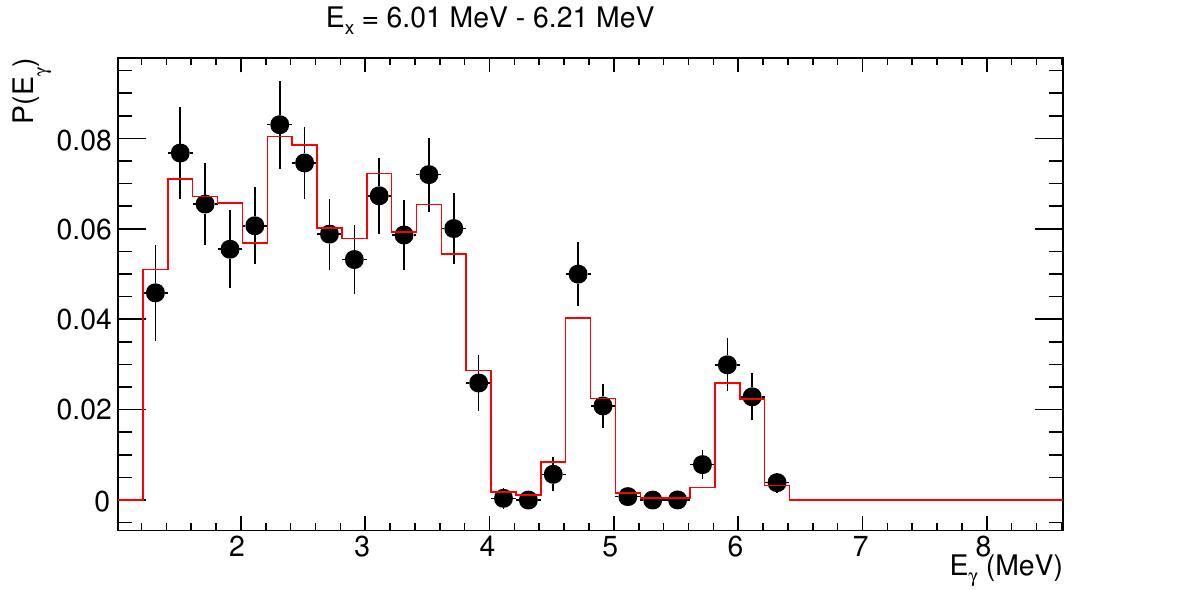}
\end{center}
\caption{Experimental (top), best fit (middle), and projection (bottom) of the probability matrix. The longest consecutive projections of $\rho(E_{x}-E_{\gamma})$ and $\mathcal{T}(E_{\gamma})$ have been highlighted in the experimental matrix. The number of data bins in these regions corresponds to the number of free parameters for calculating the reduced $\chi^{2}$. The bottom panel shows the projection of the best fit (red line) compared with the experimental data (black circles) in the energy range 6-6.2~MeV. The error bars include both statistical uncertainties and systematic uncertainties evaluated for the reduced $\chi^{2}/\mathrm{NDF} = 1$.}\label{fig:pexeg}
\end{figure}

The process of extracting the level densities and the gamma strength functions from the first-generation matrix is well described in References~\cite{Guttormsen1987, Guttormsen1996,Schiller2000, Larsen2011}. Here, we will provide a brief outline, but refer to the previous references for an in-depth description. If we denote the normalised first-generation matrix as $P(E_{x},E_{\gamma})$, then the element with an excitation energy $E_{x}$ corresponds to the probability for a nucleus with that excitation energy to decay with a $\gamma$-ray energy of $E_{\gamma}$. If we know the level densities, $\rho(E_{x})$, and the transition probabilities, $\mathcal{T}(E_{\gamma})$, this matrix can be calculated from
\begin{equation}    P(E_{x},E_{\gamma})_{\mathrm{th}}=\frac{\rho(E_{x}-E_{\gamma})\mathcal{T}(E_{\gamma})}{\sum_{E_{\gamma}}\rho(E_{x}-E_{\gamma})\mathcal{T}(E_{\gamma})},\label{eq:pexeg}
\end{equation}
and given any combination of $\rho(E_{x})$ and $\mathcal{T}(E_{\gamma})$, the probability $P(E_{x},E_{\gamma})_{\mathrm{th}}$ can be compared with the measured $P(E_{x},E_{\gamma})$ and the best solutions can be selected from a $\chi^{2}$ test. A comparison between the experimental first-generation matrix and the best fit is shown in Figure~\ref{fig:pexeg}. At this point, we evaluate the systematic uncertainties from the reduced $\chi^{2}$ by taking the \ac{NDF} into account such that $\chi^{2}/\mathrm{NDF} = 1$ and also include these uncertainties in the analysis in addition to the statistical uncertainties.
Here, one critical assumption is being introduced: that the generalised Brink-Axel hypothesis \cite{Brink1955, Axel1962} is valid. This hypothesis states that the \acp{gSF}, directly proportional to the transmission coefficients, are independent of the excitation energy in the statistical region and only depend on the energy difference between the initial and final states. This hypothesis has been tested recently for ${}^{116,120,124}$Sn \cite{Markova2021}.

\begin{figure*}[ht!]
\begin{center}
\includegraphics[width=0.49\textwidth]{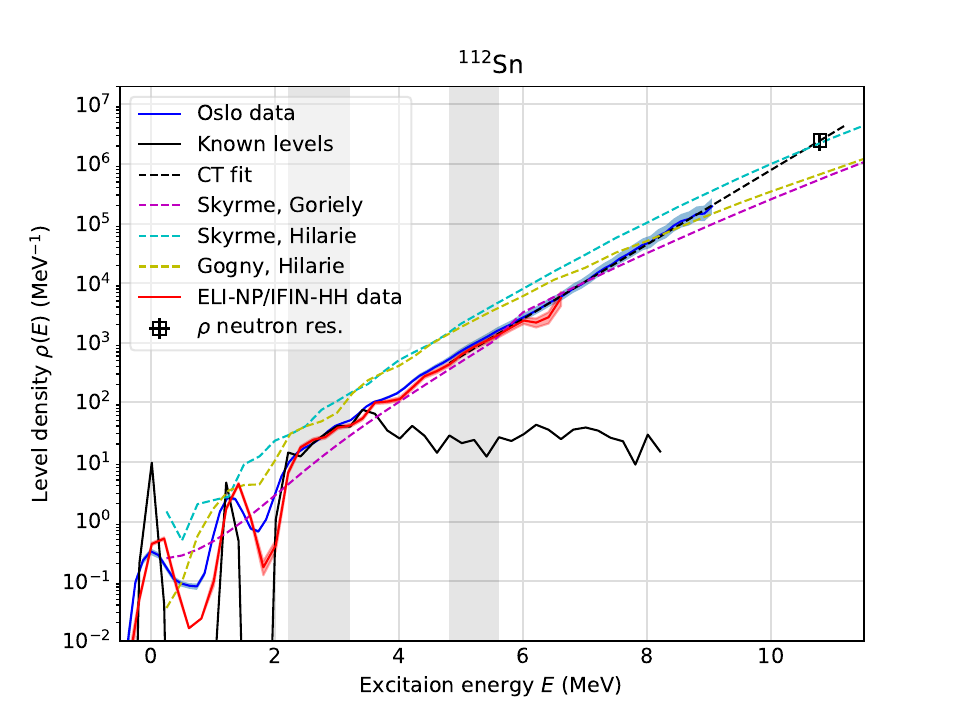}
\includegraphics[width=0.49\textwidth]{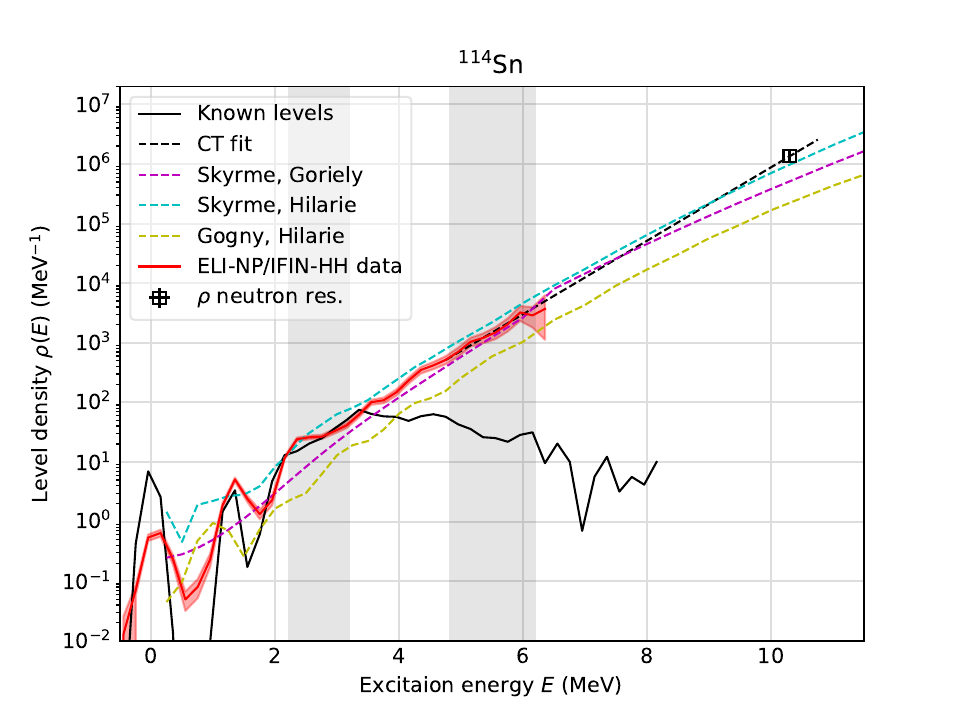}
\end{center}
\caption{Nuclear level density of ${}^{112}$Sn (left) and ${}^{114}$Sn (right) using a constant-temperature (CT) model fit compared with the results on ${}^{112}$Sn from the Oslo group \cite{Markova2023}. The left shaded region shows the range for fitting to known spectroscopic states, and the right shaded region shows the range for fitting with the CT model. The transparent bands correspond to the statistical and systematic uncertainties. Three microscopic models implemented in \talys\ are shown for comparison.}\label{fig:sn112ld}
\end{figure*}

While there are an infinite number of solutions that satisfy Equation ~(\ref{eq:pexeg}). 
It has been shown \cite{Schiller2000} that if one solution of Equation ~(\ref{eq:pexeg}) is found, one can generally construct all possible solutions. Furthermore, it has been mathematically shown in Appendix~A of Reference~\cite{Schiller2000} that these solutions must be related via a differential equation, with an exponential solution, and that these solutions, $\tilde{\rho}(E_{x})$ and $\tilde{\mathcal{T}}(E_{\gamma})$, only depend on three parameters as
\begin{align}
    \tilde{\rho}(E_{x}-E_{\gamma}) & = A_{0} \exp[\alpha(E_{x}-E_{\gamma})]\rho(E_{x}-E_{\gamma}),\label{eq:rhotilde}\\
    \tilde{\mathcal{T}}(E_{\gamma}) & = B_{0} \exp(\alpha E_{\gamma})\mathcal{T}(E_{\gamma})\label{eq:Ttilde}.
\end{align}
These parameters, $A_{0}$, $B_{0}$, and $\alpha$, can be constrained by complementary information available and define unique, physical solutions for the level densities and \acp{gSF}. The most common complementary data that is typically used to constrain equations (\ref{eq:rhotilde}) and (\ref{eq:Ttilde}) are from neutron-capture experiments using $s$-wave neutrons on the $A-1$ isotopes. In particular, the level density at the neutron threshold, $\rho(S_{\mathrm{n}})$ can be estimated from the neutron resonance spacings $D_{0}$, as
\begin{widetext}
\begin{equation}
 \rho(S_{\mathrm{n}}) = \frac{2\sigma^{2}(S_{\mathrm{n}})}{D_{0}} \left[ (J_{\mathrm{t}}+1) \exp \left( -\frac{(J_{\mathrm{t}}+1)^{2}}{2\sigma^{2}(S_{\mathrm{n}})} \right) + J_{\mathrm{t}} \exp \left( -\frac{(J_{\mathrm{t}})^{2}}{2\sigma^{2}(S_{\mathrm{n}})} \right)\right]^{-1}.   
\end{equation}
\end{widetext}
Here, $J_{\mathrm{t}}$ is the spin of the target for neutron capture, in the case of ${}^{111}$Sn $(J_{\mathrm{t}}=7/2)$ and in the case of ${}^{113}$Sn $(J_{\mathrm{t}}=1/2)$. The spin cut-off parameter, $\sigma^{2}(S_{\mathrm{n}})$, was obtained from Reference~\cite{Gilbert1965}, as
\begin{equation}
    \sigma^{2}(S_{\mathrm{n}}) = 0.0888 a \sqrt{\frac{S_{\mathrm{n}}-E_{1}}{a}} A^{2/3},
\end{equation}
where $E_{1}$ is an energy back-shift parameter and $a$ a level density parameter, with $a^{-1}$ proportional to the single-particle level spacing. See the discussion in Reference~\cite{Markova2023} for details on these relations. 
\begin{table*}[ht]
\caption{Parameters used in the Oslo method analysis for ${}^{112}$Sn and ${}^{114}$Sn. $S_{\mathrm{n}}$ corresponds to the neutron separation energy, $D_{0}$ the neutron resonance spacing at the neutron threshold for $s$-wave neutrons \cite{Mughabghab2018}, $a$ is the level density parameter at the neutron separation energy, $E_{1}$ is the Fermi-gas shift parameter, 
$\sigma(S_{\mathrm{n}})$ the spin cut-off at the neutron threshold, $T$ the temperature parameter, $E_{0}$ the constant temperature shift parameter, and $\langle \Gamma_{\gamma} \rangle$ the average $\gamma$-ray decay width at the neutron threshold \cite{Mughabghab2018}. Note that the $D_{0}$ and $\rho(S_{\mathrm{n}})$ parameters for ${}^{112}$Sn are extracted from systematics as described in Reference~\cite{Markova2023}, and $\langle \Gamma_{\gamma} \rangle$ is the reduced value as extracted in the same reference.
\label{tab:gamma_det_table}}
 \begin{tabular*}{\textwidth}{@{\extracolsep{\fill}}lccccccccc}
\hline
  & $S_{\mathrm{n}}$ & $D_{0}$ & $a$ & $E_{1}$ & 
$\sigma(S_{\mathrm{n}})$ & $\rho(S_{\mathrm{n}})$ & $T$ & $E_{0}$ & $\langle \Gamma_{\gamma} \rangle$ \\
  & (MeV) & (eV) & (MeV$^{-1}$) & (MeV) & 
& (MeV$^{-1}$) & (MeV) & (MeV) & (meV) \\
\hline											
${}^{112}$Sn & 10.788 & 3.32 & $12.53$ & $1.12$ & 
4.765 & $2.5(8) \times 10^{6}$ & 0.695 & 0.813 & 87.42 \\
${}^{114}$Sn & 10.3029  & 17.7 & 12.05 & 1.12 & 
4.676 & $1.35(29) \times 10^{6}$ & 0.705 & 0.598 & 133 \\
\hline																			
 \end{tabular*} 
\end{table*}
The $A_{0}$ and $\alpha$ can be determined from the complete spectroscopy of low-lying states and the systematic resonance density at the neutron threshold. While fitting to the first of these is straightforward, our data stops well before the neutron threshold, and we need to perform an extrapolation to these energies, which will introduce some model dependency into the results from here on. We have decided to keep with the constant temperature model \cite{Gilbert1965},
\begin{equation}
  \rho(E_{x}) = \frac{1}{T} \exp \left( \frac{E_{x}-E_{0}}{T}\right),
\end{equation}
with the parameters $T$ being the nuclear temperature, and $E_{0}$ being the constant temperature energy shift,
for consistency between this data and previous measurements in the Sn region \cite{Markova2021, Markova2022, Markova2023, Markova2024}.
The results of these fits are shown in Figures~\ref{fig:sn112ld}, and the parameters used for normalisation are listed in Table~\ref{tab:gamma_det_table}.

\begin{figure*}[ht!]
\begin{center}
\includegraphics[width=\columnwidth]{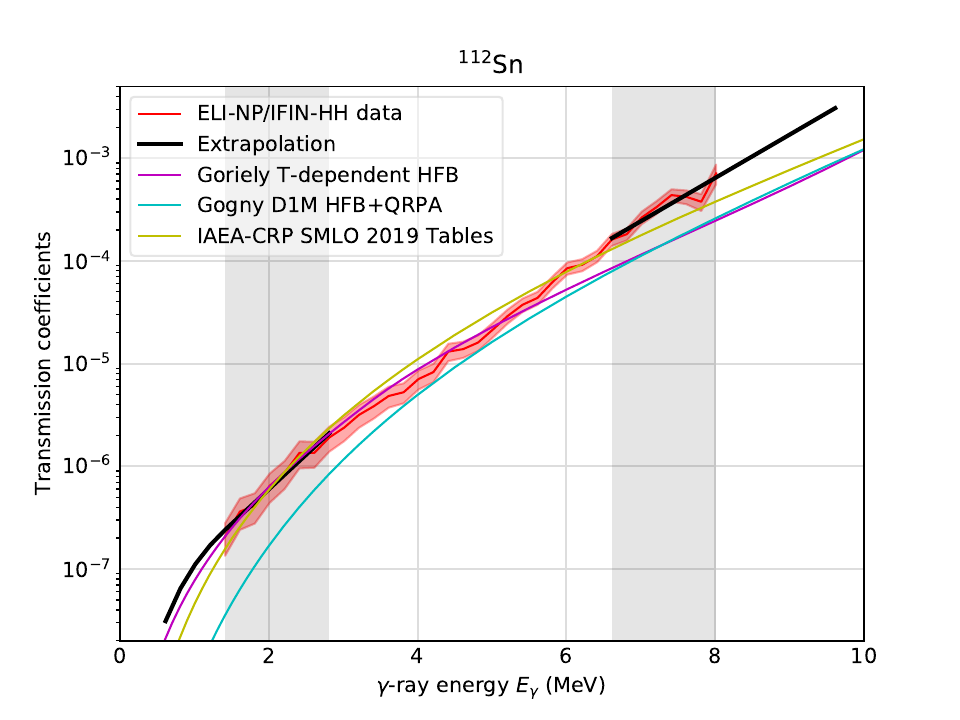}
\includegraphics[width=\columnwidth]{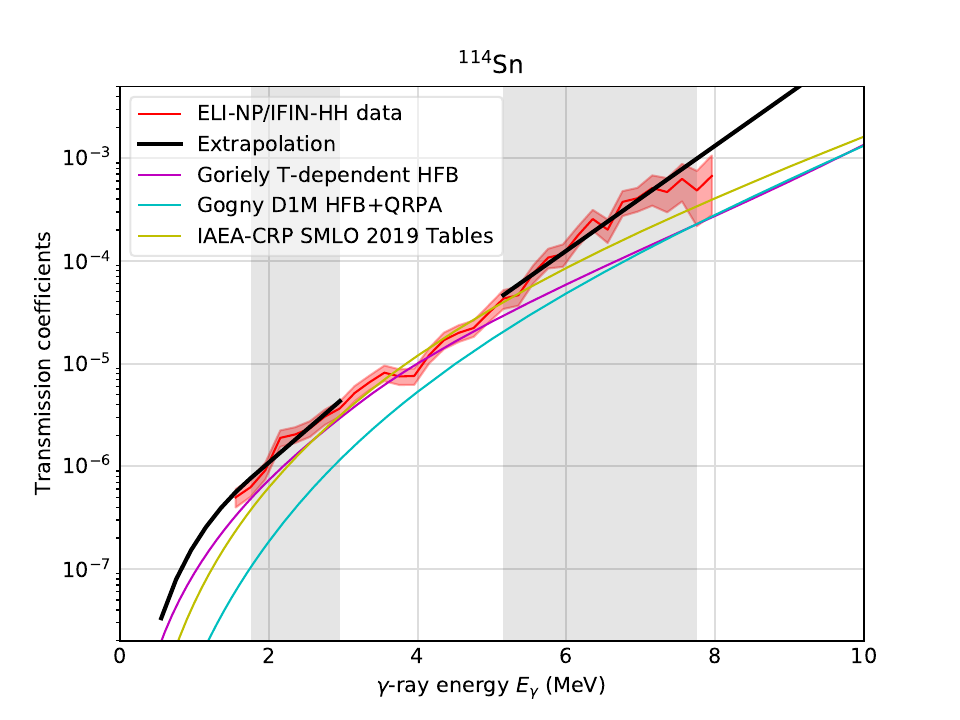}
\end{center}
\caption{Transmission coefficients as a function of $\gamma$-ray energy for ${}^{112}$Sn (left) and ${}^{114}$Sn (right). The shaded regions show the regions for the low-energy and high-energy fits to the data, and the solid lines show the extrapolation to high and low energies. For comparison, three typical parametrisations as implemented in the \talys\ code are shown}\label{fig:sn112trans}
\end{figure*}

The average experimental total radiative width, $\langle \Gamma_{\gamma} (S_{\mathrm{n}})\rangle$, can be used to normalize $B_{0}$ parameter of the \acp{gSF}, equation~(\ref{eq:Ttilde}).  For the neutron capture we need to take the spin distribution of the $A-1$ nucleus into account, as both ${}^{112}$Sn and ${}^{114}$Sn has their spin-parity as $J^{\pi}=0^{+}$ in their ground states, while ${}^{111}$Sn and ${}^{113}$Sn have $J^{\pi}=7/2^{+}$ and $J^{\pi}=1/2^{+}$, respectively. In this normalisation procedure, we do not make any difference between positive-parity and negative-parity levels but assume that they are distributed with equal density in the statistical regime or that the level densities of positive parity states and negative parity states are exactly half of the total level density, respectively \cite{Larsen2011}. Given these assumptions of dipole-dominated radiation and of parity-independent level density, the expression for the radiative decay width from Reference~\cite{Kopecky1990} can be simplified \cite{Larsen2011} into 
\begin{widetext}
\begin{equation}
    \langle \Gamma_{\gamma} \rangle = \frac{1}{2} \frac{1}{\rho(S_{\mathrm{n}},J_{\mathrm{t}}\pm1/2)} \int_{E_{\gamma}=0}^{S_{\mathrm{n}}} E_{\gamma}^{3} f(E_{\gamma})\rho(S_{\mathrm{n}}- E_{\gamma}) \mathrm{d} E_{\gamma} \sum_{J=-1}^{1} g(S_{\mathrm{n}}- E_{\gamma},J_{\mathrm{t}}\pm1/2+J),
\end{equation}
\end{widetext}
where the spin distribution function $g(E,J)$ is \cite{Gilbert1965,vonEgidy2005,Larsen2011}
\begin{equation}
    g(E,J) \approx \frac{2J+1}{2 \sigma^{2}} \exp\left[ -\frac{(J+1/2)^{2}}{2\sigma^{2}}\right].
\end{equation}
In order to make the transmission coefficients $\mathcal{T}$ continuous in the full range from zero to the neutron separation energy, the average decay width at the neutron threshold also needs to be extrapolated from the highest and lowest $\gamma$-ray energy data. This extrapolation was done by fitting an exponential function at high energies, $\mathcal{T}=\exp(aE_{\gamma}+b)$, as well as to extrapolate the low-energy region towards $E_{\gamma}=0$, as $\mathcal{T}=E_{\gamma}^{3}\exp(cE_{\gamma}+d)$, as described in Reference~\cite{Larsen2011}, and shown in Figure~\ref{fig:sn112trans}.
From here, the conversion from the transition probabilities, $\mathcal{T}$ to the \ac{gSF}, $f_{XL}$ is via the relation,
\begin{equation}
    f_{XL}(E_{\gamma}) = \frac{\mathcal{T}(E_{\gamma})}{2 \pi E_{\gamma}^{2L+1}},\label{eq:f2T}
\end{equation}
where we, in the remaining part of this paper, will assume complete dipole domination, thus, $2L+1=3$.

Thus, with these relations, we have information to completely constrain the parameters of Equations (\ref{eq:rhotilde}) and (\ref{eq:Ttilde}).

\begin{figure*}[ht!]
\begin{center}
\includegraphics[width=0.49\textwidth]{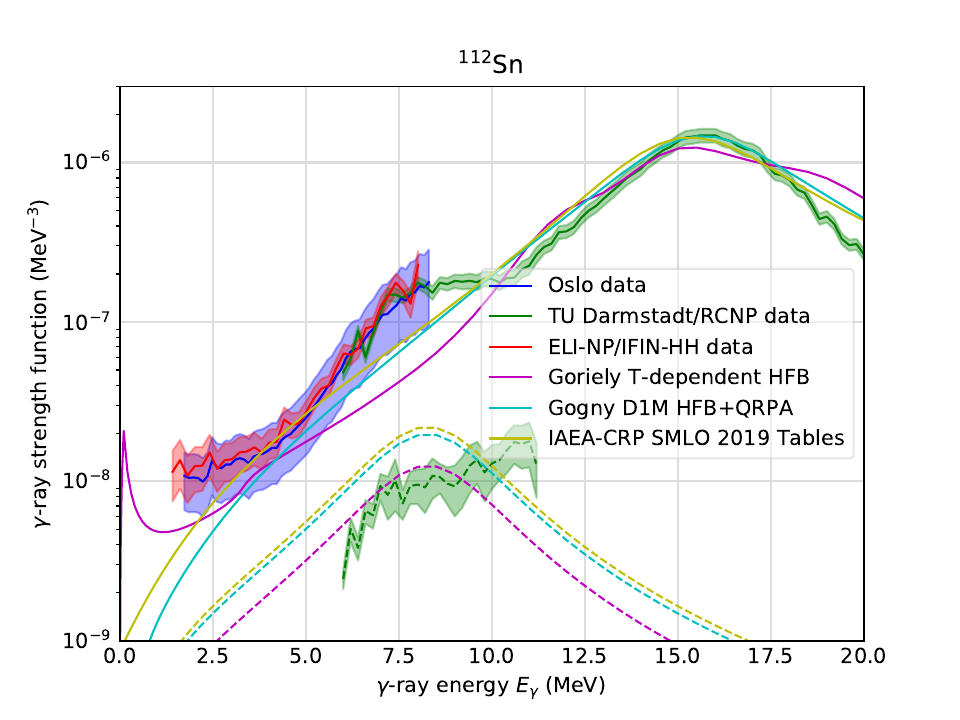}
\includegraphics[width=0.49\textwidth]{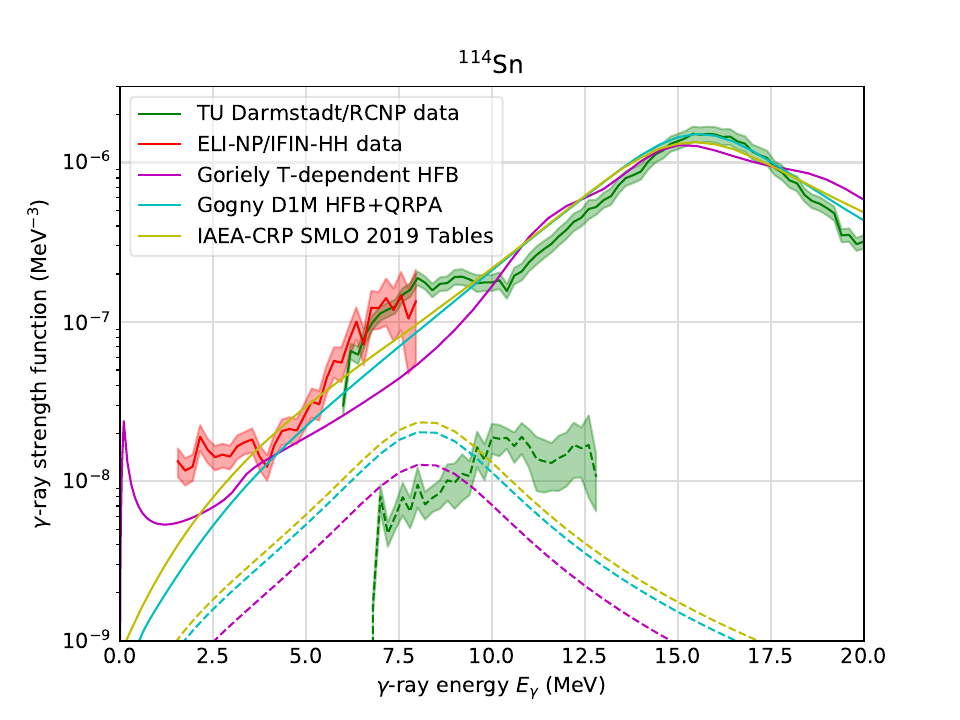}
\end{center}
\caption{Experimental $\gamma$-ray strength functions of ${}^{112}$Sn (left) and ${}^{114}$Sn (right) obtained from this work (ELI-NP/IFIN-HH), Reference~\cite{Markova2023} (Oslo), and Reference~\cite{Bassauer2020} (TU Darmstadt/RCNP). The ELI-NP/IFIN-HH and Oslo data represent the total dipole strength function, and the TU Darmstadt/RCNP data is separated into the electric and magnetic dipole strengths. For comparison, three typical parametrisations implemented in the \talys\ code are shown as solid lines for electric dipole strength and dashed lines for magnetic dipole strength.}\label{fig:sn112gsf}
\end{figure*}

To verify the results from these fits, the low-energy \acp{gSF} for ${}^{112}$Sn and ${}^{114}$Sn are compared with the high-energy total excitation strength obtained from high-energy $(\mathrm{p},\mathrm{p}')$ scattering at very forward angles performed at the \ac{RCNP}, Osaka University, and analysed and published by \ac{TUD} \cite{Bassauer2020}. As shown in Figure~\ref{fig:sn112gsf}, the overlap in terms of absolute values for both ${}^{112}$Sn and ${}^{114}$Sn between the different data sets; our data, the Oslo data, and the \ac{TUD} data are very close, suggesting that the normalisation procedures are sound. Due to the rapid cross-section variation at small angles, the zero-degree type scattering makes it possible to perform \ac{MDA} on the \ac{RCNP} data \cite{vonNeumann-Cosel2019a}, to separate the E1 and M1 strength, showing the significant electric dominance of the structures in the \ac{LEDR} region. For comparison, we also include a comparison to some of the frequently used global parametrisations of the \acp{gSF}, especially for systematic astrophysical rate calculations and similar, as implemented in the \talys\ \cite{Koning2008, Koning2012}  code. While several different \ac{gSF} models exist to choose from in this code, we have here limited the comparison to some recommended models. These \ac{gSF} models include two microscopic approaches via the temperature dependent \ac{HFB} by Goriely,  and \ac{HFB} and \ac{QRPA} calculations with the Gogny D1M interaction. In addition, the recent \ac{CRP} by the \ac{IAEA} on photoneutron reaction cross-sections has provided recommended parametrisations of the \ac{GDR} using the \ac{SMLO} function \cite{Kawano2020, Goriely2019b}, and these parametrisations are also included for comparison.

\section{Discussion} \label{sec:discussion}

Here, we will go through the implications of the measured \acp{gSF} for how to interpret the behaviour of nuclear matter in the specific cases of ${}^{112}$Sn and of ${}^{114}$Sn. We will investigate these results in three different approaches: the general gross properties of the \acp{gSF}, in particular concerning the \ac{LEDR}, the microscopic picture of the \ac{LEDR} and how the different properties of the strength can be interpreted from a fundamental picture, and the impact of the measured \acp{gSF} in terms of nucleosynthesis and the astrophysical reaction rates for $p$ nuclei.

\subsection{Data comparison}

\begin{figure*}[ht!]
\begin{center}
\includegraphics[width=0.49\textwidth]{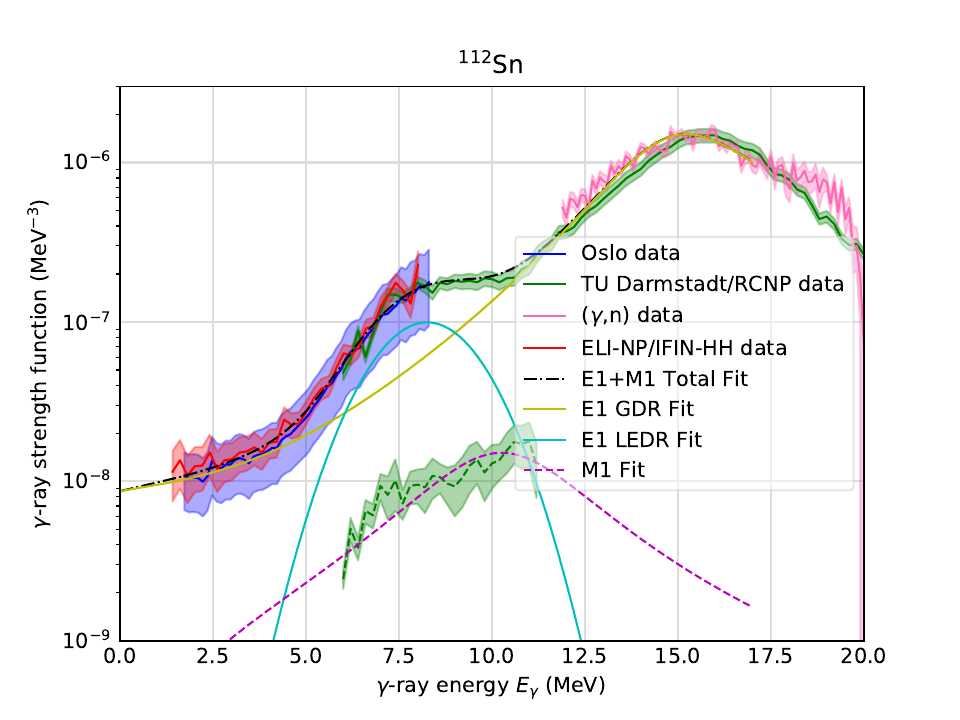}
\includegraphics[width=0.49\textwidth]{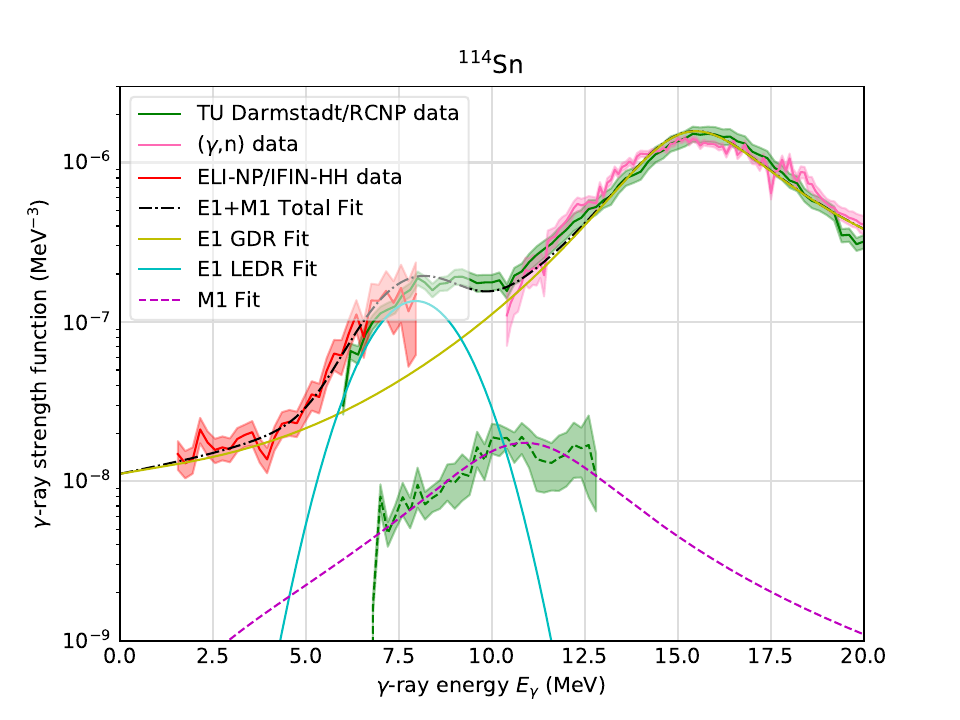}
\end{center}
\caption{Experimental $\gamma$-ray strength functions as in Figure~\ref{fig:sn112gsf}, including also the evaluated $(\gamma,\mathrm{n})$ data from Reference~\cite{Varlamov2010}. The details of the fitted functions to parameterise the data are described in detail in the text and Table~\ref{tab:fit_table}. The TU Darmstadt/RCNP data have been separated into E1 and M1 components for these fits, while the Oslo and ELI-NP/IFIN-HH data have been included in the fit as a sum of E1+M1.}\label{fig:sn_fits}
\end{figure*}

With the gross properties of the \acp{GDR}, in general, having been thoroughly investigated in the past, it is particularly interesting to investigate the systematics of the additional \ac{LEDR} from the extracted \acp{gSF}. 
To extract the gross features of the \acp{gSF}, we fitted the available data simultaneously using a combination of functions to describe the different components of the total dipole strength function. To describe the \ac{GDR} part of the strength function, an enhanced \ac{GLO} function was adopted as
\begin{widetext}
\begin{equation}
    \mathcal{F}_{\mathrm{GDR}} = \frac{\mathcal{S}_{\mathrm{GDR}}}{3 (\pi \hbar c)^{2}} = \frac{1}{3 (\pi \hbar c)^{2}} \sigma_{\mathrm{GDR}} \Gamma_{\mathrm{GDR}} \left[ E_{\gamma} \frac{\Gamma_{\mathrm{KMF}}(E_{\gamma},T_{\mathrm{f}})}{(E_{\gamma}^{2}-E_{\mathrm{GDR}}^{2})^{2}+E_{\gamma}^{2}\Gamma_{\mathrm{KMF}}^{2}(E_{\gamma},T_{\mathrm{f}})}+0.7\frac{\Gamma_{\mathrm{KMF}}(E_{\gamma},T_{\mathrm{f}}=0)}{E_{\gamma}^{3}}\right]
\end{equation}
\end{widetext}
with $\Gamma_{\mathrm{KMF}}$ being the \ac{KMF} width \cite{Kadmenskii1983}, defined as
\begin{equation}
    \Gamma_{\mathrm{KMF}} = \frac{\Gamma_{\mathrm{GDR}}}{E_{\mathrm{GDR}}^{2}}\left( E_{\gamma}^{2}-4\pi^{2}T_{\mathrm{f}}^{2}\right),
\end{equation}
where $T_{\mathrm{f}}$ represents the temperature of the final states. As the methods used in extracting these data do not differentiate between E1 and M1 transition strength, the M1 strength has been explicitly included from \ac{RCNP} data from Reference~\cite{Bassauer2020} where it has been extracted using the \ac{MDA} methods. The M1 strength was modelled as a simple Lorentzian-type distribution, namely 
\begin{equation}
    \mathcal{S}_{\mathrm{M1}} = \frac{1}{3 (\pi \hbar c)^{2}} \sigma_{\mathrm{M1}}  \frac{E_{\gamma} \Gamma_{\mathrm{M1}}^{2}}{(E_{\gamma}^{2}-E_{\mathrm{M1}}^{2})^{2}+E_{\gamma}^{2}\Gamma_{\mathrm{M1}}^{2}}.
\end{equation}
The \ac{LEDR} was included as a Gaussian function with centroid energy of $E_{\mathrm{LEDR}}$, a Gaussian width of $W_{\mathrm{LEDR}}$, and a total integral strength of $\sigma_{\mathrm{LEDR}}$. As the exact nature of the \ac{LEDR} is still an open topic, this choice was made to provide the simplest, symmetrical function that reproduces the experimental data best. In addition, a similar parametrization was used in the previous analyses of the Oslo data \cite{Markova2024}, and it is, thus, suitable for a systematic comparison of the new $^{112,114}$Sn and the earlier published data. The total fits are shown in Figure~\ref{fig:sn_fits}.

In the comparison of these data, we, for completeness, also include the $(\gamma,\mathrm{n})$ data compiled in Reference~\cite{Varlamov2010}. This compilation is based on the only direct experimental data available for ${}^{112}$Sn and ${}^{114}$Sn, and was obtained from Bremsstrahlung experiments \cite{Sorokin1972, Sorokin1975},  using BF$_{3}$ detectors. We want to stress that in the ${}^{114}$Sn case, the target had an isotopic purity of only 65.1\% and the full cross-section curves were not extracted in the original publications but as derived data in Reference~\cite{Varlamov2010}, based on the theoretical modelling of cross-section ratios. The agreement, however, is very good with the $(\mathrm{p},\mathrm{p}')$ data from \ac{RCNP} in Reference~\cite{Bassauer2020}. Here, we use the conversion between photonuclear cross-section and \ac{gSF} as
\begin{equation}
    f_{\mathrm{E1/M1}}(E_{\gamma}) = 0.1\frac{\sigma_{\mathrm{E1/M1}}}{3E(\pi \hbar c)^{2}}
\end{equation}
where the extra factor of 0.1 is from converting from fm$^{2}$ to mb.
In the case of M1 transitions, the cross section, $\sigma_{\mathrm{M1}}$, is related to the microscopic strength function, $S_{\mathrm{M1}}$, as \cite{Loens2012} 
\begin{equation}
    \sigma_{\mathrm{M1}}(E_{\gamma}) = \frac{16 \pi^{3}  E}{9 \hbar c} S_{\mathrm{M1}}(E_{\gamma}).
\end{equation}
The minimisation was done using the $\chi^{2}$ method with the Minuit package from the ROOT \cite{Brun1997, Antcheva2009} framework and the best parameters from the fits are listed in Table~\ref{tab:fit_table}.
\begin{table*}[ht]
\caption{Fit parameters for the $\gamma$-ray strength functions where $E$, $\Gamma$, and $\sigma$ correspond to the Lorentzian mean energy, width, and cross-section, respectively, of the GDR and the M1 component. The temperature parameter of the KMF modification of the Lorentzian is listed as $T_{\mathrm{f}}$, and the LEDR strength is described as a Gaussian function with centroid $E_{\mathrm{LEDR}}$, width $W_{\mathrm{LEDR}}$, and cross-section $\sigma_{\mathrm{LEDR}}$. The last column lists the fraction of the TRK sum rule exhausted by the LEDR strength. These fits include only the present results, the OCL data and the RCNP/TU~Darmstadt data.
\label{tab:fit_table}}
 \begin{tabular*}{\textwidth}{@{\extracolsep{\fill}}lccccccccccc}
\hline
  & $E_{\mathrm{GDR}}$ & $\Gamma_{\mathrm{GDR}}$ & $\sigma_{\mathrm{GDR}}$ & $T_{\mathrm{f}}$ & $E_{\mathrm{M1}}$ & $\Gamma_{\mathrm{M1}}$ & $\sigma_{\mathrm{M1}}$ & $E_{\mathrm{LEDR}}$ & $W_{\mathrm{LEDR}}$ & $\sigma_{\mathrm{LEDR}}$ & TRK \\
  & (MeV) & (MeV) & (mb) & (MeV) & (MeV) & (MeV) & (mb) & (MeV) & (MeV) & (mb) & (\%) \\
\hline											
${}^{112}$Sn \cite{Markova2024} & 16.14(9) & 5.46(31) & 265.9(95) & 0.70(5) & 10.45(43) & 4.77(53) & 1.77(21) & 8.24(9) & 1.22(8) & 3.65(27) & 1.81(15) \\
${}^{112}$Sn & 16.18(12) & 5.30(12) & 279(12) & 0.718(20) &  10.44(11) & 4.76(17) & 1.77(8) & 8.32(8) & 1.39(6) & 4.2(4) & 2.08(25) \\
${}^{114}$Sn & 15.81(12) & 3.81(8) & 318(14) & 0.941(23) &  11.15(12) & 4.94(25) & 2.22(10) & 8.11(11) & 1.20(7) & 4.8(6) & 2.15(29) \\
\hline																										 \end{tabular*} 
\end{table*}

\begin{figure}[ht!]
\begin{center}
\includegraphics[width=\columnwidth]{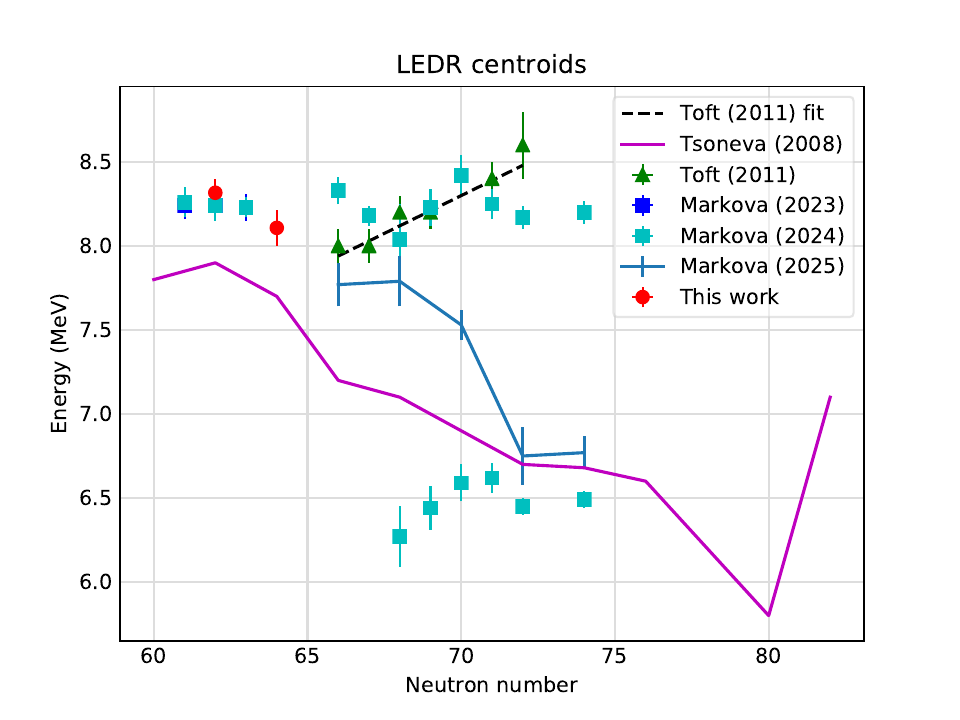}
\end{center}
\caption{Systematics of the LEDR centroids as obtained from Reference~\cite{Toft2011} (${}^{116-119,121,112}$Sn), Reference~\cite{Markova2023} (${}^{111-113}$Sn), Reference~\cite{Markova2024} (${}^{111-113,116-122,124}$Sn), and this work(${}^{112,114}$Sn). The empirical fit from Reference~\cite{Toft2011} is shown as a dashed line, the QRPA calculations from \cite{Tsoneva2008} and the RQTBA calculations from \cite{Markova2025} are shown as a solid lines.}\label{fig:pdr_systematics}
\end{figure}

For both ${}^{112}$Sn and ${}^{114}$Sn, the distribution could be described with a single Gaussian using only the \ac{RCNP} \cite{Bassauer2020}, Oslo \cite{Markova2023} and present data, with similar results as reported in Reference~\cite{Markova2024}. We omitted the $(\gamma,\mathrm{n})$ data in these fits but show them also for comparison, as the data in Reference~\cite{Varlamov2010} are, as discussed above, not direct measurements but based on derived data.
To put the evolution of the \ac{LEDR} into a systematic context, we compare the results obtained in this work with those published in References~\cite{Toft2010, Toft2011} and reanalysed in Reference~\cite{Markova2024}. In the first analysis \cite{Toft2010, Toft2011}, the systematics of the heavier tin isotopes showed an increasing trend in the \ac{LEDR} centroids with increasing neutron number. This correlation was, however, barely outside of the quoted uncertainties. One of the more remarkable aspects of those results was, however, that, despite showing just a weak correlation, they went completely against theoretical expectations. For example, the \ac{QRPA} calculations from \cite{Tsoneva2008}, which predicted significantly lower average energy of the \ac{PDR} states and a clear decreasing trend for the energy concerning neutron number. Following these results, new data on ${}^{111,112,113}$Sn were published in Reference~\cite{Markova2023}, which showed very similar values for the centroids, contradicting the picture of a linear trend in \cite{Toft2010, Toft2011}. 
 Instead, a picture of a relatively constant value across the entire tin chain started to emerge, which, however, still did not agree with the relatively significant decrease obtained from the \ac{QRPA}. A re-analysis of the data from \cite{Toft2010, Toft2011} reported in \cite{Markova2024}, however, shows a subtle splitting of the \ac{LEDR} into two separate components at higher masses, the same effect that we observe for ${}^{114}$Sn here. This splitting into a lower-energy and a higher-energy distribution stays rather constant as a function of neutron number for $N\leq66$, as shown in Figure~\ref{fig:pdr_systematics}. 
In the recently published \ac{RQTBA} calculations \cite{Markova2025}, a similar splitting was identified, with the resonant-like structures located at $\approx 7.5-7.8$ MeV and $\approx 8.8-8.9$~MeV in $^{116,118,120}$Sn. The latter group of states was suggested to be structurally different from resonant states at lower energies: the higher-energy part is dominated by the intruder \ac{RQTBA} components with prevailing proton contribution and enhanced probability of decay to excited states, while the lower-energy part is dominated by neutron oscillations and characterized by larger ground-state branching ratios. Thus, the experimentally observed splitting could be considered consisting of \ac{PDR}-like states decreasing gradually in energy and increasing in strength with increasing $N$, appearing under a constant strength distribution of different origin, which would be consistent with the pictures from both the \ac{QRPA} and \ac{RQTBA} calculations.

Another key component in understanding the electromagnetic response of the atomic nuclei is how this response is distributed relative to the \ac{TRK} \cite{Thomas1925, Reiche1925, Kuhn1925} sum rule. The \ac{TRK} sum rule is one of the \acp{EWSR} in nuclear physics, describing in a model-independent way the energy absorbed by a nucleus from an external dipole field, $F$, as 
\begin{equation}
    \sigma_{\mathrm{EWS}} = \sum_{n} E_{n} | \langle n | F |  0 \rangle |^{2},
\end{equation}
summing over all possible states, $n$, and which can be evaluated from 
\begin{equation}
    \sigma_{\mathrm{TRK}} = 60\frac{N Z}{A}
\end{equation}
The fraction of the \ac{TRK} sum-rule strength that appears in the \ac{LEDR} region is listed in Table~\ref{tab:fit_table} and is consistent with the entire known tin chain \cite{Markova2024} around 2-3\% without any apparent systematic increase or decrease.

\subsection{Thermodynamic properties}

Another interesting macroscopic property is the thermodynamic properties of the micro-canonical ensembles that correspond to the nuclear matter. A simplified way to describe the nuclear entropy in terms of \acp{NLD} is outlined in Reference~\cite{Markova2023} based on \cite{Guttormsen2001, Agvaanluvsan2009a, Nyhus2012, Roy2021}, and can be summarised as

\begin{equation}
    S(E_{x})= k_{\mathrm{B}}\ln \Omega(E_{x}),
\end{equation}
with $k_{\mathrm{B}}$ being the Boltzmann constant and $\Omega(E_{x})$ being the number of accessible states. Here, due to the large uncertainty in the spin distribution function, the simplifications that have been carried out include the omission of the dependence on the average spin as a function of excitation energy, 
\begin{equation}
  \Omega(E_{x})= \rho(E_{x})\left(2 \langle J(E_{x}) \rangle+1\right) \approx \frac{\rho(E_{x})}{\rho_{0}},
\end{equation}
by introducing a constant, $\rho_{0}$, that normalizes the entropy to $S(0)=0$. This approach, unfortunately, makes the absolute normalisation uncertain as the contribution from the degeneracy of magnetic substates is ignored. However, the relative behaviour of the entropy curves should be reliable to the first order. 

The role of entropy of thermally excited particles in hot nuclei has been extensively discussed in Reference~\cite{Guttormsen2001}. One of the main conclusions from that work is that the odd-particle entropy in hot nuclei is $\sim 1.7$~$k_{\mathrm{B}}$, and that it stays remarkably constant over a mass region ranging from $A=20$ to $A=250$, and is independent of the ground state spin, for mid-shell nuclei. However, It is also argued that this picture breaks down for very light nuclei and nuclei around closed shells due to the more significant level spacing of those cases.

To extract the odd-particle entropy is extracted from the experimental \acp{NLD} obtained in this work, compared to the experimental \acp{NLD} of ${}^{113}$Sn from \cite{Markova2023} and ${}^{115}$Sn from \cite{Roy2021}. The ${}^{115}$Sn experiment was carried out with a different methodology at the \ac{VECC} in Kolkata using the ${}^{115}$In$(\mathrm{p},\mathrm{n}){}^{115}$Sn reaction and the neutron evaporation spectra measured by liquid scintillators \cite{Roy2021}. The level density of ${}^{115}$Sn was obtained by comparing the experimental neutron differential cross-section with that obtained from \ac{HF} calculations. Thus, this level density also has some model dependency, which, unfortunately, is common in this type of data.

Another sensitive quantity in this estimation is the choice of experimental level density in the ground-state, $\rho_{0}$.

\begin{figure}[ht!]
\begin{center}
\includegraphics[width=0.49\textwidth]{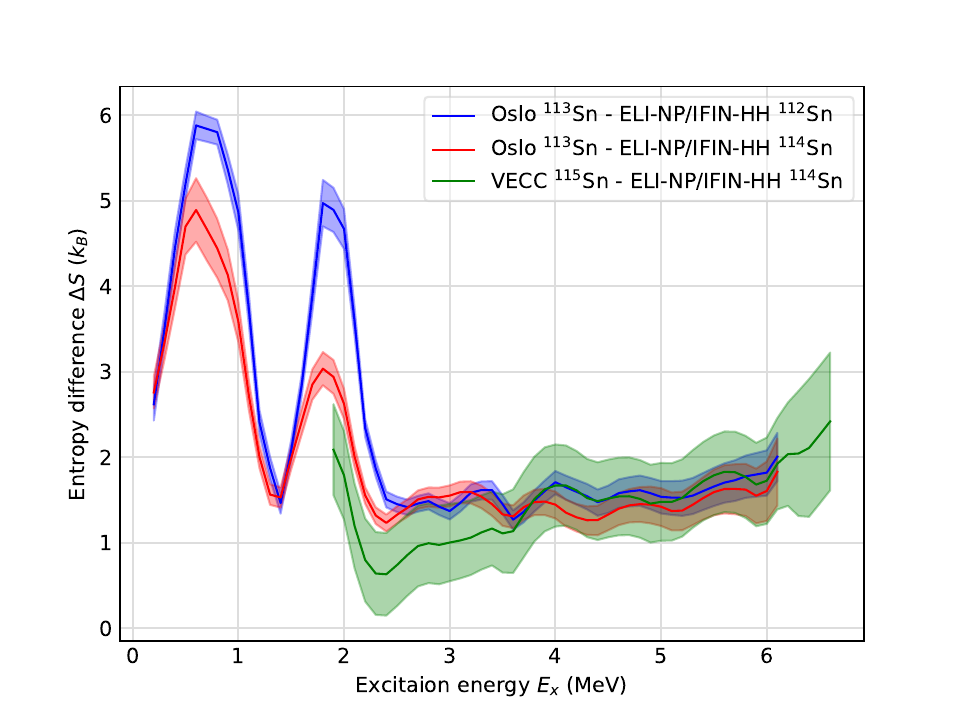}
\end{center}
\caption{Examples of entropy difference between the neighbouring even-$A$ and odd-$A$ nuclei ${}^{112,113}$Sn, ${}^{113,114}$Sn, and ${}^{114,115}$Sn. The data for the even-$A$ nuclei shown here are from this work , while the ${}^{113}$Sn data is from Reference~\cite{Markova2023} and the ${}^{115}$Sn data is from Reference~\cite{Roy2021}.}\label{fig:entropy_diff}
\end{figure}

Some of the results of these evaluations are shown in Figure~\ref{fig:entropy_diff} for ${}^{113}$Sn and ${}^{115}$Sn, which shows the discrete states at low excitation energy followed by a relatively constant $\Delta S(E_{x})$ above 2~MeV. It is worth noting here that while the \acp{NLD} for ${}^{113}$Sn and ${}^{114}$Sn were extracted using the constant temperature model and, thus, naturally exhibit a relatively flat behaviour concerning the excitation energy, ${}^{115}$Sn, was extracted with a different approach that does not inherently include that type of model dependency. Still, the behaviour of ${}^{115}$Sn is also constant, within error bars, even if a slow increase of $\Delta S(E_{x})$ with increasing $E_{x}$ can be sensed. A more in-depth picture of these more subtle aspects of the odd-particle entropy complementary experiments with alternative approaches and lower systematic and statistical uncertainties would be desired.
The systematic evolution of the odd-particle entropy is listed in Table~\ref{tab:ds_table}.
\begin{table}[ht]
\caption{Estimated single-particle entropy ($\Delta S$) from different combinations of data sets. $J^{\pi}$ is the spin and parity of the ground state and, if applicable, low-lying isomeric states. The configuration shows how the single-particle entropy originates from the even-$A$ nucleus compared to a neutron particle $(\nu)$ or hole ($\nu^{-1}$).
\label{tab:ds_table}}
 \begin{tabular*}{\columnwidth}{@{\extracolsep{\fill}}lcccc}
\hline
& $J^{\pi}$ & $\Delta S$ & Configuration & Reference \\
&  & ($k_{\mathrm{B}}$) &  &  \\
\hline									
${}^{111}$Sn & $7/2^{+}$ & 1.48($^{+4}_{-2}$) & ${}^{112}\mathrm{Sn}\otimes\nu^{-1}$ & \cite{Markova2023} \\
${}^{113}$Sn & $1/2^{+}$, $7/2^{+}$ & 1.47(2) & ${}^{112}\mathrm{Sn}\otimes\nu$ & \cite{Markova2023} \\
${}^{113}$Sn & $1/2^{+}$, $7/2^{+}$ & 1.47(2) & ${}^{112}\mathrm{Sn}\otimes\nu$ & This work, \cite{Markova2023} \\
${}^{113}$Sn & $1/2^{+}$, $7/2^{+}$ & 1.449(22) & ${}^{114}\mathrm{Sn}\otimes\nu^{-1}$ & This work, \cite{Markova2023} \\
${}^{115}$Sn & $1/2^{+}$ & 1.45(9) & ${}^{114}\mathrm{Sn}\otimes\nu$ & This work, \cite{Roy2021} \\
${}^{117}$Sn & $1/2^{+}$, $11/2^{-}$ & $\approx1.6$ & ${}^{116}\mathrm{Sn}\otimes\nu$ & \cite{Agvaanluvsan2009a} \\
${}^{119}$Sn & $1/2^{+}$, $11/2^{-}$ & 1.7(2) & ${}^{118}\mathrm{Sn}\otimes\nu$ & \cite{Toft2010} \\
\hline																										 \end{tabular*} 
\end{table}

\begin{figure}[ht!]
\begin{center}
\includegraphics[width=0.49\textwidth]{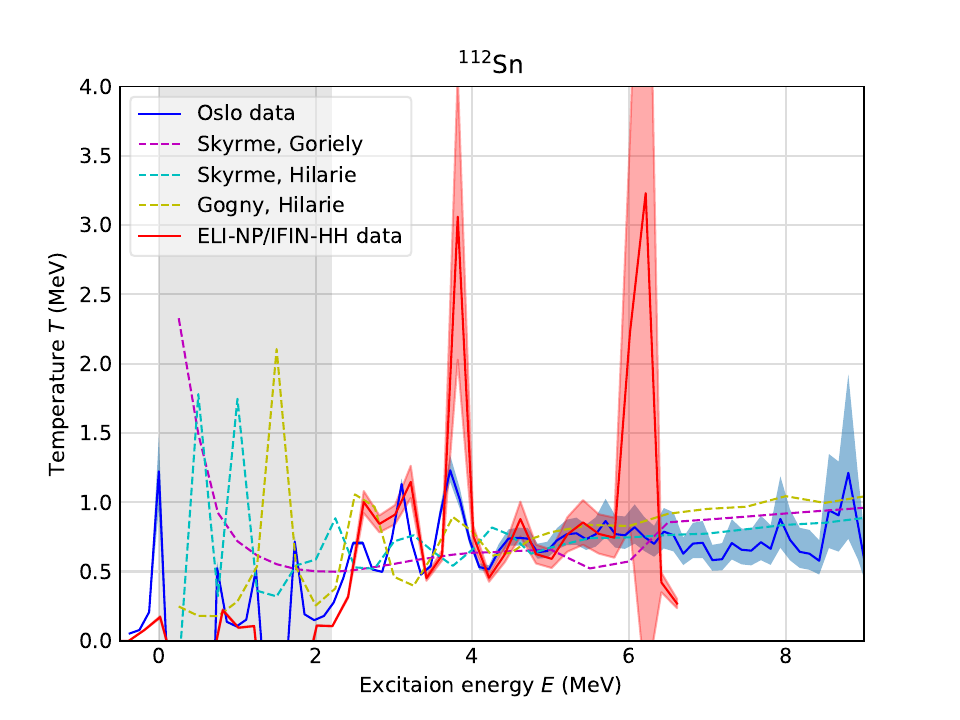}\\
\includegraphics[width=0.49\textwidth]{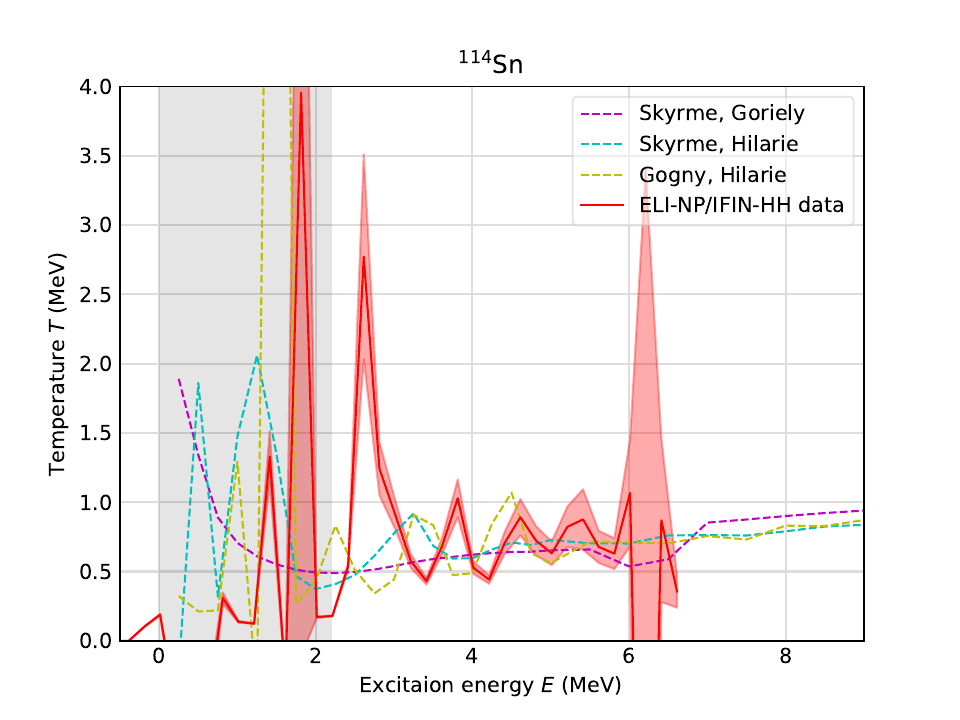}
\end{center}
\caption{Nuclear temperature as a function of excitation energy for ${}^{112}$Sn (top) from this work and Reference~\cite{Markova2023}, and ${}^{114}$Sn from this work (bottom). The shaded region corresponds to the region of discrete excited states where the entropy and temperature discussion is not applicable. For comparison, the same analysis has been performed on a selection of microscopic strength functions as implemented in the \talys\ code.}\label{fig:sn_temp}
\end{figure}

Given the discussion of the nuclear entropy, we can also calculate the microcanonical temperature, $T(E_{x})$ as
\begin{equation}
    T(E_{x}) = \left( \frac{\partial S(E_{x})}{\partial E_{x}} \right)^{-1},
\end{equation}
illustrated in Figure~\ref{fig:sn_temp}.
This temperature allows us to highlight possible structural changes in the nuclei as a function of excitation energy. In particular, with the nucleus being a small system that, furthermore, partially consists of an inert core and is not in contact with a constant-temperature heat bath, oscillations in the temperature with increasing excitation energy, $E_{x}$, are expected \cite{Agvaanluvsan2009a}. However, $T(E_{x})$ has some significant features.
In ${}^{112}$Sn, a clear structure appears at around 3.8~MeV, which is consistent with the work from Reference~\cite{Markova2023} where it was argued that a similar structure could originate from the breaking of a single Cooper pair in the nucleus, also at 3.8~MeV. A similar structure in ${}^{114}$Sn appears at a lower energy, around 2.6~MeV. Similarly, in work from Reference~\cite{Agvaanluvsan2009a}, a single Cooper-pair breaking is observed for ${}^{116}$Sn at an energy of around 2.6~MeV.

\begin{figure}[ht!]
\begin{center}
\includegraphics[width=0.49\textwidth]{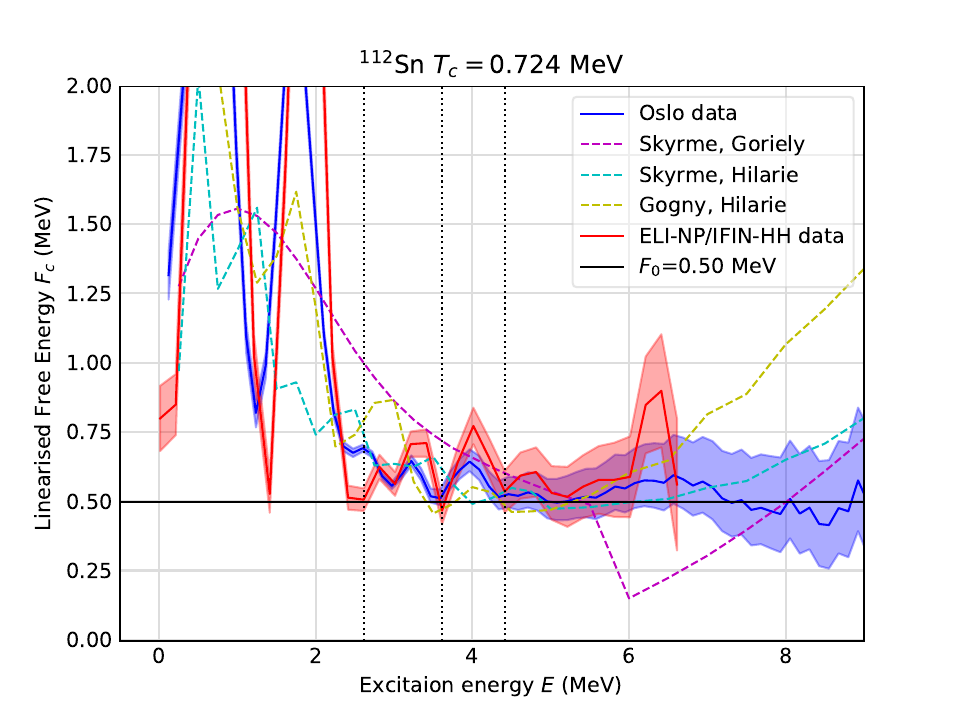}\\
\includegraphics[width=0.49\textwidth]{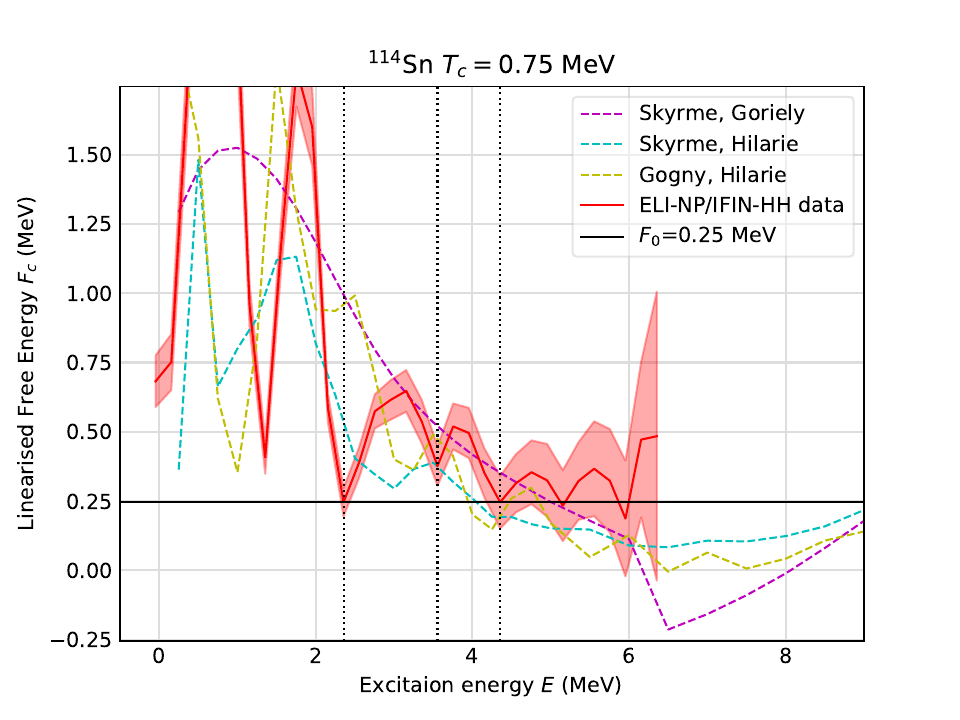}
\end{center}
\caption{Linearised Helmholtz free energy ($F_{\mathrm{c}}$) as a function of excitation energy for ${}^{112}$Sn (top) from this work and Reference~\cite{Markova2023}, and ${}^{114}$Sn from this work (bottom). the solid line shows the minimum energy, which has an arbitrary offset as listed in the legend. The critical temperature, $T_{\mathrm{c}}$ used for obtaining $F_{\mathrm{c}}$, is shown in the title of each panel. Dotted lines mark the location of the minima in $F_{\mathrm{c}}$ around the phase transition region. For comparison, the same analysis has been performed on a selection of microscopic strength functions as implemented in the \talys\ code.}\label{fig:fc}
\end{figure}

Another way to view this quantity is via the inverse relation, or the probability of finding the nucleus at a certain energy eigenstate, $E$, given a critical temperature of the system, $T$ \cite{Guttormsen2003}, as
\begin{equation}
    P(E,T_{\mathrm{c}}) = \Omega(E)\frac{\mathrm{exp}\left(-\frac{E}{k_{\mathrm{B}}T_{\mathrm{c}}}\right)}{Z(T_{\mathrm{c}})} \label{eq:pet},
\end{equation}
with $Z(T)$ being the partition function of the micro-canonical ensemble, 
\begin{equation}
  Z(T_{\mathrm{c}})=\int_{0}^{\infty}\Omega(E')\mathrm{exp}\left(-\frac{E'}{k_{\mathrm{B}}T_{\mathrm{c}}}\right)\mathrm{d}E'.
\end{equation}
This form of representing the data is very close to the approach of von~Helmholtz \cite{vonHelmholtz1888}, describing the free energy at the critical temperature, $F_{\mathrm{c}}$, of a system
\begin{equation}
    F_{\mathrm{c}} = -k_{\mathrm{B}}T_{\mathrm{c}} \ln Z. 
\end{equation}
For the entropy, the relation
\begin{equation}
    S = k_{\mathrm{B}} \ln Z+\frac{E}{T_{\mathrm{c}}},
\end{equation}
can be combined straightforwardly to an expression of the critical Helmholtz free energy of the system in terms of experimentally accessible quantities as,
\begin{equation}
    F_{\mathrm{c}} = E_{x} - T_{\mathrm{c}}S(E_{x})\label{eq:fc}.
\end{equation}
Here $F_{\mathrm{c}}$ corresponds to the linearised Helmholtz free energy at the critical temperature, $T_{\mathrm{c}}$, the excitation energy $E_{x}$ corresponding to the internal energy of the system, and $S(E_{x})$ the entropy of the system as previously defined.
It has been shown \cite{Lee1990, Lee1991} that a function $A(E,T_{\mathrm{c}}) = -\ln P(E,T_{\mathrm{c}})$ can highlight phase transitions of physical systems around the critical temperature when $A(E_{1},T_{\mathrm{c}})=A(E_{2},T_{\mathrm{c}})$ for two energies $E_{1}$ and $E_{2}$. From the relations (\ref{eq:pet})-(\ref{eq:fc}), it follows that this relation is equivalent to the same condition in the linearised free energy $F_{\mathrm{c}}(E_{1},T_{\mathrm{c}})=F_{\mathrm{c}}(E_{2},T_{\mathrm{c}})$. For both nuclei, the linearised free energy, $F_{\mathrm{c}}$, is shown for the respective critical temperatures, when the free energy is mostly flat, in Figure~\ref{fig:fc}.

What we can observe in the $F_{\mathrm{c}}$ distributions are, besides the flat behaviour at higher energies, corresponding to a smooth evolution in the nuclear structure, two distinct features in the energy regions between 2.5-4.5~MeV. As mentioned earlier, such features have been shown to be associated with first-order phase transitions \cite{Lee1990, Lee1991}. In the particular type of case discussed here, the nature of the phase transition has been suggested to be due to the transition from the paired phase to the unpaired phase with the breaking of a single neutron pair \cite{Guttormsen2003}, with a pairing energy of $2\Delta_{\mathrm{n}}=2T_{\mathrm{c}}\Delta S$. In our case this would correspond to $2\Delta_{\mathrm{n}}=2.13(10)$~MeV for ${}^{112}$Sn and $2\Delta_{\mathrm{n}}=1.95(10)$~MeV for ${}^{114}$Sn. A summary of the extracted parameters, compared with Reference~\cite{Agvaanluvsan2009a}, is listed in Table~\ref{tab:pair_table}.
\begin{table}[ht]
\caption{Parameters extracted from the linearised Helmholtz free energy ($F_{\mathrm{c}}$) from this work for ${}^{112}$Sn and ${}^{114}$Sn, and as losted in Reference~\cite{Agvaanluvsan2009a} for ${}^{116}$Sn. $T_{\mathrm{c}}$ is the critical temperature used for the calculation of $F_{\mathrm{c}}$, $E_{1}$ and $E_{2}$ the lower and upper limits of the one-pair breaking phase transition regions, respectively, $\Delta F_{\mathrm{c}}$ the height of the potential energy barrier, and $2\Delta_{\mathrm{n}}$ is twice the extracted neutron-pairing energy.
\label{tab:pair_table}}
 \begin{tabular*}{\columnwidth}{@{\extracolsep{\fill}}lcccccc}
\hline
& $T_{\mathrm{c}}$ & $E_{1}$ & $E_{2}$ & $\Delta F_{\mathrm{c}}$ & $2\Delta_{\mathrm{n}}$ & Reference \\
& ($k_{\mathrm{B}}$) & (MeV) & (MeV) & (MeV) & (MeV) & \\
\hline									
\multirow{2}{*}{${}^{112}$Sn} & \multirow{2}{*}{$0.724(_{-27}^{+33})$} & 2.6 & 3.6 & 0.21(4) & \multirow{2}{*}{2.13(10)} & \multirow{2}{*}{This work}\\
 &  & 3.6 & 4.4 & 0.28(7) & \\
\multirow{2}{*}{${}^{114}$Sn} & \multirow{2}{*}{$0.75(4)$} & 2.4 & 3.6 & 0.39(8) & \multirow{2}{*}{2.18(12)} & \multirow{2}{*}{This work}\\
 &  & 3.6 & 4.4 & 0.27(9) & \\
${}^{116}$Sn & 0.58(2) & 2.4 & 4.2 & 0.25 & 1.856 & \cite{Agvaanluvsan2009a}\\
\hline																										 \end{tabular*} 
\end{table}

Some interesting features to note in the extracted $F_{\mathrm{c}}$ is that the structure between $2.5-4.5$~MeV does not consist of a single broad feature like in Reference~\cite{Agvaanluvsan2009a}. Instead, it appears that the structure is composed of two separated features that together cover almost an identical energy range with the same potential barrier, $\Delta F_{\mathrm{c}}$, as the data from ${}^{116}$Sn. 
The prominence of these structures depend on the instrumental energy resolution, in particular of the charged particle detectors, which in the \cite{Markova2023} was around 400~keV at $E_{\mathrm{x}}=4$~MeV excitation enegy, while in this measurement around 180~keV \ac{FWHM} at $E_{\mathrm{x}}=4$~MeV.
Another interesting aspect to note is that the depression between the two features is slightly smaller in the ${}^{114}$Sn case compared to the ${}^{112}$Sn case. This could signal that a merging of the two structures with increasing $N$ is indeed happening. At the same time, as listed in Table~\ref{tab:ds_table}, there is a structural evolution in the low-lying states of the odd-$A$ nuclei in the Sn chain. This evolution consists of a transition from a $7/2^{+}$ orbital neutron in ${}^{111}$Sn to two close-lying $1/2^{+}$ and $7/2^{+}$ orbitals in ${}^{113}$Sn giving rise to a long-lived isomer, to a $1/2^{+}$ ground state in ${}^{115}$Sn, and further to a $1/2^{+}$ and $11/2^{-}$ isomeric pair in the heavier isotopes. However, further data is needed to understand these structures' nature and possible origin.

\subsection{The quasiparticle-phonon model}

\begin{figure*}[ht!]
\begin{center}
\includegraphics[width=0.49\textwidth]{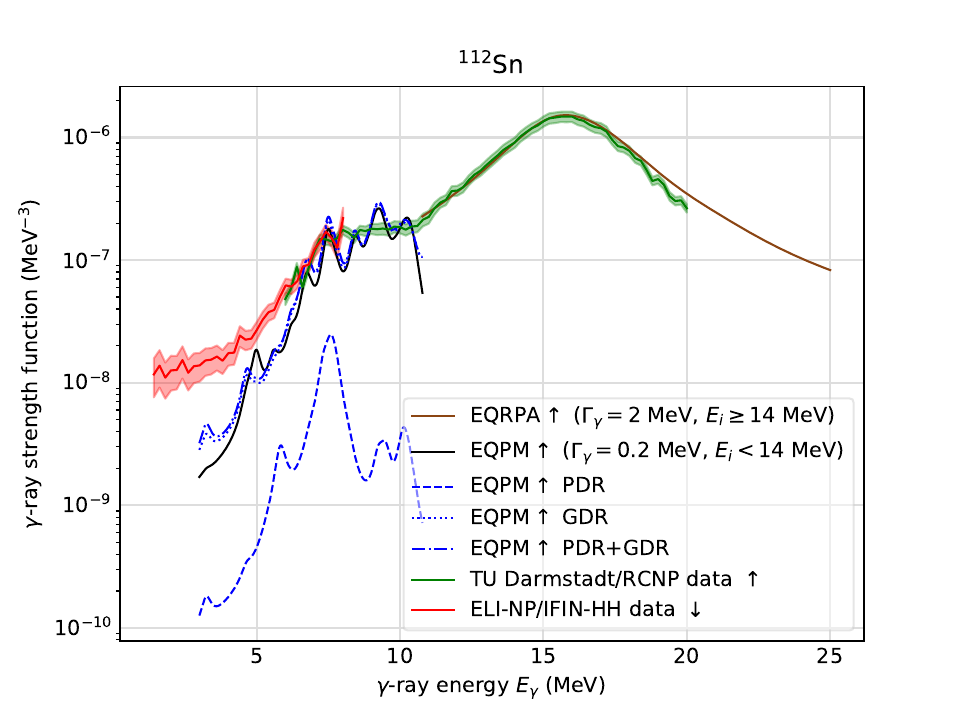}
\includegraphics[width=0.49\textwidth]{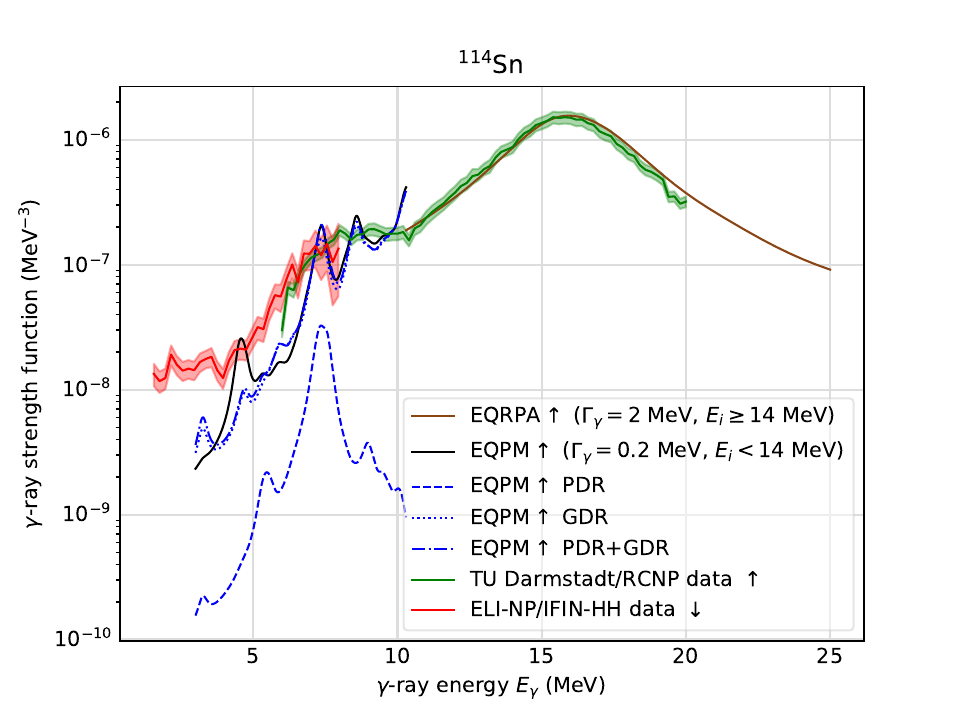}
\end{center}
\caption{Qualitative comparison of the experimental $\gamma$-ray strength functions of ${}^{112}$Sn (left) and ${}^{114}$Sn (right), including also the data reported in this work (red) as well as the data collected at RCNP by TU Darmstadt \cite{Bassauer2020} (green) together with the calculated strengths from the QRPA (brown) and EQPM (black, blue). The QRPA calculations have been smeared with a width of $\Gamma_{\gamma} = 2$~MeV, and the EQPM calculations have been smeared with a width of $\Gamma_{\gamma} = 0.2$~MeV. For the QRPA results, only states with an energy larger than $E_{i}=14$~MeV are included, and for the EQPM results, only states with an energy less than $E_{i}=14$~MeV are included. Note that the theoretical results have been renormalised for comparison to account for the long tails introduced during the Lorentzian smearing. Also, note that the theoretical and TU Darmstadt results are for excitation strength ($\uparrow$), while our data is for the decay strength ($\downarrow$). }\label{fig:qpmqtba112sn}
\end{figure*}

\begin{figure*}[ht!]
\begin{center}
\includegraphics[width=0.49\textwidth]{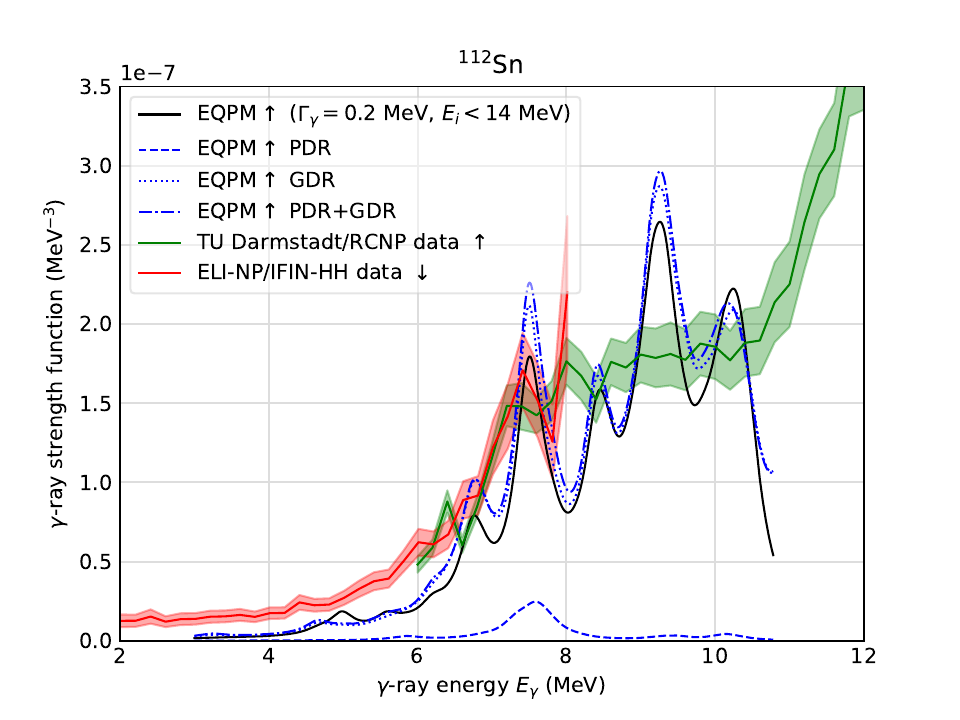}
\includegraphics[width=0.49\textwidth]{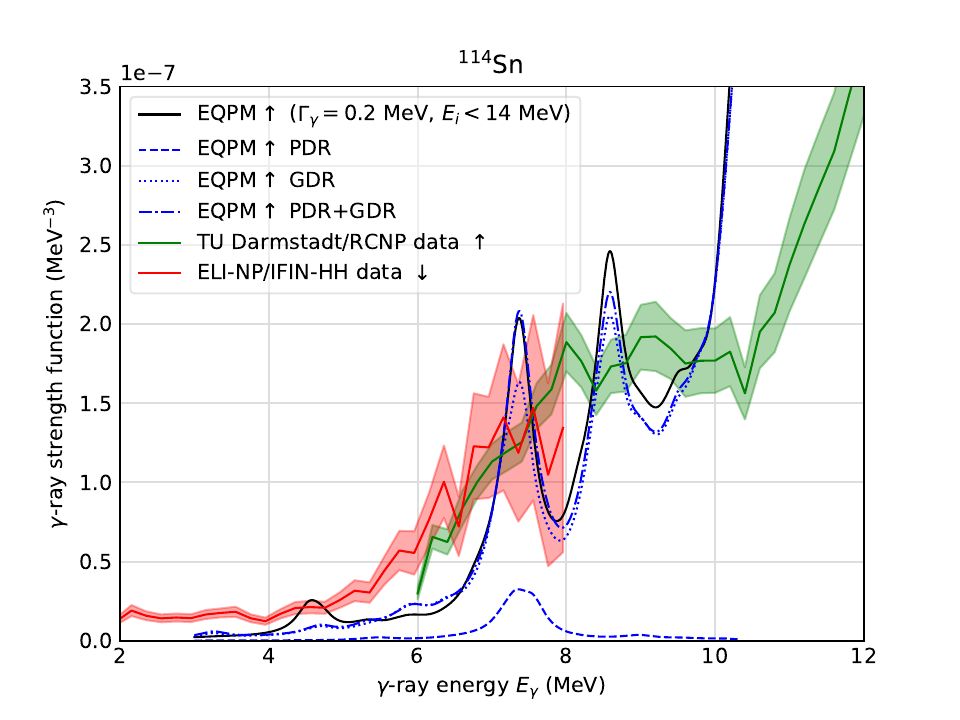}
\end{center}
\caption{Qualitative comparison of the experimental $\gamma$-ray strength functions of ${}^{112}$Sn (left) and ${}^{114}$Sn (right), including also the data reported in this work (red) as well as the data collected at RCNP by TU Darmstadt \cite{Bassauer2020} (green) together with the calculated strengths from the EQPM (black, blue). The EQPM calculations have been smeared with a width of $\Gamma_{\gamma} = 0.02$~MeV and reduced by 10 for easier comparison. Note that the theoretical results have been renormalised for comparison to account for the long tails introduced during the Lorentzian smearing. Also, note that the theoretical and TU Darmstadt results are for excitation strength ($\uparrow$), while our data is for the decay strength ($\downarrow$).}\label{fig:qpm112sn}
\end{figure*}

We have pointed out that one of the main problems with the \ac{QRPA} methods is that the calculated low-energy dipole strength does not uniquely correspond to the neutron \ac{PDR} mode but contains an admixture of the low-energy tail of the \ac{GDR}, mixed configurations between these and contributions related to the coupling between higher angular momentum multipoles. In this context, it is important in these studies to apply a theory capable of consistently accounting for these low-energy electric dipole spectra features. The \ac{EDF} theory and the three-phonon \ac{QPM} theory \cite{Tsoneva2004, Tsoneva2016} is an advanced theoretical method used in studies of the structure of electric and magnetic nuclear excitations in the \ac{PDR} region and proved to be very successful in describing the low-energy $\gamma$ strength functions and, in particular, the \ac{PDR} \,\cite{Tsoneva2016, Spieker2020, Tonchev2017, Weinert2021}.
An important advantage of the \ac{EDF}+\ac{QPM} approach, which we here call EQPM for short, is the description of nuclear excitations in terms of \ac{QRPA} phonons as building blocks of the three-phonon \ac{QPM} model \,\cite{Soloviev1976, Grinberg1998}, which provides a microscopic way to mix multi-particle configurations. Thus, EQPM allows for sufficiently large configuration spaces, which are necessary for a uniform description of low-energy single- and multi-particle states, and also the \ac{GDR}.
The current EQPM calculations include a model configuration space consisting of \ac{QRPA} states with J$^\pi$=1$^{\pm}$, 2$^{\pm}$, 3$^{\pm}$, 4$^{\pm}$, 5$^{\pm}$, taking into account one-phonon 1$^-$ states up to $E_{x}$=30~MeV, which explicitly allows the consideration of \ac{GDR} and nuclear polarisation contributions to the low-energy 1$^{-}$ state transitions. The multi-phonon constituents include two-phonon configurations with excitation energies up to $E_{x}\sim$11 MeV, which covers the experimental energy range and is effective for accounting for the gross properties of the dipole strength function below and around the neutron threshold. Choosing the EQPM configuration space allows us to perform our calculations without dynamical effective charges \cite{Tsoneva2008}.
EQPM results for the dipole response below the neutron threshold in $^{112,114}$Sn are shown in Figure~\ref{fig:qpmqtba112sn}. By comparing Figure~\ref{fig:sn112gsf} and Figure~\ref{fig:qpmqtba112sn}, it can be seen that the \ac{QRPA} strengths are strongly fragmented as soon as coupling to multi-phonon configurations is performed. Our focus is specifically on E1 strength, located between $E_{x}\sim6-8$~MeV, which resembles a resonance structure known as \ac{PDR} in neutron-excess tin isotopes \cite{Tsoneva2008, Tsoneva2016}. From the analysis of the \ac{QRPA} calculations discussed in these studies, it appears that the 1$^{-}$ states at $E_{x}\sim7-8$ MeV in $^{112,114}$Sn have a predominantly neutron structure associated with one or two major single-particle configurations and their transition densities behave as neutron oscillations along an isospin symmetric nuclear core and can be associated with \ac{PDR} modes.
In our EQPM studies, we relate the \ac{PDR} to coherent excitations of a sequence of low-energy $1^{-}$ neutron single-particle states with very similar spectroscopic properties (see, for example, References~\cite{Tsoneva2008, Tsoneva2004, Tsoneva2016}). The total $E1$ strength of the PDR, EQPM PDR in Figures~\ref{fig:qpmqtba112sn} and \ref{fig:qpm112sn}, is calculated as the sum of the individual $E1$ strengths of these 1$^-$ states, is typically less than 1\% of the TRK sum rule, which is almost completely exhausted by the \ac{GDR}. In addition, there exists a direct relation between the amount of the observed total PDR strength and the neutron excess expressed in terms of the neutron-to-proton ($N/Z$) ratio of the nucleus, which is also expressed by the neutron-skin thickness defined in our References~\cite{Tsoneva2008, Tsoneva2004,Tsoneva2016}. Considering that the $N/Z$ ratio in $^{112}$Sn ($N/Z = 1.24$) and $^{114}$Sn ($N/Z = 1.28$) is close to 1, the theoretically obtained PDR strength is relatively small. However, the observed total E1 strength at low energies contains additional counterparts due to coupling with more complex configurations and the GDR, EQPM GDR in Figures~\ref{fig:qpmqtba112sn} and \ref{fig:qpm112sn}. The full EQPM theory in Figures~\ref{fig:qpmqtba112sn} and \ref{fig:qpm112sn} explicitly accounts for multi-configuration mixing in terms of interactions between quasiparticle and phonon states. This allows us to account for the fragmentation of single-particle strength on a microscopic basis and anharmonicity effects, shifting some of the GDR's higher-lying $E1$ strength to lower energies. The EQPM calculations of the $E1$ spectral distributions up to GDR energies can currently quantitatively predict the fine and coarse properties of the total dipole strength. They can be directly compared with experimental data in $^{112}$Sn and $^{114}$Sn, which assures us of the reliability of our theoretical interpretation.
In both $^{112,114}$Sn isotopes, three-phonon EQPM calculations show that in the $E_{x}\leq 8$~MeV energy range, about 80$\%$ of the total one-phonon strength of the \ac{PDR} is depleted. 

The experimental energy-weighted strength evaluation has been performed based on the fitted functions parametrised in Table~\ref{tab:fit_table},
\begin{equation}
\sum_{n} \sigma_\mathrm{{exp}}E_{n}=\sigma_{\mathrm{LEDR}}E_{\mathrm{LEDR}}+\int_{3.2}^{S_{\mathrm{n}}} \mathcal{S}_{\mathrm{GDR}}(E_{\gamma})E_{\mathrm{GDR}}\mathrm{d} E_{\gamma},
\end{equation}
and does not include fragmentation of states; thus, it should be considered an approximate value. The total strength fraction of the \ac{TRK} sum rule from the EQPM is 7.67\% for ${}^{112}$Sn and 6.37\% for ${}^{114}$Sn, including PDR-GDR interference terms, multi-phonon contributions as well as interference terms between multi-phonon and single phonon states.
The calculated total EQPM dipole strength below the neutron separation threshold is 
$\sum_{n} \sigma_\mathrm{{EQPM}}(E1) E_{n}=127.3$~mb$\cdot$MeV in $^{112}$Sn and 
$\sum_{n} \sigma_\mathrm{{EQPM}}(E1) E_{n}=107.3$~mb$\cdot$MeV in $^{114}$Sn, which is very close to the experimentally fitted strength, 
$\sum_{n} \sigma_\mathrm{{exp}}(E1) E_{n}=110(8)$~mb$\cdot$MeV
in $^{112}$ Sn and 
$\sum_{n} \sigma_\mathrm{{exp}}(E1) E_{n}=91(8)$~mb$\cdot$MeV
in $^{114}$Sn, respectively. This strength is distributed as 4.38~mb$\cdot$MeV (0.26\% of the \ac{TRK} \ac{EWSR}) from the \ac{PDR} and 129.34~mb$\cdot$MeV (7.79\% of the \ac{TRK} \ac{EWSR}) from the \ac{GDR} in ${}^{112}$Sn between 3.2~MeV and $S_{\mathrm{n}}$. The experimental total strength, including both \ac{LEDR} and the tail of the \ac{GDR}, in this energy region, is 6.65\% of the \ac{TRK} \ac{EWSR}.
The corresponding values for ${}^{114}$Sn is 4.70~mb$\cdot$MeV (0.28\% of the \ac{TRK} \ac{EWSR}) from the \ac{PDR} and 94.30~mb$\cdot$MeV (5.60\% of the \ac{TRK} \ac{EWSR}) from the \ac{GDR} compared to the experimental total value of 5.4\% of the \ac{TRK} \ac{EWSR}.

From here, we see that the total dipole strength below the neutron threshold is smaller in ${}^{114}$Sn than it is in ${}^{114}$Sn, while we would expect the opposite from a picture of the \ac{LEDR} originating from a \ac{PDR} based on excess neutron excitations. However, from the EQPM calculations, we also see the importance of taking the composition of several components and their interferences into account when discussing the total dipole strength. These components are related to the \ac{PDR}, the core polarization effects, and the \ac{GDR}. If we calculate the integrated pure \ac{PDR} peak from the EQPM, we see that the total \ac{PDR} strength does indeed increase with increasing neutron number, from 4.38~mb$\cdot$MeV to 4.70~mb$\cdot$MeV, as expected, while the total \ac{LEDR} is decreasing due to the admixture of more complex configurations.

The good overall agreement between the EQPM calculations and the experiment in $^{112,114}$Sn suggests that in these nuclei, the properties of the low-energy dipole spectra and, in particular, of the \ac{PDR} are well determined. A similar conclusion was reached in our previous EQPM calculations in $^{124}$Sn and other tin isotopes\cite{Tsoneva2008, Tsoneva2016}. We should also emphasise that the coupling of nuclear excitations located in the quasi-continuum region, also associated with the low-energy tail of \ac{GDR}, and low-energy 1$^-$ excited states can have a substantial impact on the dipole strength below the neutron threshold. The effect becomes increasingly important in nuclei where the neutron threshold is higher and therefore approaches \ac{GDR}, such as in the lighter Sn isotopes like $^{112,114}$Sn, where the strength of \ac{PDR} is very close to the neutron threshold. The role of the quasi-continuum coupling was studied in detail in our investigations of dipole response of $^{204}$Pb \cite{Shizuma2022}. Such a situation requires much larger model spaces. This is a numerically difficult task due to the huge increase in level densities, particularly in the nuclei far from the shell closure.

\subsection{Astrophysical aspects}

The role of the \acp{NLD} and the \acp{gSF} in the nucleosynthesis can easily be illustrated with a schematic picture of the neutron-capture process, 
summarised in the Hauser-Feshbach model for radiative neutron capture \cite{Hauser1952, Goriely2007},
\begin{equation}
    \sigma(\mathrm{n},\gamma) \propto \sum_{J^{\pi},XL} \int \mathcal{T}_{XL}(E_{\gamma}) \rho(E_{x},J,\pi) \mathrm{d}E_{\gamma},
\end{equation}
where $\mathcal{T}_{XL}(E_{\gamma})$ is directly proportional to the \ac{gSF} with appropriate energy weighting, as of the equation (\ref{eq:f2T}) previously discussed. Thus, only these two quantities should determine the neutron-capture rate directly.

\begin{figure}[ht!]
\begin{center}
\includegraphics[width=\columnwidth]{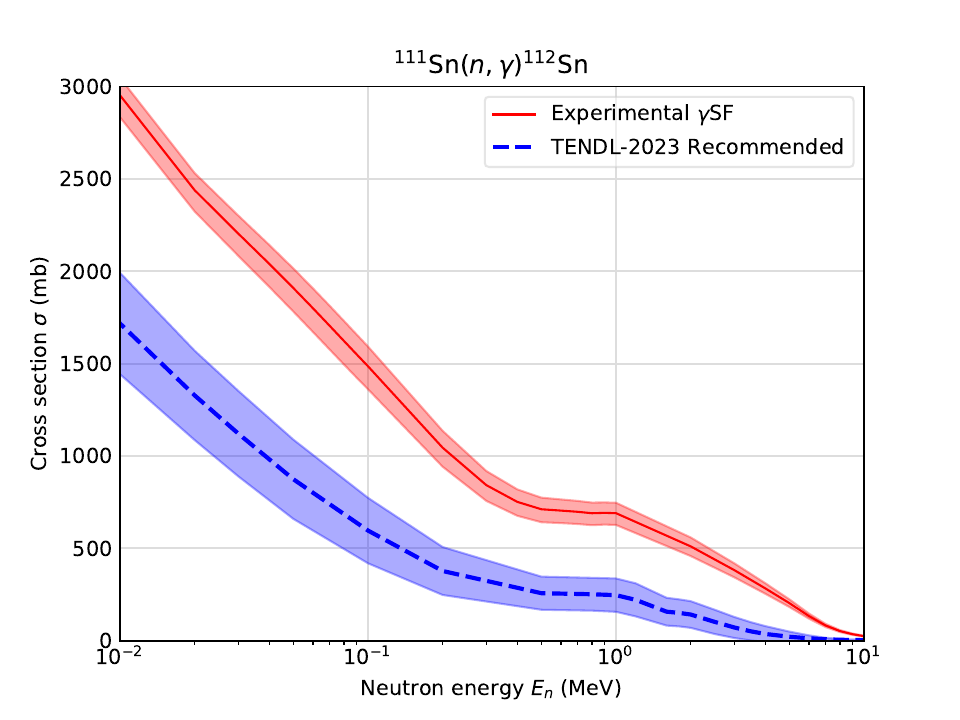}\\
\includegraphics[width=\columnwidth]{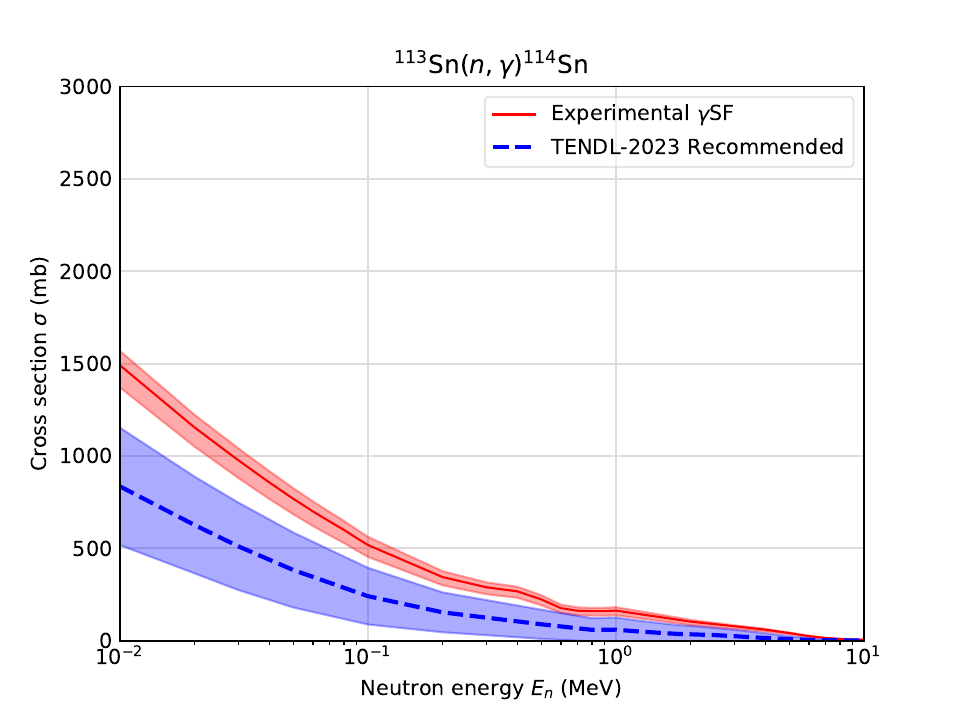}
\end{center}
\caption{Radiative neutron-capture cross sections calculated from the experimental $\gamma$ strength functions and constant temperature parameters adjusted to the nuclear level densities presented in this work, compared to the recommended values and uncertainties from ten models in the TENDL-2023 library. The experimental uncertainties only contain the statistical uncertainties from the $\gamma$SF.}
\label{fig:talys_xsec}
\end{figure}

To evaluate the 
impact of the measurements presented here, the cross-section was calculated for the ${}^{111,113}\mathrm{Sn}(\mathrm{n},\gamma)$ reactions using the \talys\ \cite{Koning2008, Koning2012} code. The maximum $\gamma$-ray multipolarity was set to $\ell=4$. For the level densities, the Constant temperature + Fermi gas model was used with the level density parameter at the neutron separation energy for ${}^{112}$Sn as 12.530~MeV$^{-1}$, the temperature of the Gilbert-Cameron formula as 0.695~MeV, and the back-shift energy of the four-component formula as 0.813~MeV. An extra pairing shift to adjust the Fermi Gas level density was manually introduced as $-1.906$~MeV instead of being automatically calculated by \talys. The pre-equilibrium reaction mechanism was enabled for all incident energies. The upper limit for including the transmission coefficients in the Hauser-Feshbach calculations was reduced to $10^{-10}$, and also the upper limit for considering cross sections in the calculation was reduced to $10^{-10}$~mb, to increase the precision in the output. Reaction cross-sections were calculated for 33 incident energies between $0.001-20$~MeV in the laboratory frame of reference, and the results are shown in Figure~\ref{fig:talys_xsec} together with the \ac{TENDL} 2023 recommendations \cite{Koning2019, Rochman2025}.
The calculations were performed for 
the weighted average of the experimental data that has been presented in this work
and the \ac{RCNP} data.

\begin{figure}[ht!]
\begin{center}
\includegraphics[width=\columnwidth]{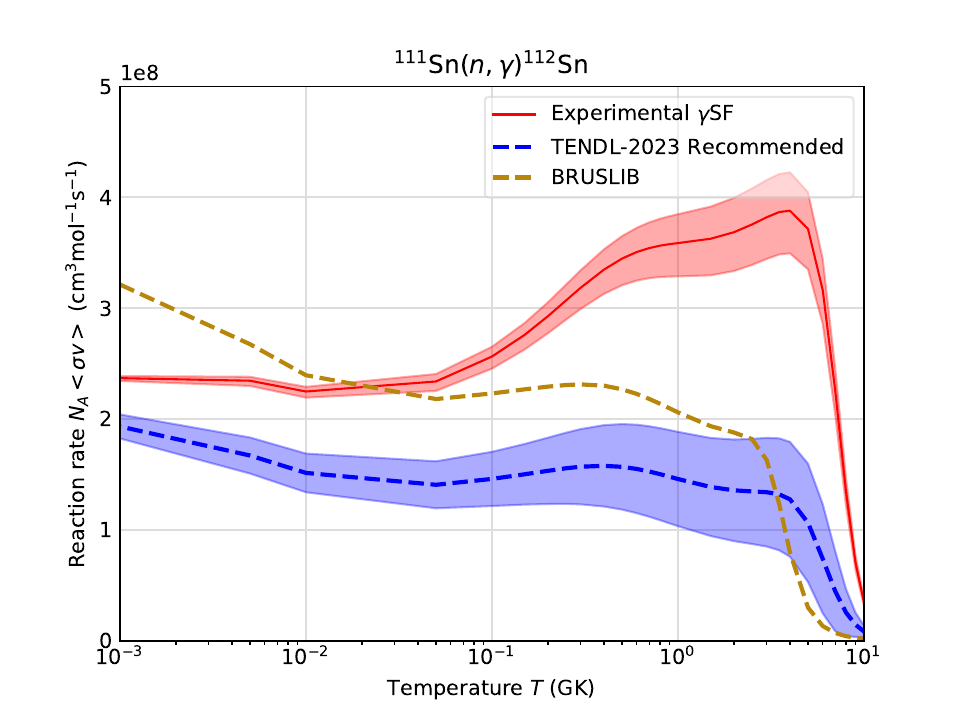}\\
\includegraphics[width=\columnwidth]{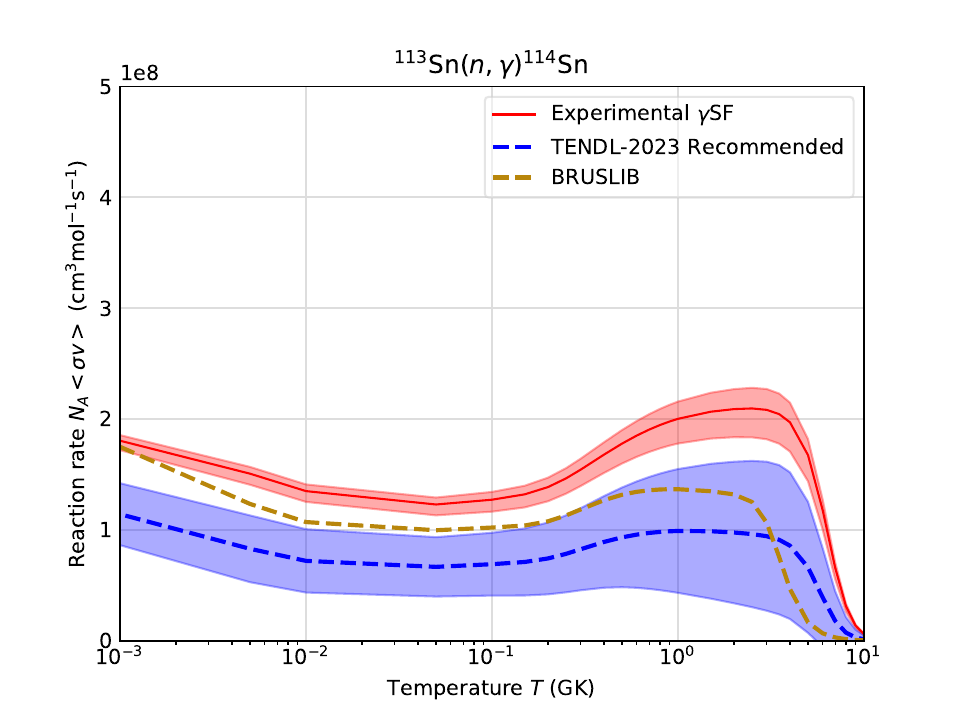}
\end{center}
\caption{Reaction rates for the radiative neutron capture calculated from the experimental data, compared to the  
astrophysical data library BRUSLIB, and the recommended values and uncertainties from ten models in the TENDL-2023 library. The experimental uncertainties only contain the statistical uncertainties from the $\gamma$SF.}\label{fig:talys_rate}
\end{figure}

From the neutron capture cross-section, we can calculate the neutron-capture reaction rate $N_{\mathrm{A}}\langle \sigma v \rangle (T)$ using
\begin{equation}
N_{\mathrm{A}}\langle \sigma v \rangle (T) = \int_{0}^{\infty} \sigma^{\ast}(E)\Phi(E,T)\mathrm{d}E,
\end{equation}
with $\sigma^{\ast}(E)$ being the neutron-capture cross-sections and $\Phi(E,T)$ the Maxwell-Boltzmann energy distribution \cite{Goriely2008}. The resulting $N_{\mathrm{A}}\langle \sigma v \rangle (T)$ rates are shown in Figure~\ref{fig:talys_rate} togethery with the compilations from \ac{BRUSLIB} \cite{Goriely2004,Xu2013} and for \ac{TENDL}-astro \cite{Rochman2025}. While the ${}^{113}\mathrm{Sn}(\mathrm{n},\gamma){}^{114}\mathrm{Sn}$ reaction rates using the experimental \ac{gSF} are still rather close to the \ac{TENDL} and \ac{BRUSLIB} recommendations, while for the the ${}^{111}\mathrm{Sn}(\mathrm{n},\gamma){}^{112}\mathrm{Sn}$ reaction the rates using the experimental \ac{gSF} are significanlty higher. 

\section{Conclusions} \label{sec:conclusions}

We have performed the first measurements of \acp{NLD} and \acp{gSF} at the 9~MV facilities at IFIN-HH, combining the ELIGANT-GN large volume \labr\ and \cebr\ detectors with the ROSPHERE array. In the case of ${}^{112}$Sn we have shown that we can reproduce the results from the Oslo Cyclotron Laboratory. The \ac{NLD} and \ac{gSF} data have been further extended to ${}^{114}$Sn that was measured for the first time, completing the data on the even-even stable tin isotopic chain. We compare the obtained data to high-energy (p,p') scattering at zero degrees from the RCNP facilities in Osaka, as well as to statistical model calculations and fully microscopic quasiparticle-phonon model calculations. Lastly, we investigate the impact of these new data on the reaction rates in astrophysical $p$-process environments, showing an increased reaction rate for $(\mathrm{n},\gamma)$ neutron capture reactions compared to existing astrophysical libraries.

\section*{Acknowledgements} \label{sec:acknowledgements}

    PAS, AK, SA, RSB, RB, MB, CCo, NMF, RL, CM, TP, AS, DAT, AT, GVT, and SU were supported by the ELI-RO program funded by the Institute of Atomic Physics, M\u{a}gurele, Romania, contract number ELI-RO/RDI/2024-002 CIPHERS and DLB, and AR contract number ELI-RO/RDI/2024-007 ELITE. NT was supported by the Romanian Ministry of Research, Innovation and Digitization, CNCS - UEFISCDI, project number PN-III-P4-PCE-2021-0595, within PNCDI III. 
    YX, MC, AD, VL, DN, HP, IPP, and TT acknowledge the support of the Romanian Ministry of Research and Innovation under research contract PN~23~21~01~06. The experiment was further supported by EUROpean Laboratories for Accelerator Based Sciences, EURO-LABS. We would also like to thank M.~Guttormsen from the University of Oslo for his support.

\acrodef{AGATA}{Advanced GAmma Tracking Array}
\acrodef{ALBA}{African LaBr Array}
\acrodef{AMD}{antisymmetrized molecular dynamics}
\acrodef{BGO}{bismuth germanate}
\acrodef{BNC}{Bayonet Neill-Concelman}
\acrodef{BRIKEN}{Beta-delayed neutrons at RIKEN}
\acrodef{BRUSLIB}{BRUSsels Nuclear LIBrary}
\acrodef{CAKE}{Coincidence Array for K600 Experiment}
\acrodef{CAD}{computer-aided design}
\acrodef{CCB}{Centrum Cyklotronowe Bronowice}
\acrodef{CFD}{constant-fraction discriminator}
\acrodef{CLHEP}{Class Library for High Energy Physics}
\acrodef{CMB}{cosmic microwave-background}
\acrodef{CRP}{Coordinated Research Projects}
\acrodef{DAQ}{data acquisition}
\acrodef{DC}{direct current}
\acrodef{DELILA}{Digital ELI List-mode Acquisition}
\acrodef{DPP}{Digital Pulse-Processing}
\acrodef{DPP-PHA}{Digital Pulse Processing for Pulse Height Analysis}
\acrodef{DPP-PSD}{Digital Pulse Processing for Charge Integration and Pulse Shape Discrimination}
\acrodef{DSSSD}{double-sided silicon strip detector}
\acrodef{D-sub}{D-subminiature}
\acrodef{E8}{Experimental Area 8}
\acrodef{E9}{Experimental Area 9}
\acrodef{ECL}{emitter-coupled logic}
\acrodef{EDF}{Energy Density Functional}
\acrodef{ELI}{Extreme Light Infrastructure}
\acrodef{ELIADE}{ELI Array of DEtectors}
\acrodef{ELI-BIC}{ELI Bragg ionization chamber}
\acrodef{ELIGANT}{ELI Gamma Above Neutron Threshold}
\acrodef{ELIGANT-GG}{ELIGANT Gamma Gamma}
\acrodef{ELIGANT-GN}{ELIGANT Gamma Neutron}
\acrodef{ELIGANT-TN}{ELIGANT Thermal Neutron}
\acrodef{ELI-NP}{Extreme Light Infrastructure -- Nuclear Physics}
\acrodef{ELISSA}{ELI Silicon Strip Array}
\acrodef{EWSR}{energy-weighted sum rule}
\acrodef{FADC}{flash ADC}
\acrodef{FASTER}{Fast Acquisition System for nuclEar Research}
\acrodef{FATIMA}{FAst TIming Array}
\acrodef{FFT}{fast Fourier transform}
\acrodef{FOM}{figure-of-merit}
\acrodef{FREYA}{Fission Reaction Event Yield Algorithm}
\acrodef{FWHM}{full width at half maximum}
\acrodef{GDR}{giant dipole-resonance}
\acrodef{GECO2020}{GEneral COntrol Software}
\acrodef{GLO}{Generalized Lorentzian}
\acrodef{GRAF}{Grand RAiden Forward}
\acrodef{GROOT}{\geant\ and ROOT Object-Oriented Toolkit}
\acrodef{gSF}[$\gamma$SF]{$\gamma$-ray strength function}
\acrodef{GSI}{Gesellschaft f\"{u}r Schwerionenforschung}
\acrodef{HDPE}{high-density polyethylene}
\acrodef{HF}{Hartree-Fock}
\acrodef{HFB}{Hartree-Fock-Bogolyubov}
\acrodef{HIgS}[HI$\gamma$S]{High Intensity $\gamma$-ray Source}
\acrodef{HPGe}{high-purity germanium}
\acrodef{HPLS}{high-power laser system}
\acrodef{HV}{high voltage}
\acrodef{IAEA}{International Atomic Energy Agency}
\acrodef{IFIN-HH}{Horia Hulubei Institute for Physics and Nuclear Engineering}
\acrodef{iThemba LABS}{iThemba Laboratory for Accelerator Based Sciences}
\acrodef{JINR}{Joint Institute for Nuclear Research}
\acrodef{KMF}{Kadmenskii-Markushev-Furman}
\acrodef{KRATTA}{Krak\'{o}w Triple Telescope Array}
\acrodef{LANL}{Los Alamos National Laboratory}
\acrodef{LCS}{laser Compton backscattering}
\acrodef{LED}{leading-edge discriminator}
\acrodef{LEDR}{low-energy electric dipole response}
\acrodef{LLNL}{Lawrence Livermore National Laboratory}
\acrodef{MDA}{multipole decomposition analysis}
\acrodef{MDR}{magnetic dipole-resonance}
\acrodef{MCA}{multi-channel analyzer}
\acrodef{MCNP}{Monte Carlo N-Particle Transport Code}
\acrodef{MCX}{micro coaxial connector}
\acrodef{MIDAS}{Multi Instance Data Acquisition System}
\acrodef{MONSTER}{MOdular Neutron time-of-flight SpectromeTER}
\acrodef{MWDC}{Multi-Wire Drift Chambers}
\acrodef{NDF}{number of degrees of freedom}
\acrodef{NEDA}{NEutron Detector Array}
\acrodef{NIM}{Nuclear Instrumentation Module}
\acrodef{NLD}{nuclear level density}
\acrodefplural{NLD}[NLDs]{nuclear level densities}
\acrodef{NNDC}{National Nuclear Data Center}
\acrodef{NRF}{nuclear resonance fluorescence}
\acrodef{OCL}{Oslo Cyclotron Laboratory}
\acrodef{ODeSA}{Oak Ridge National Laboratory Deuterated Spectroscopic Array}
\acrodef{ORNL}{Oak Ridge National Laboratory}
\acrodef{OSCAR}{Oslo Scintillator Array}
\acrodef{PANDORA}{Photo-Absorption of Nuclei and Decay Observables for Reactions in Astrophysics}
\acrodef{PARIS}{Photon Array for Studies with Radioactive Ion and Stable Beams}
\acrodef{PCB}{printed circuit board}
\acrodef{PCIe}{Peripheral Component Interconnect Express}
\acrodef{PDR}{pygmy dipole resonance}
\acrodef{PLL}{phase-locked loop}
\acrodef{PMT}{photomultiplier tube}
\acrodef{PRD}{prompt-response distribution}
\acrodef{PSD}{pulse-shape discrimination}
\acrodef{PuBe}{plutonium-beryllium}
\acrodef{QED}{quantum electrodynamics}
\acrodef{QPM}{quasiparticle-phonon model}
\acrodef{QRPA}{Quasiparticle Random Phase Approximation}
\acrodef{RC}{resistor-capacitor}
\acrodef{RCNP}{Research Center for Nuclear Physics}
\acrodef{RF}{radio frequency}
\acrodef{RMS}{root-mean-square}
\acrodef{ROSPHERE}{ROmanian array for SPectroscopy in HEavy ion REactions}
\acrodef{RQRPA}{Relativistic Quasiparticle Random Phase Approximation}
\acrodef{RQTBA}{Relativistic Quasiparticle Time-Blocking Approximation}
\acrodef{SHV}{safe high voltage}
\acrodef{SAKRA}{Si Array developed by Kyoto and osaka for Research into Alpha cluster states}
\acrodef{SMLO}{Simplified version of the modified Lorentzian}
\acrodef{SORCERER}{SOlaR CElls for Reaction Experiments at ROSPHERE}
\acrodef{SSC}{Separated Sector Cyclotron}
\acrodef{TENDL}{TALYS-based Evaluated Nuclear Data Library}
\acrodef{TCP/IP}{Transmission Control Protocol/Internet Protocol}
\acrodef{TDR}{technical design report}
\acrodef{TRK}{Thomas-Reiche-Kuhn}
\acrodef{TOF}{time-of-flight}
\acrodef{TUD}[TU Darmstadt]{Technische Universität Darmstadt}
\acrodef{TUNL}{Triangle Universities Nuclear Laboratory}
\acrodef{UHECR}{ultra-high energy cosmic-ray}
\acrodef{UPC}{Universitat Polit\`{e}cnica de Catalunya}
\acrodef{USB}{Universal Serial Bus}
\acrodef{VECC}{Variable Energy Cyclotron Centre}
\acrodef{VEGA}{Variable Energy Gamma-ray}
\acrodef{VME}{Versa Module Europa}


\begin{thebibliography}{99}%
\makeatletter
\providecommand \@ifxundefined [1]{%
 \@ifx{#1\undefined}
}%
\providecommand \@ifnum [1]{%
 \ifnum #1\expandafter \@firstoftwo
 \else \expandafter \@secondoftwo
 \fi
}%
\providecommand \@ifx [1]{%
 \ifx #1\expandafter \@firstoftwo
 \else \expandafter \@secondoftwo
 \fi
}%
\providecommand \natexlab [1]{#1}%
\providecommand \enquote  [1]{``#1''}%
\providecommand \bibnamefont  [1]{#1}%
\providecommand \bibfnamefont [1]{#1}%
\providecommand \citenamefont [1]{#1}%
\providecommand \href@noop [0]{\@secondoftwo}%
\providecommand \href [0]{\begingroup \@sanitize@url \@href}%
\providecommand \@href[1]{\@@startlink{#1}\@@href}%
\providecommand \@@href[1]{\endgroup#1\@@endlink}%
\providecommand \@sanitize@url [0]{\catcode `\\12\catcode `\$12\catcode
  `\&12\catcode `\#12\catcode `\^12\catcode `\_12\catcode `\%12\relax}%
\providecommand \@@startlink[1]{}%
\providecommand \@@endlink[0]{}%
\providecommand \url  [0]{\begingroup\@sanitize@url \@url }%
\providecommand \@url [1]{\endgroup\@href {#1}{\urlprefix }}%
\providecommand \urlprefix  [0]{URL }%
\providecommand \Eprint [0]{\href }%
\providecommand \doibase [0]{http://dx.doi.org/}%
\providecommand \selectlanguage [0]{\@gobble}%
\providecommand \bibinfo  [0]{\@secondoftwo}%
\providecommand \bibfield  [0]{\@secondoftwo}%
\providecommand \translation [1]{[#1]}%
\providecommand \BibitemOpen [0]{}%
\providecommand \bibitemStop [0]{}%
\providecommand \bibitemNoStop [0]{.\EOS\space}%
\providecommand \EOS [0]{\spacefactor3000\relax}%
\providecommand \BibitemShut  [1]{\csname bibitem#1\endcsname}%
\let\auto@bib@innerbib\@empty
\bibitem [{\citenamefont {Goriely}\ \emph {et~al.}(2019)\citenamefont
  {Goriely}, \citenamefont {Dimitriou}, \citenamefont {Wiedeking},
  \citenamefont {Belgya}, \citenamefont {Firestone}, \citenamefont {Kopecky},
  \citenamefont {Krticka}, \citenamefont {Plujko}, \citenamefont {Schwengner},
  \citenamefont {Siem}, \citenamefont {Utsunomiya}, \citenamefont {Hilaire},
  \citenamefont {P\'eru}, \citenamefont {Cho}, \citenamefont {Filipescu},
  \citenamefont {Iwamoto}, \citenamefont {Kawano}, \citenamefont {Varlamov},\
  and\ \citenamefont {Xu}}]{Goriely2019b}%
  \BibitemOpen
  \bibfield  {author} {\bibinfo {author} {\bibfnamefont {S.}~\bibnamefont
  {Goriely}}, \bibinfo {author} {\bibfnamefont {P.}~\bibnamefont {Dimitriou}},
  \bibinfo {author} {\bibfnamefont {M.}~\bibnamefont {Wiedeking}}, \bibinfo
  {author} {\bibfnamefont {T.}~\bibnamefont {Belgya}}, \bibinfo {author}
  {\bibfnamefont {R.}~\bibnamefont {Firestone}}, \bibinfo {author}
  {\bibfnamefont {J.}~\bibnamefont {Kopecky}}, \bibinfo {author} {\bibfnamefont
  {M.}~\bibnamefont {Krticka}}, \bibinfo {author} {\bibfnamefont
  {V.}~\bibnamefont {Plujko}}, \bibinfo {author} {\bibfnamefont
  {R.}~\bibnamefont {Schwengner}}, \bibinfo {author} {\bibfnamefont
  {S.}~\bibnamefont {Siem}}, {\it{et al.}},\ }\href@noop {} {\bibfield  {journal} {\bibinfo
  {journal} {Eur. Phys. J. A}\ }\textbf {\bibinfo {volume} {55}},\ \bibinfo
  {pages} {172} (\bibinfo {year} {2019})}\BibitemShut {NoStop}%
\bibitem [{\citenamefont {Kawano}\ \emph {et~al.}(2020)\citenamefont {Kawano},
  \citenamefont {Cho}, \citenamefont {Dimitriou}, \citenamefont {Filipescu},
  \citenamefont {Iwamoto}, \citenamefont {Plujko}, \citenamefont {Tao},
  \citenamefont {Utsunomiya}, \citenamefont {Varlamov}, \citenamefont {Xu},
  \citenamefont {Capote}, \citenamefont {Gheorghe}, \citenamefont
  {Gorbachenko}, \citenamefont {Jin}, \citenamefont {Renstrøm}, \citenamefont
  {Sin}, \citenamefont {Stopani}, \citenamefont {Tian}, \citenamefont {Tveten},
  \citenamefont {Wang}, \citenamefont {Belgya}, \citenamefont {Firestone},
  \citenamefont {Goriely}, \citenamefont {Kopecky}, \citenamefont {Krtička},
  \citenamefont {Schwengner}, \citenamefont {Siem},\ and\ \citenamefont
  {Wiedeking}}]{Kawano2020}%
  \BibitemOpen
  \bibfield  {author} {\bibinfo {author} {\bibfnamefont {T.}~\bibnamefont
  {Kawano}}, \bibinfo {author} {\bibfnamefont {Y.}~\bibnamefont {Cho}},
  \bibinfo {author} {\bibfnamefont {P.}~\bibnamefont {Dimitriou}}, \bibinfo
  {author} {\bibfnamefont {D.}~\bibnamefont {Filipescu}}, \bibinfo {author}
  {\bibfnamefont {N.}~\bibnamefont {Iwamoto}}, \bibinfo {author} {\bibfnamefont
  {V.}~\bibnamefont {Plujko}}, \bibinfo {author} {\bibfnamefont
  {X.}~\bibnamefont {Tao}}, \bibinfo {author} {\bibfnamefont {H.}~\bibnamefont
  {Utsunomiya}}, \bibinfo {author} {\bibfnamefont {V.}~\bibnamefont
  {Varlamov}}, \bibinfo {author} {\bibfnamefont {R.}~\bibnamefont {Xu}},
  {\it{et al.}},\ }\href@noop {} {\bibfield  {journal}
  {\bibinfo  {journal} {Nucl. Data Sheets}\ }\textbf {\bibinfo {volume}
  {163}},\ \bibinfo {pages} {109} (\bibinfo {year} {2020})}\BibitemShut
  {NoStop}%
\bibitem [{\citenamefont {Harakeh}(2001)}]{Harakeh2001}%
  \BibitemOpen
  \bibfield  {author} {\bibinfo {author} {\bibfnamefont {M.}~\bibnamefont
  {Harakeh}},\ }\href@noop {} {\emph {\bibinfo {title} {Giant Resonances:
  Fundamental High-frequency Modes of Nuclear Excitation}}}\ (\bibinfo
  {publisher} {Oxford University Press},\ \bibinfo {year} {2001})\BibitemShut
  {NoStop}%
\bibitem [{\citenamefont {Tamii}\ \emph {et~al.}(2011)\citenamefont {Tamii},
  \citenamefont {Poltoratska}, \citenamefont {von Neumann-Cosel}, \citenamefont
  {Fujita}, \citenamefont {Adachi}, \citenamefont {Bertulani}, \citenamefont
  {Carter}, \citenamefont {Dozono}, \citenamefont {Fujita}, \citenamefont
  {Fujita}, \citenamefont {Hatanaka}, \citenamefont {Ishikawa}, \citenamefont
  {Itoh}, \citenamefont {Kawabata}, \citenamefont {Kalmykov}, \citenamefont
  {Krumbholz}, \citenamefont {Litvinova}, \citenamefont {Matsubara},
  \citenamefont {Nakanishi}, \citenamefont {Neveling}, \citenamefont {Okamura},
  \citenamefont {Ong}, \citenamefont {\"Ozel-Tashenov}, \citenamefont
  {Ponomarev}, \citenamefont {Richter}, \citenamefont {Rubio}, \citenamefont
  {Sakaguchi}, \citenamefont {Sakemi}, \citenamefont {Sasamoto}, \citenamefont
  {Shimbara}, \citenamefont {Shimizu}, \citenamefont {Smit}, \citenamefont
  {Suzuki}, \citenamefont {Tameshige}, \citenamefont {Wambach}, \citenamefont
  {Yamada}, \citenamefont {Yosoi},\ and\ \citenamefont {Zenihiro}}]{Tamii2011}%
  \BibitemOpen
  \bibfield  {author} {\bibinfo {author} {\bibfnamefont {A.}~\bibnamefont
  {Tamii}}, \bibinfo {author} {\bibfnamefont {I.}~\bibnamefont {Poltoratska}},
  \bibinfo {author} {\bibfnamefont {P.}~\bibnamefont {von Neumann-Cosel}},
  \bibinfo {author} {\bibfnamefont {Y.}~\bibnamefont {Fujita}}, \bibinfo
  {author} {\bibfnamefont {T.}~\bibnamefont {Adachi}}, \bibinfo {author}
  {\bibfnamefont {C.~A.}\ \bibnamefont {Bertulani}}, \bibinfo {author}
  {\bibfnamefont {J.}~\bibnamefont {Carter}}, \bibinfo {author} {\bibfnamefont
  {M.}~\bibnamefont {Dozono}}, \bibinfo {author} {\bibfnamefont
  {H.}~\bibnamefont {Fujita}}, \bibinfo {author} {\bibfnamefont
  {K.}~\bibnamefont {Fujita}}, {\it{et al.}},\ }\href@noop {} {\bibfield
  {journal} {\bibinfo  {journal} {Phys. Rev. Lett.}\ }\textbf {\bibinfo
  {volume} {107}},\ \bibinfo {pages} {062502} (\bibinfo {year}
  {2011})}\BibitemShut {NoStop}%
\bibitem [{\citenamefont {Heyde}\ \emph {et~al.}(2010)\citenamefont {Heyde},
  \citenamefont {{von Neumann-Cosel}},\ and\ \citenamefont
  {Richter}}]{Heyde2010}%
  \BibitemOpen
  \bibfield  {author} {\bibinfo {author} {\bibfnamefont {K.}~\bibnamefont
  {Heyde}}, \bibinfo {author} {\bibfnamefont {P.}~\bibnamefont {{von
  Neumann-Cosel}}}, \ and\ \bibinfo {author} {\bibfnamefont {A.}~\bibnamefont
  {Richter}},\ }\href@noop {} {\bibfield  {journal} {\bibinfo  {journal} {Rev.
  Mod. Phys.}\ }\textbf {\bibinfo {volume} {82}},\ \bibinfo {pages} {2365}
  (\bibinfo {year} {2010})}\BibitemShut {NoStop}%
\bibitem [{\citenamefont {Schwengner}\ \emph {et~al.}(2013)\citenamefont
  {Schwengner}, \citenamefont {Frauendorf},\ and\ \citenamefont
  {Larsen}}]{Schwengner2013}%
  \BibitemOpen
  \bibfield  {author} {\bibinfo {author} {\bibfnamefont {R.}~\bibnamefont
  {Schwengner}}, \bibinfo {author} {\bibfnamefont {S.}~\bibnamefont
  {Frauendorf}}, \ and\ \bibinfo {author} {\bibfnamefont {A.~C.}\ \bibnamefont
  {Larsen}},\ }\href@noop {} {\bibfield  {journal} {\bibinfo  {journal} {Phys.
  Rev. Lett.}\ }\textbf {\bibinfo {volume} {111}},\ \bibinfo {pages} {232504}
  (\bibinfo {year} {2013})}\BibitemShut {NoStop}%
\bibitem [{\citenamefont {Paar}\ \emph {et~al.}(2007)\citenamefont {Paar},
  \citenamefont {Vretenar}, \citenamefont {Khan},\ and\ \citenamefont
  {Col\`{o}}}]{Paar2007}%
  \BibitemOpen
  \bibfield  {author} {\bibinfo {author} {\bibfnamefont {N.}~\bibnamefont
  {Paar}}, \bibinfo {author} {\bibfnamefont {D.}~\bibnamefont {Vretenar}},
  \bibinfo {author} {\bibfnamefont {E.}~\bibnamefont {Khan}}, \ and\ \bibinfo
  {author} {\bibfnamefont {G.}~\bibnamefont {Col\`{o}}},\ }\href@noop {}
  {\bibfield  {journal} {\bibinfo  {journal} {Rep. Prog. Phys.}\ }\textbf
  {\bibinfo {volume} {70}},\ \bibinfo {pages} {R02} (\bibinfo {year}
  {2007})}\BibitemShut {NoStop}%
\bibitem [{\citenamefont {Savran}\ \emph {et~al.}(2013)\citenamefont {Savran},
  \citenamefont {Aumann},\ and\ \citenamefont {Zilges}}]{Savran2013}%
  \BibitemOpen
  \bibfield  {author} {\bibinfo {author} {\bibfnamefont {D.}~\bibnamefont
  {Savran}}, \bibinfo {author} {\bibfnamefont {T.}~\bibnamefont {Aumann}}, \
  and\ \bibinfo {author} {\bibfnamefont {A.}~\bibnamefont {Zilges}},\
  }\href@noop {} {\bibfield  {journal} {\bibinfo  {journal} {Prog. Part. Nucl.
  Phys.}\ }\textbf {\bibinfo {volume} {70}},\ \bibinfo {pages} {210} (\bibinfo
  {year} {2013})}\BibitemShut {NoStop}%
\bibitem [{\citenamefont {Bracco}\ \emph {et~al.}(2019)\citenamefont {Bracco},
  \citenamefont {Lanza},\ and\ \citenamefont {Tamii}}]{Bracco2019b}%
  \BibitemOpen
  \bibfield  {author} {\bibinfo {author} {\bibfnamefont {A.}~\bibnamefont
  {Bracco}}, \bibinfo {author} {\bibfnamefont {E.}~\bibnamefont {Lanza}}, \
  and\ \bibinfo {author} {\bibfnamefont {A.}~\bibnamefont {Tamii}},\
  }\href@noop {} {\bibfield  {journal} {\bibinfo  {journal} {Prog. Part. Nucl.
  Phys.}\ }\textbf {\bibinfo {volume} {106}},\ \bibinfo {pages} {360} (\bibinfo
  {year} {2019})}\BibitemShut {NoStop}%
\bibitem [{\citenamefont {Lanza}\ \emph {et~al.}(2023)\citenamefont {Lanza},
  \citenamefont {Pellegri}, \citenamefont {Vitturi},\ and\ \citenamefont
  {Andr\'{e}s}}]{Lanza2023}%
  \BibitemOpen
  \bibfield  {author} {\bibinfo {author} {\bibfnamefont {E.~G.}\ \bibnamefont
  {Lanza}}, \bibinfo {author} {\bibfnamefont {L.}~\bibnamefont {Pellegri}},
  \bibinfo {author} {\bibfnamefont {A.}~\bibnamefont {Vitturi}}, \ and\
  \bibinfo {author} {\bibfnamefont {M.~V.}\ \bibnamefont {Andr\'{e}s}},\
  }\href@noop {} {\bibfield  {journal} {\bibinfo  {journal} {Prog. Part. Nucl.
  Phys.}\ }\textbf {\bibinfo {volume} {129}},\ \bibinfo {pages} {104006}
  (\bibinfo {year} {2023})}\BibitemShut {NoStop}%
\bibitem [{\citenamefont {Spieker}\ \emph {et~al.}(2020)\citenamefont
  {Spieker}, \citenamefont {Heusler}, \citenamefont {Brown}, \citenamefont
  {Faestermann}, \citenamefont {Hertenberger}, \citenamefont {Potel},
  \citenamefont {Scheck}, \citenamefont {Tsoneva}, \citenamefont {Weinert},
  \citenamefont {Wirth},\ and\ \citenamefont {Zilges}}]{Spieker2020}%
  \BibitemOpen
  \bibfield  {author} {\bibinfo {author} {\bibfnamefont {M.}~\bibnamefont
  {Spieker}}, \bibinfo {author} {\bibfnamefont {A.}~\bibnamefont {Heusler}},
  \bibinfo {author} {\bibfnamefont {B.~A.}\ \bibnamefont {Brown}}, \bibinfo
  {author} {\bibfnamefont {T.}~\bibnamefont {Faestermann}}, \bibinfo {author}
  {\bibfnamefont {R.}~\bibnamefont {Hertenberger}}, \bibinfo {author}
  {\bibfnamefont {G.}~\bibnamefont {Potel}}, \bibinfo {author} {\bibfnamefont
  {M.}~\bibnamefont {Scheck}}, \bibinfo {author} {\bibfnamefont
  {N.}~\bibnamefont {Tsoneva}}, \bibinfo {author} {\bibfnamefont
  {M.}~\bibnamefont {Weinert}}, \bibinfo {author} {\bibfnamefont {H.-F.}\
  \bibnamefont {Wirth}}, \ and\ \bibinfo {author} {\bibfnamefont
  {A.}~\bibnamefont {Zilges}},\ }\href@noop {} {\bibfield  {journal} {\bibinfo
  {journal} {Phys. Rev. Lett.}\ }\textbf {\bibinfo {volume} {125}},\ \bibinfo
  {pages} {102503} (\bibinfo {year} {2020})}\BibitemShut {NoStop}%
\bibitem [{\citenamefont {Weinert}\ \emph {et~al.}(2021)\citenamefont
  {Weinert}, \citenamefont {Spieker}, \citenamefont {Potel}, \citenamefont
  {Tsoneva}, \citenamefont {M\"uscher}, \citenamefont {Wilhelmy},\ and\
  \citenamefont {Zilges}}]{Weinert2021}%
  \BibitemOpen
  \bibfield  {author} {\bibinfo {author} {\bibfnamefont {M.}~\bibnamefont
  {Weinert}}, \bibinfo {author} {\bibfnamefont {M.}~\bibnamefont {Spieker}},
  \bibinfo {author} {\bibfnamefont {G.}~\bibnamefont {Potel}}, \bibinfo
  {author} {\bibfnamefont {N.}~\bibnamefont {Tsoneva}}, \bibinfo {author}
  {\bibfnamefont {M.}~\bibnamefont {M\"uscher}}, \bibinfo {author}
  {\bibfnamefont {J.}~\bibnamefont {Wilhelmy}}, \ and\ \bibinfo {author}
  {\bibfnamefont {A.}~\bibnamefont {Zilges}},\ }\href@noop {} {\bibfield
  {journal} {\bibinfo  {journal} {Phys. Rev. Lett.}\ }\textbf {\bibinfo
  {volume} {127}},\ \bibinfo {pages} {242501} (\bibinfo {year}
  {2021})}\BibitemShut {NoStop}%
\bibitem [{\citenamefont {Guttormsen}\ \emph {et~al.}(1987)\citenamefont
  {Guttormsen}, \citenamefont {Rams\o{}y},\ and\ \citenamefont
  {Rekstad}}]{Guttormsen1987}%
  \BibitemOpen
  \bibfield  {author} {\bibinfo {author} {\bibfnamefont {M.}~\bibnamefont
  {Guttormsen}}, \bibinfo {author} {\bibfnamefont {T.}~\bibnamefont
  {Rams\o{}y}}, \ and\ \bibinfo {author} {\bibfnamefont {J.}~\bibnamefont
  {Rekstad}},\ }\href@noop {} {\bibfield  {journal} {\bibinfo  {journal} {Nucl.
  Instrum. Methods Phys. Res. A}\ }\textbf {\bibinfo {volume} {255}},\ \bibinfo
  {pages} {518} (\bibinfo {year} {1987})}\BibitemShut {NoStop}%
\bibitem [{\citenamefont {Guttormsen}\ \emph {et~al.}(1996)\citenamefont
  {Guttormsen}, \citenamefont {Tveter}, \citenamefont {Bergholt}, \citenamefont
  {Ingebretsen},\ and\ \citenamefont {Rekstad}}]{Guttormsen1996}%
  \BibitemOpen
  \bibfield  {author} {\bibinfo {author} {\bibfnamefont {M.}~\bibnamefont
  {Guttormsen}}, \bibinfo {author} {\bibfnamefont {T.~S.}\ \bibnamefont
  {Tveter}}, \bibinfo {author} {\bibfnamefont {L.}~\bibnamefont {Bergholt}},
  \bibinfo {author} {\bibfnamefont {F.}~\bibnamefont {Ingebretsen}}, \ and\
  \bibinfo {author} {\bibfnamefont {J.}~\bibnamefont {Rekstad}},\ }\href@noop
  {} {\bibfield  {journal} {\bibinfo  {journal} {Nucl. Instrum. Methods Phys.
  Res. A}\ }\textbf {\bibinfo {volume} {374}},\ \bibinfo {pages} {371}
  (\bibinfo {year} {1996})}\BibitemShut {NoStop}%
\bibitem [{\citenamefont {Schiller}\ \emph {et~al.}(2000)\citenamefont
  {Schiller}, \citenamefont {Bergholt}, \citenamefont {Guttormsen},
  \citenamefont {Melby}, \citenamefont {Rekstad},\ and\ \citenamefont
  {Siem}}]{Schiller2000}%
  \BibitemOpen
  \bibfield  {author} {\bibinfo {author} {\bibfnamefont {A.}~\bibnamefont
  {Schiller}}, \bibinfo {author} {\bibfnamefont {L.}~\bibnamefont {Bergholt}},
  \bibinfo {author} {\bibfnamefont {M.}~\bibnamefont {Guttormsen}}, \bibinfo
  {author} {\bibfnamefont {E.}~\bibnamefont {Melby}}, \bibinfo {author}
  {\bibfnamefont {J.}~\bibnamefont {Rekstad}}, \ and\ \bibinfo {author}
  {\bibfnamefont {S.}~\bibnamefont {Siem}},\ }\href@noop {} {\bibfield
  {journal} {\bibinfo  {journal} {Nucl. Instrum. Methods Phys. Res. A}\
  }\textbf {\bibinfo {volume} {447}},\ \bibinfo {pages} {498} (\bibinfo {year}
  {2000})}\BibitemShut {NoStop}%
\bibitem [{\citenamefont {Larsen}\ \emph {et~al.}(2011)\citenamefont {Larsen},
  \citenamefont {Guttormsen}, \citenamefont {Krti\ifmmode~\check{c}\else
  \v{c}\fi{}ka}, \citenamefont {B\ifmmode~\check{e}\else \v{e}\fi{}t\'ak},
  \citenamefont {B\"urger}, \citenamefont {G\"orgen}, \citenamefont {Nyhus},
  \citenamefont {Rekstad}, \citenamefont {Schiller}, \citenamefont {Siem},
  \citenamefont {Toft}, \citenamefont {Tveten}, \citenamefont {Voinov},\ and\
  \citenamefont {Wikan}}]{Larsen2011}%
  \BibitemOpen
  \bibfield  {author} {\bibinfo {author} {\bibfnamefont {A.~C.}\ \bibnamefont
  {Larsen}}, \bibinfo {author} {\bibfnamefont {M.}~\bibnamefont {Guttormsen}},
  \bibinfo {author} {\bibfnamefont {M.}~\bibnamefont
  {Krti\ifmmode~\check{c}\else \v{c}\fi{}ka}}, \bibinfo {author} {\bibfnamefont
  {E.}~\bibnamefont {B\ifmmode~\check{e}\else \v{e}\fi{}t\'ak}}, \bibinfo
  {author} {\bibfnamefont {A.}~\bibnamefont {B\"urger}}, \bibinfo {author}
  {\bibfnamefont {A.}~\bibnamefont {G\"orgen}}, \bibinfo {author}
  {\bibfnamefont {H.~T.}\ \bibnamefont {Nyhus}}, \bibinfo {author}
  {\bibfnamefont {J.}~\bibnamefont {Rekstad}}, \bibinfo {author} {\bibfnamefont
  {A.}~\bibnamefont {Schiller}}, \bibinfo {author} {\bibfnamefont
  {S.}~\bibnamefont {Siem}}, {\it{et al.}},\
  }\href@noop {} {\bibfield  {journal} {\bibinfo  {journal} {Phys. Rev. C}\
  }\textbf {\bibinfo {volume} {83}},\ \bibinfo {pages} {034315} (\bibinfo
  {year} {2011})}\BibitemShut {NoStop}%
\bibitem [{\citenamefont {Agvaanluvsan}\ \emph
  {et~al.}(2009{\natexlab{a}})\citenamefont {Agvaanluvsan}, \citenamefont
  {Larsen}, \citenamefont {Guttormsen}, \citenamefont {Chankova}, \citenamefont
  {Mitchell}, \citenamefont {Schiller}, \citenamefont {Siem},\ and\
  \citenamefont {Voinov}}]{Agvaanluvsan2009a}%
  \BibitemOpen
  \bibfield  {author} {\bibinfo {author} {\bibfnamefont {U.}~\bibnamefont
  {Agvaanluvsan}}, \bibinfo {author} {\bibfnamefont {A.~C.}\ \bibnamefont
  {Larsen}}, \bibinfo {author} {\bibfnamefont {M.}~\bibnamefont {Guttormsen}},
  \bibinfo {author} {\bibfnamefont {R.}~\bibnamefont {Chankova}}, \bibinfo
  {author} {\bibfnamefont {G.~E.}\ \bibnamefont {Mitchell}}, \bibinfo {author}
  {\bibfnamefont {A.}~\bibnamefont {Schiller}}, \bibinfo {author}
  {\bibfnamefont {S.}~\bibnamefont {Siem}}, \ and\ \bibinfo {author}
  {\bibfnamefont {A.}~\bibnamefont {Voinov}},\ }\href@noop {} {\bibfield
  {journal} {\bibinfo  {journal} {Phys. Rev. C}\ }\textbf {\bibinfo {volume}
  {79}},\ \bibinfo {pages} {014320} (\bibinfo {year}
  {2009}{\natexlab{a}})}\BibitemShut {NoStop}%
\bibitem [{\citenamefont {Agvaanluvsan}\ \emph
  {et~al.}(2009{\natexlab{b}})\citenamefont {Agvaanluvsan}, \citenamefont
  {Larsen}, \citenamefont {Chankova}, \citenamefont {Guttormsen}, \citenamefont
  {Mitchell}, \citenamefont {Schiller}, \citenamefont {Siem},\ and\
  \citenamefont {Voinov}}]{Agvaanluvsan2009b}%
  \BibitemOpen
  \bibfield  {author} {\bibinfo {author} {\bibfnamefont {U.}~\bibnamefont
  {Agvaanluvsan}}, \bibinfo {author} {\bibfnamefont {A.~C.}\ \bibnamefont
  {Larsen}}, \bibinfo {author} {\bibfnamefont {R.}~\bibnamefont {Chankova}},
  \bibinfo {author} {\bibfnamefont {M.}~\bibnamefont {Guttormsen}}, \bibinfo
  {author} {\bibfnamefont {G.~E.}\ \bibnamefont {Mitchell}}, \bibinfo {author}
  {\bibfnamefont {A.}~\bibnamefont {Schiller}}, \bibinfo {author}
  {\bibfnamefont {S.}~\bibnamefont {Siem}}, \ and\ \bibinfo {author}
  {\bibfnamefont {A.}~\bibnamefont {Voinov}},\ }\href@noop {} {\bibfield
  {journal} {\bibinfo  {journal} {Phys. Rev. Lett.}\ }\textbf {\bibinfo
  {volume} {102}},\ \bibinfo {pages} {162504} (\bibinfo {year}
  {2009}{\natexlab{b}})}\BibitemShut {NoStop}%
\bibitem [{\citenamefont {Toft}\ \emph {et~al.}(2010)\citenamefont {Toft},
  \citenamefont {Larsen}, \citenamefont {Agvaanluvsan}, \citenamefont
  {B\"urger}, \citenamefont {Guttormsen}, \citenamefont {Mitchell},
  \citenamefont {Nyhus}, \citenamefont {Schiller}, \citenamefont {Siem},
  \citenamefont {Syed},\ and\ \citenamefont {Voinov}}]{Toft2010}%
  \BibitemOpen
  \bibfield  {author} {\bibinfo {author} {\bibfnamefont {H.~K.}\ \bibnamefont
  {Toft}}, \bibinfo {author} {\bibfnamefont {A.~C.}\ \bibnamefont {Larsen}},
  \bibinfo {author} {\bibfnamefont {U.}~\bibnamefont {Agvaanluvsan}}, \bibinfo
  {author} {\bibfnamefont {A.}~\bibnamefont {B\"urger}}, \bibinfo {author}
  {\bibfnamefont {M.}~\bibnamefont {Guttormsen}}, \bibinfo {author}
  {\bibfnamefont {G.~E.}\ \bibnamefont {Mitchell}}, \bibinfo {author}
  {\bibfnamefont {H.~T.}\ \bibnamefont {Nyhus}}, \bibinfo {author}
  {\bibfnamefont {A.}~\bibnamefont {Schiller}}, \bibinfo {author}
  {\bibfnamefont {S.}~\bibnamefont {Siem}}, \bibinfo {author} {\bibfnamefont
  {N.~U.~H.}\ \bibnamefont {Syed}}, \ and\ \bibinfo {author} {\bibfnamefont
  {A.}~\bibnamefont {Voinov}},\ }\href@noop {} {\bibfield  {journal} {\bibinfo
  {journal} {Phys. Rev. C}\ }\textbf {\bibinfo {volume} {81}},\ \bibinfo
  {pages} {064311} (\bibinfo {year} {2010})}\BibitemShut {NoStop}%
\bibitem [{\citenamefont {Toft}\ \emph {et~al.}(2011)\citenamefont {Toft},
  \citenamefont {Larsen}, \citenamefont {B\"urger}, \citenamefont {Guttormsen},
  \citenamefont {G\"orgen}, \citenamefont {Nyhus}, \citenamefont {Renstr\o{}m},
  \citenamefont {Siem}, \citenamefont {Tveten},\ and\ \citenamefont
  {Voinov}}]{Toft2011}%
  \BibitemOpen
  \bibfield  {author} {\bibinfo {author} {\bibfnamefont {H.~K.}\ \bibnamefont
  {Toft}}, \bibinfo {author} {\bibfnamefont {A.~C.}\ \bibnamefont {Larsen}},
  \bibinfo {author} {\bibfnamefont {A.}~\bibnamefont {B\"urger}}, \bibinfo
  {author} {\bibfnamefont {M.}~\bibnamefont {Guttormsen}}, \bibinfo {author}
  {\bibfnamefont {A.}~\bibnamefont {G\"orgen}}, \bibinfo {author}
  {\bibfnamefont {H.~T.}\ \bibnamefont {Nyhus}}, \bibinfo {author}
  {\bibfnamefont {T.}~\bibnamefont {Renstr\o{}m}}, \bibinfo {author}
  {\bibfnamefont {S.}~\bibnamefont {Siem}}, \bibinfo {author} {\bibfnamefont
  {G.~M.}\ \bibnamefont {Tveten}}, \ and\ \bibinfo {author} {\bibfnamefont
  {A.}~\bibnamefont {Voinov}},\ }\href@noop {} {\bibfield  {journal} {\bibinfo
  {journal} {Phys. Rev. C}\ }\textbf {\bibinfo {volume} {83}},\ \bibinfo
  {pages} {044320} (\bibinfo {year} {2011})}\BibitemShut {NoStop}%
\bibitem [{\citenamefont {Markova}\ \emph {et~al.}(2021)\citenamefont
  {Markova}, \citenamefont {von Neumann-Cosel}, \citenamefont {Larsen},
  \citenamefont {Bassauer}, \citenamefont {G\"orgen}, \citenamefont
  {Guttormsen}, \citenamefont {Bello~Garrote}, \citenamefont {Berg},
  \citenamefont {Bj\o{}r\o{}en}, \citenamefont {Dahl-Jacobsen}, \citenamefont
  {Eriksen}, \citenamefont {Gjestvang}, \citenamefont {Isaak}, \citenamefont
  {Mbabane}, \citenamefont {Paulsen}, \citenamefont {Pedersen}, \citenamefont
  {Pettersen}, \citenamefont {Richter}, \citenamefont {Sahin}, \citenamefont
  {Scholz}, \citenamefont {Siem}, \citenamefont {Tveten}, \citenamefont
  {Valsdottir}, \citenamefont {Wiedeking},\ and\ \citenamefont
  {Zeiser}}]{Markova2021}%
  \BibitemOpen
  \bibfield  {author} {\bibinfo {author} {\bibfnamefont {M.}~\bibnamefont
  {Markova}}, \bibinfo {author} {\bibfnamefont {P.}~\bibnamefont {von
  Neumann-Cosel}}, \bibinfo {author} {\bibfnamefont {A.~C.}\ \bibnamefont
  {Larsen}}, \bibinfo {author} {\bibfnamefont {S.}~\bibnamefont {Bassauer}},
  \bibinfo {author} {\bibfnamefont {A.}~\bibnamefont {G\"orgen}}, \bibinfo
  {author} {\bibfnamefont {M.}~\bibnamefont {Guttormsen}}, \bibinfo {author}
  {\bibfnamefont {F.~L.}\ \bibnamefont {Bello~Garrote}}, \bibinfo {author}
  {\bibfnamefont {H.~C.}\ \bibnamefont {Berg}}, \bibinfo {author}
  {\bibfnamefont {M.~M.}\ \bibnamefont {Bj\o{}r\o{}en}}, \bibinfo {author}
  {\bibfnamefont {T.}~\bibnamefont {Dahl-Jacobsen}}, {\it{et al.}},\ }\href@noop {} {\bibfield
  {journal} {\bibinfo  {journal} {Phys. Rev. Lett.}\ }\textbf {\bibinfo
  {volume} {127}},\ \bibinfo {pages} {182501} (\bibinfo {year}
  {2021})}\BibitemShut {NoStop}%
\bibitem [{\citenamefont {Markova}\ \emph {et~al.}(2022)\citenamefont
  {Markova}, \citenamefont {Larsen}, \citenamefont {von Neumann-Cosel},
  \citenamefont {Bassauer}, \citenamefont {G\"orgen}, \citenamefont
  {Guttormsen}, \citenamefont {Garrote}, \citenamefont {Berg}, \citenamefont
  {Bj\o{}r\o{}en}, \citenamefont {Eriksen}, \citenamefont {Gjestvang},
  \citenamefont {Isaak}, \citenamefont {Mbabane}, \citenamefont {Paulsen},
  \citenamefont {Pedersen}, \citenamefont {Pettersen}, \citenamefont {Richter},
  \citenamefont {Sahin}, \citenamefont {Scholz}, \citenamefont {Siem},
  \citenamefont {Tveten}, \citenamefont {Valsdottir},\ and\ \citenamefont
  {Wiedeking}}]{Markova2022}%
  \BibitemOpen
  \bibfield  {author} {\bibinfo {author} {\bibfnamefont {M.}~\bibnamefont
  {Markova}}, \bibinfo {author} {\bibfnamefont {A.~C.}\ \bibnamefont {Larsen}},
  \bibinfo {author} {\bibfnamefont {P.}~\bibnamefont {von Neumann-Cosel}},
  \bibinfo {author} {\bibfnamefont {S.}~\bibnamefont {Bassauer}}, \bibinfo
  {author} {\bibfnamefont {A.}~\bibnamefont {G\"orgen}}, \bibinfo {author}
  {\bibfnamefont {M.}~\bibnamefont {Guttormsen}}, \bibinfo {author}
  {\bibfnamefont {F.~L.~B.}\ \bibnamefont {Garrote}}, \bibinfo {author}
  {\bibfnamefont {H.~C.}\ \bibnamefont {Berg}}, \bibinfo {author}
  {\bibfnamefont {M.~M.}\ \bibnamefont {Bj\o{}r\o{}en}}, \bibinfo {author}
  {\bibfnamefont {T.~K.}\ \bibnamefont {Eriksen}}, {\it{et al.}},\ }\href@noop {} {\bibfield
  {journal} {\bibinfo  {journal} {Phys. Rev. C}\ }\textbf {\bibinfo {volume}
  {106}},\ \bibinfo {pages} {034322} (\bibinfo {year} {2022})}\BibitemShut
  {NoStop}%
\bibitem [{\citenamefont {Markova}\ \emph {et~al.}(2023)\citenamefont
  {Markova}, \citenamefont {Larsen}, \citenamefont {Tveten}, \citenamefont {von
  Neumann-Cosel}, \citenamefont {Eriksen}, \citenamefont {Bello~Garrote},
  \citenamefont {Crespo~Campo}, \citenamefont {Giacoppo}, \citenamefont
  {G\"orgen}, \citenamefont {Guttormsen}, \citenamefont {Hadynska-Klek},
  \citenamefont {Klintefjord}, \citenamefont {Renstr\o{}m}, \citenamefont
  {Sahin}, \citenamefont {Siem},\ and\ \citenamefont {Tornyi}}]{Markova2023}%
  \BibitemOpen
  \bibfield  {author} {\bibinfo {author} {\bibfnamefont {M.}~\bibnamefont
  {Markova}}, \bibinfo {author} {\bibfnamefont {A.~C.}\ \bibnamefont {Larsen}},
  \bibinfo {author} {\bibfnamefont {G.~M.}\ \bibnamefont {Tveten}}, \bibinfo
  {author} {\bibfnamefont {P.}~\bibnamefont {von Neumann-Cosel}}, \bibinfo
  {author} {\bibfnamefont {T.~K.}\ \bibnamefont {Eriksen}}, \bibinfo {author}
  {\bibfnamefont {F.~L.}\ \bibnamefont {Bello~Garrote}}, \bibinfo {author}
  {\bibfnamefont {L.}~\bibnamefont {Crespo~Campo}}, \bibinfo {author}
  {\bibfnamefont {F.}~\bibnamefont {Giacoppo}}, \bibinfo {author}
  {\bibfnamefont {A.}~\bibnamefont {G\"orgen}}, \bibinfo {author}
  {\bibfnamefont {M.}~\bibnamefont {Guttormsen}}, {\it{et al.}},\ }\href@noop {} {\bibfield  {journal} {\bibinfo
  {journal} {Phys. Rev. C}\ }\textbf {\bibinfo {volume} {108}},\ \bibinfo
  {pages} {014315} (\bibinfo {year} {2023})}\BibitemShut {NoStop}%
\bibitem [{\citenamefont {Markova}\ \emph {et~al.}(2024)\citenamefont
  {Markova}, \citenamefont {Larsen}, \citenamefont {von Neumann-Cosel},
  \citenamefont {Litvinova}, \citenamefont {Choplin}, \citenamefont {Goriely},
  \citenamefont {Martinet}, \citenamefont {Siess}, \citenamefont {Guttormsen},
  \citenamefont {Pogliano},\ and\ \citenamefont {Siem}}]{Markova2024}%
  \BibitemOpen
  \bibfield  {author} {\bibinfo {author} {\bibfnamefont {M.}~\bibnamefont
  {Markova}}, \bibinfo {author} {\bibfnamefont {A.~C.}\ \bibnamefont {Larsen}},
  \bibinfo {author} {\bibfnamefont {P.}~\bibnamefont {von Neumann-Cosel}},
  \bibinfo {author} {\bibfnamefont {E.}~\bibnamefont {Litvinova}}, \bibinfo
  {author} {\bibfnamefont {A.}~\bibnamefont {Choplin}}, \bibinfo {author}
  {\bibfnamefont {S.}~\bibnamefont {Goriely}}, \bibinfo {author} {\bibfnamefont
  {S.}~\bibnamefont {Martinet}}, \bibinfo {author} {\bibfnamefont
  {L.}~\bibnamefont {Siess}}, \bibinfo {author} {\bibfnamefont
  {M.}~\bibnamefont {Guttormsen}}, \bibinfo {author} {\bibfnamefont
  {F.}~\bibnamefont {Pogliano}}, {\it{et al.}},\ }\href@noop {} {\bibfield  {journal} {\bibinfo
  {journal} {Phys. Rev. C}\ }\textbf {\bibinfo {volume} {109}},\ \bibinfo
  {pages} {054311} (\bibinfo {year} {2024})}\BibitemShut {NoStop}%
\bibitem [{\citenamefont {Ingeberg}\ \emph
  {et~al.}(2022{\natexlab{a}})\citenamefont {Ingeberg}, \citenamefont {Siem},
  \citenamefont {Wiedeking}, \citenamefont {Sieja}, \citenamefont {Bleuel},
  \citenamefont {Brits}, \citenamefont {Bucher}, \citenamefont {Dinoko},
  \citenamefont {Easton}, \citenamefont {Görgen}, \citenamefont {Guttormsen},
  \citenamefont {Jones}, \citenamefont {Kheswa}, \citenamefont {Khumalo},
  \citenamefont {Larsen}, \citenamefont {Lawrie}, \citenamefont {Lawrie},
  \citenamefont {Majola}, \citenamefont {Malatji}, \citenamefont {Makhathini},
  \citenamefont {Maqabuka}, \citenamefont {Negi}, \citenamefont {Noncolela},
  \citenamefont {Papka}, \citenamefont {Sahin}, \citenamefont {Schwengner},
  \citenamefont {Tveten}, \citenamefont {Zeiser},\ and\ \citenamefont
  {Zikhali}}]{Ingeberg2020}%
  \BibitemOpen
  \bibfield  {author} {\bibinfo {author} {\bibfnamefont {V.~W.}\ \bibnamefont
  {Ingeberg}}, \bibinfo {author} {\bibfnamefont {S.}~\bibnamefont {Siem}},
  \bibinfo {author} {\bibfnamefont {M.}~\bibnamefont {Wiedeking}}, \bibinfo
  {author} {\bibfnamefont {K.}~\bibnamefont {Sieja}}, \bibinfo {author}
  {\bibfnamefont {D.~L.}\ \bibnamefont {Bleuel}}, \bibinfo {author}
  {\bibfnamefont {C.~P.}\ \bibnamefont {Brits}}, \bibinfo {author}
  {\bibfnamefont {T.~D.}\ \bibnamefont {Bucher}}, \bibinfo {author}
  {\bibfnamefont {T.~S.}\ \bibnamefont {Dinoko}}, \bibinfo {author}
  {\bibfnamefont {J.~L.}\ \bibnamefont {Easton}}, \bibinfo {author}
  {\bibfnamefont {A.}~\bibnamefont {Görgen}}, {\it{et al.}},\ }\href@noop {} {\bibfield  {journal} {\bibinfo  {journal} {Eur.
  Phys. J A}\ }\textbf {\bibinfo {volume} {56}},\ \bibinfo {pages} {68}
  (\bibinfo {year} {2022}{\natexlab{a}})}\BibitemShut {NoStop}%
\bibitem [{\citenamefont {Ingeberg}\ \emph
  {et~al.}(2022{\natexlab{b}})\citenamefont {Ingeberg}, \citenamefont {Jones},
  \citenamefont {Msebi}, \citenamefont {Siem}, \citenamefont {Wiedeking},
  \citenamefont {Avaa}, \citenamefont {Chisapi}, \citenamefont {Lawrie},
  \citenamefont {Malatji}, \citenamefont {Makhathini}, \citenamefont
  {Noncolela},\ and\ \citenamefont {Shirinda}}]{Ingeberg2022}%
  \BibitemOpen
  \bibfield  {author} {\bibinfo {author} {\bibfnamefont {V.~W.}\ \bibnamefont
  {Ingeberg}}, \bibinfo {author} {\bibfnamefont {P.}~\bibnamefont {Jones}},
  \bibinfo {author} {\bibfnamefont {L.}~\bibnamefont {Msebi}}, \bibinfo
  {author} {\bibfnamefont {S.}~\bibnamefont {Siem}}, \bibinfo {author}
  {\bibfnamefont {M.}~\bibnamefont {Wiedeking}}, \bibinfo {author}
  {\bibfnamefont {A.~A.}\ \bibnamefont {Avaa}}, \bibinfo {author}
  {\bibfnamefont {M.~V.}\ \bibnamefont {Chisapi}}, \bibinfo {author}
  {\bibfnamefont {E.~A.}\ \bibnamefont {Lawrie}}, \bibinfo {author}
  {\bibfnamefont {K.~L.}\ \bibnamefont {Malatji}}, \bibinfo {author}
  {\bibfnamefont {L.}~\bibnamefont {Makhathini}}, \bibinfo {author}
  {\bibfnamefont {S.~P.}\ \bibnamefont {Noncolela}}, \ and\ \bibinfo {author}
  {\bibfnamefont {O.}~\bibnamefont {Shirinda}},\ }\href@noop {} {\bibfield
  {journal} {\bibinfo  {journal} {Phys. Rev. C}\ }\textbf {\bibinfo {volume}
  {106}},\ \bibinfo {pages} {054315} (\bibinfo {year}
  {2022}{\natexlab{b}})}\BibitemShut {NoStop}%
\bibitem [{\citenamefont {Liddick}\ \emph {et~al.}(2019)\citenamefont
  {Liddick}, \citenamefont {Larsen}, \citenamefont {Guttormsen}, \citenamefont
  {Spyrou}, \citenamefont {Crider}, \citenamefont {Naqvi}, \citenamefont
  {Midtb\o{}}, \citenamefont {Bello~Garrote}, \citenamefont {Bleuel},
  \citenamefont {Crespo~Campo}, \citenamefont {Couture}, \citenamefont
  {Dombos}, \citenamefont {Giacoppo}, \citenamefont {G\"orgen}, \citenamefont
  {Hadynska-Klek}, \citenamefont {Hagen}, \citenamefont {Ingeberg},
  \citenamefont {Kheswa}, \citenamefont {Lewis}, \citenamefont {Mosby},
  \citenamefont {Perdikakis}, \citenamefont {Prokop}, \citenamefont {Quinn},
  \citenamefont {Renstr\o{}m}, \citenamefont {Rose}, \citenamefont {Sahin},
  \citenamefont {Siem}, \citenamefont {Tveten}, \citenamefont {Wiedeking},\
  and\ \citenamefont {Zeiser}}]{Liddick2019}%
  \BibitemOpen
  \bibfield  {author} {\bibinfo {author} {\bibfnamefont {S.~N.}\ \bibnamefont
  {Liddick}}, \bibinfo {author} {\bibfnamefont {A.~C.}\ \bibnamefont {Larsen}},
  \bibinfo {author} {\bibfnamefont {M.}~\bibnamefont {Guttormsen}}, \bibinfo
  {author} {\bibfnamefont {A.}~\bibnamefont {Spyrou}}, \bibinfo {author}
  {\bibfnamefont {B.~P.}\ \bibnamefont {Crider}}, \bibinfo {author}
  {\bibfnamefont {F.}~\bibnamefont {Naqvi}}, \bibinfo {author} {\bibfnamefont
  {J.~E.}\ \bibnamefont {Midtb\o{}}}, \bibinfo {author} {\bibfnamefont {F.~L.}\
  \bibnamefont {Bello~Garrote}}, \bibinfo {author} {\bibfnamefont {D.~L.}\
  \bibnamefont {Bleuel}}, \bibinfo {author} {\bibfnamefont {L.}~\bibnamefont
  {Crespo~Campo}}, {\it{et al.}},\ }\href@noop {} {\bibfield
  {journal} {\bibinfo  {journal} {Phys. Rev. C}\ }\textbf {\bibinfo {volume}
  {100}},\ \bibinfo {pages} {024624} (\bibinfo {year} {2019})}\BibitemShut
  {NoStop}%
\bibitem [{\citenamefont {Filipescu}\ \emph {et~al.}(2015)\citenamefont
  {Filipescu}, \citenamefont {Anzalone}, \citenamefont {Balabanski},
  \citenamefont {Belyshev}, \citenamefont {Camera}, \citenamefont {Cognata},
  \citenamefont {Constantin}, \citenamefont {Csige}, \citenamefont {Cuong},
  \citenamefont {Cwiok}, \citenamefont {Derya}, \citenamefont {Dominik},
  \citenamefont {Gai}, \citenamefont {Gales}, \citenamefont {Gheorghe},
  \citenamefont {Ishkhanov}, \citenamefont {Krasznahorkay}, \citenamefont
  {Kuznetsov}, \citenamefont {Mazzocchi}, \citenamefont {Orlin}, \citenamefont
  {Pietralla}, \citenamefont {Sin}, \citenamefont {Spitaleri}, \citenamefont
  {Stopani}, \citenamefont {Tesileanu}, \citenamefont {Ur}, \citenamefont
  {Ursu}, \citenamefont {Utsunomiya}, \citenamefont {Varlamov}, \citenamefont
  {Weller}, \citenamefont {Zamfir},\ and\ \citenamefont
  {Zilges}}]{Filipescu2015}%
  \BibitemOpen
  \bibfield  {author} {\bibinfo {author} {\bibfnamefont {D.}~\bibnamefont
  {Filipescu}}, \bibinfo {author} {\bibfnamefont {A.}~\bibnamefont {Anzalone}},
  \bibinfo {author} {\bibfnamefont {D.~L.}\ \bibnamefont {Balabanski}},
  \bibinfo {author} {\bibfnamefont {S.~S.}\ \bibnamefont {Belyshev}}, \bibinfo
  {author} {\bibfnamefont {F.}~\bibnamefont {Camera}}, \bibinfo {author}
  {\bibfnamefont {M.~L.}\ \bibnamefont {Cognata}}, \bibinfo {author}
  {\bibfnamefont {P.}~\bibnamefont {Constantin}}, \bibinfo {author}
  {\bibfnamefont {L.}~\bibnamefont {Csige}}, \bibinfo {author} {\bibfnamefont
  {P.~V.}\ \bibnamefont {Cuong}}, \bibinfo {author} {\bibfnamefont
  {M.}~\bibnamefont {Cwiok}}, {\it{et al.}},\ }\href@noop {} {\bibfield
  {journal} {\bibinfo  {journal} {Eur. Phys. J.}\ }\textbf {\bibinfo {volume}
  {A51}},\ \bibinfo {pages} {185} (\bibinfo {year} {2015})}\BibitemShut
  {NoStop}%
\bibitem [{\citenamefont {Gales}\ \emph {et~al.}(2016)\citenamefont {Gales},
  \citenamefont {Balabanski}, \citenamefont {Negoita}, \citenamefont
  {Tesileanu}, \citenamefont {Ur}, \citenamefont {Ursescu},\ and\ \citenamefont
  {Zamfir}}]{Gales2016}%
  \BibitemOpen
  \bibfield  {author} {\bibinfo {author} {\bibfnamefont {S.}~\bibnamefont
  {Gales}}, \bibinfo {author} {\bibfnamefont {D.~L.}\ \bibnamefont
  {Balabanski}}, \bibinfo {author} {\bibfnamefont {F.}~\bibnamefont {Negoita}},
  \bibinfo {author} {\bibfnamefont {O.}~\bibnamefont {Tesileanu}}, \bibinfo
  {author} {\bibfnamefont {C.~A.}\ \bibnamefont {Ur}}, \bibinfo {author}
  {\bibfnamefont {D.}~\bibnamefont {Ursescu}}, \ and\ \bibinfo {author}
  {\bibfnamefont {N.~V.}\ \bibnamefont {Zamfir}},\ }\href@noop {} {\bibfield
  {journal} {\bibinfo  {journal} {Phys. Scr.}\ }\textbf {\bibinfo {volume}
  {91}},\ \bibinfo {pages} {093004} (\bibinfo {year} {2016})}\BibitemShut
  {NoStop}%
\bibitem [{\citenamefont {Gales}\ \emph {et~al.}(2018)\citenamefont {Gales},
  \citenamefont {Tanaka}, \citenamefont {Balabanski}, \citenamefont {Negoita},
  \citenamefont {Stutman}, \citenamefont {Tesileanu}, \citenamefont {Ur},
  \citenamefont {Ursescu}, \citenamefont {Andrei}, \citenamefont {Ataman},
  \citenamefont {Cernaianu}, \citenamefont {D’Alessi}, \citenamefont
  {Dancus}, \citenamefont {Diaconescu}, \citenamefont {Djourelov},
  \citenamefont {Filipescu}, \citenamefont {Ghenuche}, \citenamefont {Ghita},
  \citenamefont {Matei}, \citenamefont {Seto}, \citenamefont {Zeng},\ and\
  \citenamefont {Zamfir}}]{Gales2018}%
  \BibitemOpen
  \bibfield  {author} {\bibinfo {author} {\bibfnamefont {S.}~\bibnamefont
  {Gales}}, \bibinfo {author} {\bibfnamefont {K.~A.}\ \bibnamefont {Tanaka}},
  \bibinfo {author} {\bibfnamefont {D.~L.}\ \bibnamefont {Balabanski}},
  \bibinfo {author} {\bibfnamefont {F.}~\bibnamefont {Negoita}}, \bibinfo
  {author} {\bibfnamefont {D.}~\bibnamefont {Stutman}}, \bibinfo {author}
  {\bibfnamefont {O.}~\bibnamefont {Tesileanu}}, \bibinfo {author}
  {\bibfnamefont {C.~A.}\ \bibnamefont {Ur}}, \bibinfo {author} {\bibfnamefont
  {D.}~\bibnamefont {Ursescu}}, \bibinfo {author} {\bibfnamefont
  {I.}~\bibnamefont {Andrei}}, \bibinfo {author} {\bibfnamefont
  {S.}~\bibnamefont {Ataman}}, {\it{et al.}},\ }\href@noop {} {\bibfield  {journal} {\bibinfo  {journal} {Rep.
  Prog. Phys.}\ }\textbf {\bibinfo {volume} {81}},\ \bibinfo {pages} {094301}
  (\bibinfo {year} {2018})}\BibitemShut {NoStop}%
\bibitem [{\citenamefont {Tanaka}\ \emph {et~al.}(2020)\citenamefont {Tanaka},
  \citenamefont {Spohr}, \citenamefont {Balabanski}, \citenamefont {Balascuta},
  \citenamefont {Capponi}, \citenamefont {Cernaianu}, \citenamefont {Cuciuc},
  \citenamefont {Cucoanes}, \citenamefont {Dancus}, \citenamefont {Dhal},
  \citenamefont {Doria}, \citenamefont {Ghenuche}, \citenamefont {Ghita},
  \citenamefont {Kisyov}, \citenamefont {Nastasa}, \citenamefont {Ong},
  \citenamefont {Rotaru}, \citenamefont {Sangwan}, \citenamefont
  {Söderström}, \citenamefont {Stutman}, \citenamefont {Suliman},
  \citenamefont {Tesileanu}, \citenamefont {Tudor}, \citenamefont {Tsoneva},
  \citenamefont {Ur}, \citenamefont {Ursescu},\ and\ \citenamefont
  {Zamfir}}]{Tanaka2020}%
  \BibitemOpen
  \bibfield  {author} {\bibinfo {author} {\bibfnamefont {K.~A.}\ \bibnamefont
  {Tanaka}}, \bibinfo {author} {\bibfnamefont {K.~M.}\ \bibnamefont {Spohr}},
  \bibinfo {author} {\bibfnamefont {D.~L.}\ \bibnamefont {Balabanski}},
  \bibinfo {author} {\bibfnamefont {S.}~\bibnamefont {Balascuta}}, \bibinfo
  {author} {\bibfnamefont {L.}~\bibnamefont {Capponi}}, \bibinfo {author}
  {\bibfnamefont {M.~O.}\ \bibnamefont {Cernaianu}}, \bibinfo {author}
  {\bibfnamefont {M.}~\bibnamefont {Cuciuc}}, \bibinfo {author} {\bibfnamefont
  {A.}~\bibnamefont {Cucoanes}}, \bibinfo {author} {\bibfnamefont
  {I.}~\bibnamefont {Dancus}}, \bibinfo {author} {\bibfnamefont
  {A.}~\bibnamefont {Dhal}}, {\it{et al.}},\ }\href@noop {} {\bibfield  {journal} {\bibinfo  {journal} {Matter
  Radiat. Extremes}\ }\textbf {\bibinfo {volume} {5}},\ \bibinfo {pages}
  {024402} (\bibinfo {year} {2020})}\BibitemShut {NoStop}%
\bibitem [{\citenamefont {Constantin}\ \emph {et~al.}(2024)\citenamefont
  {Constantin}, \citenamefont {Matei},\ and\ \citenamefont
  {Ur}}]{Constantin2024}%
  \BibitemOpen
  \bibfield  {author} {\bibinfo {author} {\bibfnamefont {P.}~\bibnamefont
  {Constantin}}, \bibinfo {author} {\bibfnamefont {C.}~\bibnamefont {Matei}}, \
  and\ \bibinfo {author} {\bibfnamefont {C.~A.}\ \bibnamefont {Ur}},\
  }\href@noop {} {\bibfield  {journal} {\bibinfo  {journal} {Phys. Rev. Accel.
  Beams}\ }\textbf {\bibinfo {volume} {27}},\ \bibinfo {pages} {021601}
  (\bibinfo {year} {2024})}\BibitemShut {NoStop}%
\bibitem [{\citenamefont {Isaak}\ \emph {et~al.}(2019)\citenamefont {Isaak},
  \citenamefont {Savran}, \citenamefont {Löher}, \citenamefont {Beck},
  \citenamefont {Bhike}, \citenamefont {Gayer}, \citenamefont
  {\relax{Krishichayan}}, \citenamefont {Pietralla}, \citenamefont {Scheck},
  \citenamefont {Tornow}, \citenamefont {Werner}, \citenamefont {Zilges},\ and\
  \citenamefont {Zweidinger}}]{Isaak2019}%
  \BibitemOpen
  \bibfield  {author} {\bibinfo {author} {\bibfnamefont {J.}~\bibnamefont
  {Isaak}}, \bibinfo {author} {\bibfnamefont {D.}~\bibnamefont {Savran}},
  \bibinfo {author} {\bibfnamefont {B.}~\bibnamefont {Löher}}, \bibinfo
  {author} {\bibfnamefont {T.}~\bibnamefont {Beck}}, \bibinfo {author}
  {\bibfnamefont {M.}~\bibnamefont {Bhike}}, \bibinfo {author} {\bibfnamefont
  {U.}~\bibnamefont {Gayer}}, \bibinfo {author} {\bibnamefont
  {\relax{Krishichayan}}}, \bibinfo {author} {\bibfnamefont {N.}~\bibnamefont
  {Pietralla}}, \bibinfo {author} {\bibfnamefont {M.}~\bibnamefont {Scheck}},
  \bibinfo {author} {\bibfnamefont {W.}~\bibnamefont {Tornow}}, \bibinfo
  {author} {\bibfnamefont {V.}~\bibnamefont {Werner}}, \bibinfo {author}
  {\bibfnamefont {A.}~\bibnamefont {Zilges}}, \ and\ \bibinfo {author}
  {\bibfnamefont {M.}~\bibnamefont {Zweidinger}},\ }\href@noop {} {\bibfield
  {journal} {\bibinfo  {journal} {Phys. Lett. B}\ }\textbf {\bibinfo {volume}
  {788}},\ \bibinfo {pages} {225} (\bibinfo {year} {2019})}\BibitemShut
  {NoStop}%
\bibitem [{\citenamefont {Camera}\ \emph {et~al.}(2016)\citenamefont {Camera},
  \citenamefont {Utsunomiya}, \citenamefont {Varlamov}, \citenamefont
  {Filipescu}, \citenamefont {Baran}, \citenamefont {Bracco}, \citenamefont
  {Colo}, \citenamefont {Gheorghe}, \citenamefont {Glodariu}, \citenamefont
  {Matei},\ and\ \citenamefont {Wieland}}]{Camera2016}%
  \BibitemOpen
  \bibfield  {author} {\bibinfo {author} {\bibfnamefont {F.}~\bibnamefont
  {Camera}}, \bibinfo {author} {\bibfnamefont {H.}~\bibnamefont {Utsunomiya}},
  \bibinfo {author} {\bibfnamefont {V.}~\bibnamefont {Varlamov}}, \bibinfo
  {author} {\bibfnamefont {D.}~\bibnamefont {Filipescu}}, \bibinfo {author}
  {\bibfnamefont {V.}~\bibnamefont {Baran}}, \bibinfo {author} {\bibfnamefont
  {A.}~\bibnamefont {Bracco}}, \bibinfo {author} {\bibfnamefont
  {G.}~\bibnamefont {Colo}}, \bibinfo {author} {\bibfnamefont {I.}~\bibnamefont
  {Gheorghe}}, \bibinfo {author} {\bibfnamefont {T.}~\bibnamefont {Glodariu}},
  \bibinfo {author} {\bibfnamefont {C.}~\bibnamefont {Matei}}, \ and\ \bibinfo
  {author} {\bibfnamefont {O.}~\bibnamefont {Wieland}},\ }\href@noop {}
  {\bibfield  {journal} {\bibinfo  {journal} {Rom. Rep. Phys.}\ }\textbf
  {\bibinfo {volume} {68}},\ \bibinfo {pages} {S539} (\bibinfo {year}
  {2016})}\BibitemShut {NoStop}%
\bibitem [{\citenamefont {Krzysiek}\ \emph {et~al.}(2019)\citenamefont
  {Krzysiek}, \citenamefont {Camera}, \citenamefont {Filipescu}, \citenamefont
  {Utsunomiya}, \citenamefont {Col\`{o}}, \citenamefont {Gheorghe},\ and\
  \citenamefont {Niu}}]{Krzysiek2019a}%
  \BibitemOpen
  \bibfield  {author} {\bibinfo {author} {\bibfnamefont {M.}~\bibnamefont
  {Krzysiek}}, \bibinfo {author} {\bibfnamefont {F.}~\bibnamefont {Camera}},
  \bibinfo {author} {\bibfnamefont {D.~M.}\ \bibnamefont {Filipescu}}, \bibinfo
  {author} {\bibfnamefont {H.}~\bibnamefont {Utsunomiya}}, \bibinfo {author}
  {\bibfnamefont {G.}~\bibnamefont {Col\`{o}}}, \bibinfo {author}
  {\bibfnamefont {I.}~\bibnamefont {Gheorghe}}, \ and\ \bibinfo {author}
  {\bibfnamefont {Y.}~\bibnamefont {Niu}},\ }\href@noop {} {\bibfield
  {journal} {\bibinfo  {journal} {Nucl. Instrum. Methods Phys. Res. A}\
  }\textbf {\bibinfo {volume} {916}},\ \bibinfo {pages} {257} (\bibinfo {year}
  {2019})}\BibitemShut {NoStop}%
\bibitem [{\citenamefont {S\"{o}derstr\"{o}m}\ \emph
  {et~al.}(2022)\citenamefont {S\"{o}derstr\"{o}m}, \citenamefont
  {A\c{c}{\i}ks\"{o}z}, \citenamefont {Balabanski}, \citenamefont {Camera},
  \citenamefont {Capponi}, \citenamefont {Ciocan}, \citenamefont {Cuciuc},
  \citenamefont {Filipescu}, \citenamefont {Gheorghe}, \citenamefont
  {Glodariu}, \citenamefont {Kaur}, \citenamefont {Krzysiek}, \citenamefont
  {Matei}, \citenamefont {Roman}, \citenamefont {Rotaru}, \citenamefont
  {\textcommabelow{S}erban}, \citenamefont {State}, \citenamefont
  {Utsunomiya},\ and\ \citenamefont {Vasilca}}]{Soderstrom2022}%
  \BibitemOpen
  \bibfield  {author} {\bibinfo {author} {\bibfnamefont {P.-A.}\ \bibnamefont
  {S\"{o}derstr\"{o}m}}, \bibinfo {author} {\bibfnamefont {E.}~\bibnamefont
  {A\c{c}{\i}ks\"{o}z}}, \bibinfo {author} {\bibfnamefont {D.~L.}\ \bibnamefont
  {Balabanski}}, \bibinfo {author} {\bibfnamefont {F.}~\bibnamefont {Camera}},
  \bibinfo {author} {\bibfnamefont {L.}~\bibnamefont {Capponi}}, \bibinfo
  {author} {\bibfnamefont {G.}~\bibnamefont {Ciocan}}, \bibinfo {author}
  {\bibfnamefont {M.}~\bibnamefont {Cuciuc}}, \bibinfo {author} {\bibfnamefont
  {D.~M.}\ \bibnamefont {Filipescu}}, \bibinfo {author} {\bibfnamefont
  {I.}~\bibnamefont {Gheorghe}}, \bibinfo {author} {\bibfnamefont
  {T.}~\bibnamefont {Glodariu}}, \bibinfo {author} {\bibfnamefont
  {J.}~\bibnamefont {Kaur}}, {\it{et al.}},\ }\href@noop {} {\bibfield
  {journal} {\bibinfo  {journal} {Nucl. Instrum. Methods Phys. Res. A}\
  }\textbf {\bibinfo {volume} {1027}},\ \bibinfo {pages} {166171} (\bibinfo
  {year} {2022})}\BibitemShut {NoStop}%
\bibitem [{\citenamefont {Clisu}\ \emph {et~al.}(2023)\citenamefont {Clisu},
  \citenamefont {Gheorghe}, \citenamefont {Filipescu}, \citenamefont
  {Renstr\o{}m}, \citenamefont {Aciksoz}, \citenamefont {Boromiza},
  \citenamefont {Florea}, \citenamefont {Gosta}, \citenamefont {Ionescu},
  \citenamefont {Krzysiek}, \citenamefont {Maj}, \citenamefont {Mihai},
  \citenamefont {Negret}, \citenamefont {Nita}, \citenamefont {Olacel},
  \citenamefont {Petrone}, \citenamefont {Serban}, \citenamefont {Sotty},
  \citenamefont {Stiru}, \citenamefont {Stan}, \citenamefont {Suvaila},
  \citenamefont {Toma}, \citenamefont {Turturica}, \citenamefont {Tveten},
  \citenamefont {Ujeniuc}, \citenamefont {Wieland}, \citenamefont {Zeiser},
  \citenamefont {Camera},\ and\ \citenamefont {Utsunomiya}}]{Clisu2023}%
  \BibitemOpen
  \bibfield  {author} {\bibinfo {author} {\bibfnamefont {C.}~\bibnamefont
  {Clisu}}, \bibinfo {author} {\bibfnamefont {I.}~\bibnamefont {Gheorghe}},
  \bibinfo {author} {\bibfnamefont {D.}~\bibnamefont {Filipescu}}, \bibinfo
  {author} {\bibfnamefont {T.}~\bibnamefont {Renstr\o{}m}}, \bibinfo {author}
  {\bibfnamefont {E.}~\bibnamefont {Aciksoz}}, \bibinfo {author} {\bibfnamefont
  {M.}~\bibnamefont {Boromiza}}, \bibinfo {author} {\bibfnamefont
  {N.}~\bibnamefont {Florea}}, \bibinfo {author} {\bibfnamefont
  {G.}~\bibnamefont {Gosta}}, \bibinfo {author} {\bibfnamefont
  {A.}~\bibnamefont {Ionescu}}, \bibinfo {author} {\bibfnamefont
  {M.}~\bibnamefont {Krzysiek}}, \bibinfo {author} {\bibfnamefont
  {A.}~\bibnamefont {Maj}}, {\it{et al.}},\ }\href@noop {} {\bibfield  {journal}
  {\bibinfo  {journal} {EPJ Web Conf.}\ }\textbf {\bibinfo {volume} {284}},\
  \bibinfo {pages} {01015} (\bibinfo {year} {2023})}\BibitemShut {NoStop}%
\bibitem [{\citenamefont {S\"{o}derstr\"{o}m}\ \emph
  {et~al.}(2023{\natexlab{a}})\citenamefont {S\"{o}derstr\"{o}m}, \citenamefont
  {Ku\c{s}o\u{g}lu},\ and\ \citenamefont {Testov}}]{Soderstrom2023b}%
  \BibitemOpen
  \bibfield  {author} {\bibinfo {author} {\bibfnamefont {P.-A.}\ \bibnamefont
  {S\"{o}derstr\"{o}m}}, \bibinfo {author} {\bibfnamefont {A.}~\bibnamefont
  {Ku\c{s}o\u{g}lu}}, \ and\ \bibinfo {author} {\bibfnamefont {D.}~\bibnamefont
  {Testov}},\ }\href@noop {} {\bibfield  {journal} {\bibinfo  {journal} {Front.
  Astron. Space Sci.}\ }\textbf {\bibinfo {volume} {10}},\ \bibinfo {pages}
  {1248834} (\bibinfo {year} {2023}{\natexlab{a}})}\BibitemShut {NoStop}%
\bibitem [{\citenamefont {Mohr}(2004)}]{Mohr2004}%
  \BibitemOpen
  \bibfield  {author} {\bibinfo {author} {\bibfnamefont {P.}~\bibnamefont
  {Mohr}},\ }\href@noop {} {\bibfield  {journal} {\bibinfo  {journal} {AIP
  Conf. Proc.}\ }\textbf {\bibinfo {volume} {704}},\ \bibinfo {pages} {532}
  (\bibinfo {year} {2004})}\BibitemShut {NoStop}%
\bibitem [{\citenamefont {Utsunomiya}\ \emph {et~al.}(2006)\citenamefont
  {Utsunomiya}, \citenamefont {Mohr}, \citenamefont {Zilges},\ and\
  \citenamefont {Rayet}}]{Utsunomiya2006}%
  \BibitemOpen
  \bibfield  {author} {\bibinfo {author} {\bibfnamefont {H.}~\bibnamefont
  {Utsunomiya}}, \bibinfo {author} {\bibfnamefont {P.}~\bibnamefont {Mohr}},
  \bibinfo {author} {\bibfnamefont {A.}~\bibnamefont {Zilges}}, \ and\ \bibinfo
  {author} {\bibfnamefont {M.}~\bibnamefont {Rayet}},\ }\href@noop {}
  {\bibfield  {journal} {\bibinfo  {journal} {Nucl. Phys. A}\ }\textbf
  {\bibinfo {volume} {777}},\ \bibinfo {pages} {459} (\bibinfo {year}
  {2006})}\BibitemShut {NoStop}%
\bibitem [{\citenamefont {Rauscher}(2012)}]{Rauscher2012}%
  \BibitemOpen
  \bibfield  {author} {\bibinfo {author} {\bibfnamefont {T.}~\bibnamefont
  {Rauscher}},\ }\href@noop {} {\bibfield  {journal} {\bibinfo  {journal}
  {Astrophys. J., Suppl. Ser.}\ }\textbf {\bibinfo {volume} {291}},\ \bibinfo
  {pages} {26} (\bibinfo {year} {2012})}\BibitemShut {NoStop}%
\bibitem [{\citenamefont {Rauscher}(2013)}]{Rauscher2013a}%
  \BibitemOpen
  \bibfield  {author} {\bibinfo {author} {\bibfnamefont {T.}~\bibnamefont
  {Rauscher}},\ }\href@noop {} {\bibfield  {journal} {\bibinfo  {journal} {J.
  Phys. Conf. Ser.}\ }\textbf {\bibinfo {volume} {420}},\ \bibinfo {pages}
  {012138} (\bibinfo {year} {2013})}\BibitemShut {NoStop}%
\bibitem [{\citenamefont {Rauscher}(2014)}]{Rauscher2014}%
  \BibitemOpen
  \bibfield  {author} {\bibinfo {author} {\bibfnamefont {T.}~\bibnamefont
  {Rauscher}},\ }\href@noop {} {\bibfield  {journal} {\bibinfo  {journal} {AIP
  Adv.}\ }\textbf {\bibinfo {volume} {4}},\ \bibinfo {pages} {041012} (\bibinfo
  {year} {2014})}\BibitemShut {NoStop}%
\bibitem [{\citenamefont {Nishimura}\ \emph {et~al.}(2018)\citenamefont
  {Nishimura}, \citenamefont {Rauscher}, \citenamefont {Hirschi}, \citenamefont
  {\relax{A. St. J.} Murphy}, \citenamefont {Cescutti},\ and\ \citenamefont
  {Travaglio}}]{Nishimura2018}%
  \BibitemOpen
  \bibfield  {author} {\bibinfo {author} {\bibfnamefont {N.}~\bibnamefont
  {Nishimura}}, \bibinfo {author} {\bibfnamefont {T.}~\bibnamefont {Rauscher}},
  \bibinfo {author} {\bibfnamefont {R.}~\bibnamefont {Hirschi}}, \bibinfo
  {author} {\bibnamefont {\relax{A. St. J.} Murphy}}, \bibinfo {author}
  {\bibfnamefont {G.}~\bibnamefont {Cescutti}}, \ and\ \bibinfo {author}
  {\bibfnamefont {C.}~\bibnamefont {Travaglio}},\ }\href@noop {} {\bibfield
  {journal} {\bibinfo  {journal} {Mon. Not. R. Astron. Soc.}\ }\textbf
  {\bibinfo {volume} {474}},\ \bibinfo {pages} {3133} (\bibinfo {year}
  {2018})}\BibitemShut {NoStop}%
\bibitem [{\citenamefont {Bucurescu}\ \emph {et~al.}(2016)\citenamefont
  {Bucurescu}, \citenamefont {C\u{a}ta-Danil}, \citenamefont {Ciocan},
  \citenamefont {Costache}, \citenamefont {Deleanu}, \citenamefont {Dima},
  \citenamefont {Filipescu}, \citenamefont {Florea}, \citenamefont
  {Ghi\textcommabelow{t}\u{a}}, \citenamefont {Glodariu}, \citenamefont
  {Iva\textcommabelow{s}cu}, \citenamefont {Lic\u{a}}, \citenamefont
  {M\u{a}rginean}, \citenamefont {M\u{a}rginean}, \citenamefont {Mihai},
  \citenamefont {Negret}, \citenamefont {Ni\textcommabelow{t}\u{a}},
  \citenamefont {Ol\u{a}cel}, \citenamefont {Pascu}, \citenamefont {Sava},
  \citenamefont {Stroe}, \citenamefont {\textcommabelow{S}erban}, \citenamefont
  {\textcommabelow{S}uv\u{a}il\u{a}}, \citenamefont {Toma}, \citenamefont
  {Zamfir}, \citenamefont {C\u{a}ta-Danil}, \citenamefont {Gheorghe},
  \citenamefont {Mitu}, \citenamefont {Suliman}, \citenamefont {Ur},
  \citenamefont {Braunroth}, \citenamefont {Dewald}, \citenamefont {Fransen},
  \citenamefont {Bruce}, \citenamefont {Podoly\'{a}k}, \citenamefont {Regan},\
  and\ \citenamefont {Roberts}}]{Bucurescu2016}%
  \BibitemOpen
  \bibfield  {author} {\bibinfo {author} {\bibfnamefont {D.}~\bibnamefont
  {Bucurescu}}, \bibinfo {author} {\bibfnamefont {I.}~\bibnamefont
  {C\u{a}ta-Danil}}, \bibinfo {author} {\bibfnamefont {G.}~\bibnamefont
  {Ciocan}}, \bibinfo {author} {\bibfnamefont {C.}~\bibnamefont {Costache}},
  \bibinfo {author} {\bibfnamefont {D.}~\bibnamefont {Deleanu}}, \bibinfo
  {author} {\bibfnamefont {R.}~\bibnamefont {Dima}}, \bibinfo {author}
  {\bibfnamefont {D.}~\bibnamefont {Filipescu}}, \bibinfo {author}
  {\bibfnamefont {N.}~\bibnamefont {Florea}}, \bibinfo {author} {\bibfnamefont
  {D.~G.}\ \bibnamefont {Ghi\textcommabelow{t}\u{a}}}, \bibinfo {author}
  {\bibfnamefont {T.}~\bibnamefont {Glodariu}}, \bibinfo {author}
  {\bibfnamefont {M.}~\bibnamefont {Iva\textcommabelow{s}cu}}, {\it{et al.}},\ }\href@noop {} {\bibfield  {journal} {\bibinfo
  {journal} {Nucl. Instrum. Methods Phys. Res. A}\ }\textbf {\bibinfo {volume}
  {837}},\ \bibinfo {pages} {1} (\bibinfo {year} {2016})}\BibitemShut {NoStop}%
\bibitem [{\citenamefont {Weller}\ \emph {et~al.}(2016)\citenamefont {Weller},
  \citenamefont {Ur}, \citenamefont {Matei}, \citenamefont {Mueller},
  \citenamefont {Sikora}, \citenamefont {Suliman}, \citenamefont {Iancu},\ and\
  \citenamefont {Yasin}}]{Weller2016}%
  \BibitemOpen
  \bibfield  {author} {\bibinfo {author} {\bibfnamefont {H.~R.}\ \bibnamefont
  {Weller}}, \bibinfo {author} {\bibfnamefont {C.~A.}\ \bibnamefont {Ur}},
  \bibinfo {author} {\bibfnamefont {C.}~\bibnamefont {Matei}}, \bibinfo
  {author} {\bibfnamefont {J.~M.}\ \bibnamefont {Mueller}}, \bibinfo {author}
  {\bibfnamefont {M.~H.}\ \bibnamefont {Sikora}}, \bibinfo {author}
  {\bibfnamefont {G.}~\bibnamefont {Suliman}}, \bibinfo {author} {\bibfnamefont
  {V.}~\bibnamefont {Iancu}}, \ and\ \bibinfo {author} {\bibfnamefont
  {Z.}~\bibnamefont {Yasin}},\ }\href@noop {} {\bibfield  {journal} {\bibinfo
  {journal} {Rom. Rep. Phys.}\ }\textbf {\bibinfo {volume} {68}},\ \bibinfo
  {pages} {S447} (\bibinfo {year} {2016})}\BibitemShut {NoStop}%
\bibitem [{\citenamefont {Ku\c{s}o\u{g}lu}\ \emph
  {et~al.}(2024{\natexlab{a}})\citenamefont {Ku\c{s}o\u{g}lu}, \citenamefont
  {Constantin}, \citenamefont {S\"{o}derstr\"{o}m}, \citenamefont {Balabanski},
  \citenamefont {Cuciuc}, \citenamefont {Aogaki}, \citenamefont {Ban},
  \citenamefont {Borcea}, \citenamefont {Corbu}, \citenamefont {Costache},
  \citenamefont {Covali}, \citenamefont {Dinescu}, \citenamefont {Florea},
  \citenamefont {Iancu}, \citenamefont {Ionescu}, \citenamefont {Marginean},
  \citenamefont {Mihai}, \citenamefont {Mihai}, \citenamefont {Nedelcu},
  \citenamefont {Coman}, \citenamefont {Pai}, \citenamefont {Pappalardo},
  \citenamefont {Sirbu}, \citenamefont {~}, \citenamefont {Sotty},
  \citenamefont {Testov}, \citenamefont {Tozar}, \citenamefont {Turturica},
  \citenamefont {Turturica}, \citenamefont {Ujeniuc}, \citenamefont {Ur},
  \citenamefont {Vasilca},\ and\ \citenamefont {Zhu}}]{Kusoglu2024a}%
  \BibitemOpen
  \bibfield  {author} {\bibinfo {author} {\bibfnamefont {A.}~\bibnamefont
  {Ku\c{s}o\u{g}lu}}, \bibinfo {author} {\bibfnamefont {P.}~\bibnamefont
  {Constantin}}, \bibinfo {author} {\bibfnamefont {P.-A.}\ \bibnamefont
  {S\"{o}derstr\"{o}m}}, \bibinfo {author} {\bibfnamefont {D.~L.}\ \bibnamefont
  {Balabanski}}, \bibinfo {author} {\bibfnamefont {M.}~\bibnamefont {Cuciuc}},
  \bibinfo {author} {\bibfnamefont {S.}~\bibnamefont {Aogaki}}, \bibinfo
  {author} {\bibfnamefont {R.~S.}\ \bibnamefont {Ban}}, \bibinfo {author}
  {\bibfnamefont {R.}~\bibnamefont {Borcea}}, \bibinfo {author} {\bibfnamefont
  {R.}~\bibnamefont {Corbu}}, \bibinfo {author} {\bibfnamefont
  {C.}~\bibnamefont {Costache}}, \bibinfo {author} {\bibfnamefont
  {A.}~\bibnamefont {Covali}}, {\it{et al.}},\ }\href@noop {} {\bibfield  {journal} {\bibinfo  {journal} {Nuovo
  Cimento C}\ }\textbf {\bibinfo {volume} {47}},\ \bibinfo {pages} {47}
  (\bibinfo {year} {2024}{\natexlab{a}})}\BibitemShut {NoStop}%
\bibitem [{\citenamefont {Ku\c{s}o\u{g}lu}\ \emph
  {et~al.}(2024{\natexlab{b}})\citenamefont {Ku\c{s}o\u{g}lu}, \citenamefont
  {Balabanski}, \citenamefont {Hu}, \citenamefont {Fan}, \citenamefont {Xu},
  \citenamefont {Constantin}, \citenamefont {S\"oderstr\"om}, \citenamefont
  {Cuciuc}, \citenamefont {Aogaki}, \citenamefont {Ban}, \citenamefont
  {Borcea}, \citenamefont {Coman}, \citenamefont {Corbu}, \citenamefont
  {Costache}, \citenamefont {Covali}, \citenamefont {Dinescu}, \citenamefont
  {Florea}, \citenamefont {Iancu}, \citenamefont {Ionescu}, \citenamefont
  {M\ifmmode~\u{a}\else \u{a}\fi{}rginean}, \citenamefont {Mihai},
  \citenamefont {Mihai}, \citenamefont {Nedelcu}, \citenamefont {Petruse},
  \citenamefont {Pai}, \citenamefont {Pappalardo}, \citenamefont {Sirbu},
  \citenamefont {Sotty}, \citenamefont {Stan}, \citenamefont {State},
  \citenamefont {Testov}, \citenamefont {Tozar}, \citenamefont {Turturica},
  \citenamefont {Turturica}, \citenamefont {Ujeniuc}, \citenamefont {Ur},
  \citenamefont {Vasilca},\ and\ \citenamefont {Zhu}}]{Kusoglu2024b}%
  \BibitemOpen
  \bibfield  {author} {\bibinfo {author} {\bibfnamefont {A.}~\bibnamefont
  {Ku\c{s}o\u{g}lu}}, \bibinfo {author} {\bibfnamefont {D.~L.}\ \bibnamefont
  {Balabanski}}, \bibinfo {author} {\bibfnamefont {R.~Z.}\ \bibnamefont {Hu}},
  \bibinfo {author} {\bibfnamefont {S.~Q.}\ \bibnamefont {Fan}}, \bibinfo
  {author} {\bibfnamefont {F.~R.}\ \bibnamefont {Xu}}, \bibinfo {author}
  {\bibfnamefont {P.}~\bibnamefont {Constantin}}, \bibinfo {author}
  {\bibfnamefont {P.-A.}\ \bibnamefont {S\"oderstr\"om}}, \bibinfo {author}
  {\bibfnamefont {M.}~\bibnamefont {Cuciuc}}, \bibinfo {author} {\bibfnamefont
  {S.}~\bibnamefont {Aogaki}}, \bibinfo {author} {\bibfnamefont {R.~S.}\
  \bibnamefont {Ban}}, \bibinfo {author} {\bibfnamefont {R.}~\bibnamefont
  {Borcea}}, {\it{et al.}},\ }\href@noop {} {\bibfield  {journal} {\bibinfo  {journal} {Phys.
  Rev. Lett.}\ }\textbf {\bibinfo {volume} {133}},\ \bibinfo {pages} {072502}
  (\bibinfo {year} {2024}{\natexlab{b}})}\BibitemShut {NoStop}%
\bibitem [{\citenamefont {Wieland}\ \emph {et~al.}(2024)\citenamefont
  {Wieland}, \citenamefont {Bracco}, \citenamefont {Camera}, \citenamefont
  {Aogaki}, \citenamefont {Balabanski}, \citenamefont {Boicu(3)}, \citenamefont
  {Borcea}, \citenamefont {Boromiza}, \citenamefont {Burducea}, \citenamefont
  {Calinescu}, \citenamefont {Coman}, \citenamefont {Constantin}, \citenamefont
  {Costache}, \citenamefont {Ciemala}, \citenamefont {Ciocan}, \citenamefont
  {Clisu}, \citenamefont {Crespi}, \citenamefont {Cuciuc}, \citenamefont
  {Dhal}, \citenamefont {Djourelov}, \citenamefont {Florea}, \citenamefont
  {Gheorghe}, \citenamefont {Giaz}, \citenamefont {Iancu}, \citenamefont
  {Kahl}, \citenamefont {Kmiecik}, \citenamefont {Ku\c{s}o\u{g}lu},
  \citenamefont {Lica}, \citenamefont {M\u{a}rginean}, \citenamefont {Maj},
  \citenamefont {Marginean}, \citenamefont {Mihai}, \citenamefont {Mihai(5)},
  \citenamefont {Million}, \citenamefont {Neacsu}, \citenamefont {Nichita},
  \citenamefont {Nit¸aˇ}, \citenamefont {Pai}, \citenamefont {Pappalardo},
  \citenamefont {Petruse}, \citenamefont {Rotaru}, \citenamefont {Serban},
  \citenamefont {S\"{o}derstr\"{o}om}, \citenamefont {Sotty}, \citenamefont
  {Stan}, \citenamefont {State}, \citenamefont {Stiru}, \citenamefont {Stoica},
  \citenamefont {Testov}, \citenamefont {Toma}, \citenamefont {Tozar},
  \citenamefont {Turturic\u{a}}, \citenamefont {Turturic\u{a}}, \citenamefont
  {Ujeniuc}, \citenamefont {Vasilca},\ and\ \citenamefont {Xu}}]{Wieland2024a}%
  \BibitemOpen
  \bibfield  {author} {\bibinfo {author} {\bibfnamefont {O.}~\bibnamefont
  {Wieland}}, \bibinfo {author} {\bibfnamefont {A.}~\bibnamefont {Bracco}},
  \bibinfo {author} {\bibfnamefont {F.}~\bibnamefont {Camera}}, \bibinfo
  {author} {\bibfnamefont {S.}~\bibnamefont {Aogaki}}, \bibinfo {author}
  {\bibfnamefont {D.~L.}\ \bibnamefont {Balabanski}}, \bibinfo {author}
  {\bibfnamefont {E.}~\bibnamefont {Boicu}}, \bibinfo {author}
  {\bibfnamefont {R.}~\bibnamefont {Borcea}}, \bibinfo {author} {\bibfnamefont
  {M.}~\bibnamefont {Boromiza}}, \bibinfo {author} {\bibfnamefont
  {I.}~\bibnamefont {Burducea}}, \bibinfo {author} {\bibfnamefont
  {S.}~\bibnamefont {Calinescu}}, {\it{et al.}},\ }\href@noop {} {\bibfield  {journal}
  {\bibinfo  {journal} {Nuovo Cimento C}\ }\textbf {\bibinfo {volume} {47}},\
  \bibinfo {pages} {24} (\bibinfo {year} {2024})}\BibitemShut {NoStop}%
\bibitem [{\citenamefont {S\"{o}derstr\"{o}m}\ \emph
  {et~al.}(2024)\citenamefont {S\"{o}derstr\"{o}m}, \citenamefont
  {Ku\c{s}o\u{g}lu}, \citenamefont {Balabanski}, \citenamefont {Brezeanu},
  \citenamefont {Choudhury}, \citenamefont {Gavrilescu}, \citenamefont
  {Gu\textcommabelow{t}oiu}, \citenamefont {Ioannidis}, \citenamefont
  {Lorusso}, \citenamefont {Markova}, \citenamefont {Roy}, \citenamefont
  {Testov}, \citenamefont {Adachi}, \citenamefont {Aogaki}, \citenamefont
  {Borcea}, \citenamefont {Camera}, \citenamefont {Constantin}, \citenamefont
  {Costache}, \citenamefont {Cuciuc}, \citenamefont {Crespi}, \citenamefont
  {Florea}, \citenamefont {Fujikawa}, \citenamefont {Furuno}, \citenamefont
  {Giaz}, \citenamefont {Kawabata}, \citenamefont {Mihai}, \citenamefont
  {Mihai}, \citenamefont {Million}, \citenamefont {Nichita}, \citenamefont
  {Niina}, \citenamefont {Okamoto}, \citenamefont {Sakanashi}, \citenamefont
  {Stan}, \citenamefont {Tamii}, \citenamefont {Turturic\u{a}}, \citenamefont
  {Ujeniuc}, \citenamefont {Wieland}, \citenamefont {Ban}, \citenamefont
  {Ciemała}, \citenamefont {Ciocan}, \citenamefont {Clisu}, \citenamefont
  {Dinescu}, \citenamefont {Iancu}, \citenamefont {Kmiecik}, \citenamefont
  {Lelasseux}, \citenamefont {Lica}, \citenamefont {M\u{a}rginean},
  \citenamefont {Neac\textcommabelow{s}u}, \citenamefont {Pai}, \citenamefont
  {P\^{a}rlea}, \citenamefont {Petruse}, \citenamefont {Rotaru}, \citenamefont
  {Sotty}, \citenamefont {Sp\u{a}taru}, \citenamefont {State}, \citenamefont
  {Straticiuc}, \citenamefont {Tofan}, \citenamefont {Toma}, \citenamefont
  {Tozar}, \citenamefont {Turturic\u{a}},\ and\ \citenamefont
  {Ur}}]{Soderstrom2024a}%
  \BibitemOpen
  \bibfield  {author} {\bibinfo {author} {\bibfnamefont {P.-A.}\ \bibnamefont
  {S\"{o}derstr\"{o}m}}, \bibinfo {author} {\bibfnamefont {A.}~\bibnamefont
  {Ku\c{s}o\u{g}lu}}, \bibinfo {author} {\bibfnamefont {D.~L.}\ \bibnamefont
  {Balabanski}}, \bibinfo {author} {\bibfnamefont {M.}~\bibnamefont
  {Brezeanu}}, \bibinfo {author} {\bibfnamefont {D.}~\bibnamefont {Choudhury}},
  \bibinfo {author} {\bibfnamefont {A.}~\bibnamefont {Gavrilescu}}, \bibinfo
  {author} {\bibfnamefont {R.~A.}\ \bibnamefont {Gu\textcommabelow{t}oiu}},
  \bibinfo {author} {\bibfnamefont {S.}~\bibnamefont {Ioannidis}}, \bibinfo
  {author} {\bibfnamefont {G.}~\bibnamefont {Lorusso}}, \bibinfo {author}
  {\bibfnamefont {M.}~\bibnamefont {Markova}}, {\it{et al.}},\ }\href@noop {} {\bibfield
  {journal} {\bibinfo  {journal} {Nuovo Cimento C}\ }\textbf {\bibinfo {volume}
  {47}},\ \bibinfo {pages} {58} (\bibinfo {year} {2024})}\BibitemShut {NoStop}%
\bibitem [{\citenamefont {Aogaki}\ \emph {et~al.}(2023)\citenamefont {Aogaki},
  \citenamefont {Balabanski}, \citenamefont {Borcea}, \citenamefont
  {Constantin}, \citenamefont {Costache}, \citenamefont {Cuciuc}, \citenamefont
  {Ku\c{s}o\u{g}lu}, \citenamefont {Mihai}, \citenamefont {Mihai},
  \citenamefont {Stan}, \citenamefont {Söderström}, \citenamefont {Testov},
  \citenamefont {Turturic\u{a}}, \citenamefont {Ujeniuc}, \citenamefont
  {Adachi}, \citenamefont {Camera}, \citenamefont {Ciocan}, \citenamefont
  {Crespi}, \citenamefont {Florea}, \citenamefont {Fujikawa}, \citenamefont
  {Furuno}, \citenamefont {Gamba}, \citenamefont {Gu\textcommabelow{t}oiu},
  \citenamefont {Kawabata}, \citenamefont {Million}, \citenamefont {Nichita},
  \citenamefont {Niina}, \citenamefont {Okamoto}, \citenamefont {Pai},
  \citenamefont {Pappalardo}, \citenamefont {Sakanashi}, \citenamefont
  {Tamii},\ and\ \citenamefont {Wieland}}]{Aogaki2023}%
  \BibitemOpen
  \bibfield  {author} {\bibinfo {author} {\bibfnamefont {S.}~\bibnamefont
  {Aogaki}}, \bibinfo {author} {\bibfnamefont {D.~L.}\ \bibnamefont
  {Balabanski}}, \bibinfo {author} {\bibfnamefont {R.}~\bibnamefont {Borcea}},
  \bibinfo {author} {\bibfnamefont {P.}~\bibnamefont {Constantin}}, \bibinfo
  {author} {\bibfnamefont {C.}~\bibnamefont {Costache}}, \bibinfo {author}
  {\bibfnamefont {M.}~\bibnamefont {Cuciuc}}, \bibinfo {author} {\bibfnamefont
  {A.}~\bibnamefont {Ku\c{s}o\u{g}lu}}, \bibinfo {author} {\bibfnamefont
  {C.}~\bibnamefont {Mihai}}, \bibinfo {author} {\bibfnamefont {R.~E.}\
  \bibnamefont {Mihai}}, \bibinfo {author} {\bibfnamefont {L.}~\bibnamefont
  {Stan}}, {\it{et al.}},\
  }\href@noop {} {\bibfield  {journal} {\bibinfo  {journal} {Nucl. Instrum.
  Methods Phys. Res. A}\ }\textbf {\bibinfo {volume} {1056}},\ \bibinfo {pages}
  {168628} (\bibinfo {year} {2023})}\BibitemShut {NoStop}%
\bibitem [{\citenamefont {S\"{o}derstr\"{o}m}\ \emph
  {et~al.}(2021)\citenamefont {S\"{o}derstr\"{o}m}, \citenamefont {Matei},
  \citenamefont {Capponi}, \citenamefont {A\c{c}{\i}ks\"{o}z}, \citenamefont
  {Balabanski},\ and\ \citenamefont {Mitu}}]{Soderstrom2021}%
  \BibitemOpen
  \bibfield  {author} {\bibinfo {author} {\bibfnamefont {P.-A.}\ \bibnamefont
  {S\"{o}derstr\"{o}m}}, \bibinfo {author} {\bibfnamefont {C.}~\bibnamefont
  {Matei}}, \bibinfo {author} {\bibfnamefont {L.}~\bibnamefont {Capponi}},
  \bibinfo {author} {\bibfnamefont {E.}~\bibnamefont {A\c{c}{\i}ks\"{o}z}},
  \bibinfo {author} {\bibfnamefont {D.~L.}\ \bibnamefont {Balabanski}}, \ and\
  \bibinfo {author} {\bibfnamefont {I.~O.}\ \bibnamefont {Mitu}},\ }\href@noop
  {} {\bibfield  {journal} {\bibinfo  {journal} {Appl. Radiat. Isot.}\ }\textbf
  {\bibinfo {volume} {167}},\ \bibinfo {pages} {109441} (\bibinfo {year}
  {2021})}\BibitemShut {NoStop}%
\bibitem [{\citenamefont {S\"{o}derstr\"{o}m}\ \emph
  {et~al.}(2023{\natexlab{b}})\citenamefont {S\"{o}derstr\"{o}m}, \citenamefont
  {Balabanski}, \citenamefont {Ban}, \citenamefont {Ciocan}, \citenamefont
  {Cuciuc}, \citenamefont {Dhal}, \citenamefont {Fugaru}, \citenamefont
  {Iancu}, \citenamefont {Rotaru}, \citenamefont {\textcommabelow{S}erban},
  \citenamefont {State}, \citenamefont {Testov}, \citenamefont
  {Turturic\u{a}},\ and\ \citenamefont {Vasilca}}]{Soderstrom2023a}%
  \BibitemOpen
  \bibfield  {author} {\bibinfo {author} {\bibfnamefont {P.-A.}\ \bibnamefont
  {S\"{o}derstr\"{o}m}}, \bibinfo {author} {\bibfnamefont {D.~L.}\ \bibnamefont
  {Balabanski}}, \bibinfo {author} {\bibfnamefont {R.~S.}\ \bibnamefont {Ban}},
  \bibinfo {author} {\bibfnamefont {G.}~\bibnamefont {Ciocan}}, \bibinfo
  {author} {\bibfnamefont {M.}~\bibnamefont {Cuciuc}}, \bibinfo {author}
  {\bibfnamefont {A.}~\bibnamefont {Dhal}}, \bibinfo {author} {\bibfnamefont
  {V.}~\bibnamefont {Fugaru}}, \bibinfo {author} {\bibfnamefont
  {V.}~\bibnamefont {Iancu}}, \bibinfo {author} {\bibfnamefont
  {A.}~\bibnamefont {Rotaru}}, \bibinfo {author} {\bibfnamefont {A.~B.}\
  \bibnamefont {\textcommabelow{S}erban}}, {\it{et al.}},\ }\href@noop {} {\bibfield  {journal} {\bibinfo
   {journal} {Appl. Radiat. Isot.}\ }\textbf {\bibinfo {volume} {191}},\
  \bibinfo {pages} {110559} (\bibinfo {year} {2023}{\natexlab{b}})}\BibitemShut
  {NoStop}%
\bibitem [{\citenamefont {{Agostinelli}}\ \emph {et~al.}(2003)\citenamefont
  {{Agostinelli}}, \citenamefont {{Allison}}, \citenamefont {{Amako}},
  \citenamefont {{Apostolakis}}, \citenamefont {{Araujo}}, \citenamefont
  {{Arce}}, \citenamefont {{Asai}}, \citenamefont {{Axen}}, \citenamefont
  {{Banerjee}}, \citenamefont {{Barrand}}, \citenamefont {{Behner}},
  \citenamefont {{Bellagamba}}, \citenamefont {{Boudreau}}, \citenamefont
  {{Broglia}}, \citenamefont {{Brunengo}}, \citenamefont {{Burkhardt}},
  \citenamefont {{Chauvie}}, \citenamefont {{Chuma}}, \citenamefont
  {{Chytracek}}, \citenamefont {{Cooperman}}, \citenamefont {{Cosmo}},
  \citenamefont {{Degtyarenko}}, \citenamefont {{dell'Acqua}}, \citenamefont
  {{Depaola}}, \citenamefont {{Dietrich}}, \citenamefont {{Enami}},
  \citenamefont {{Feliciello}}, \citenamefont {{Ferguson}}, \citenamefont
  {{Fesefeldt}}, \citenamefont {{Folger}}, \citenamefont {{Foppiano}},
  \citenamefont {{Forti}}, \citenamefont {{Garelli}}, \citenamefont {{Giani}},
  \citenamefont {{Giannitrapani}}, \citenamefont {{Gibin}}, \citenamefont
  {{G{\'o}mez Cadenas}}, \citenamefont {{Gonz{\'a}lez}}, \citenamefont {{Gracia
  Abril}}, \citenamefont {{Greeniaus}}, \citenamefont {{Greiner}},
  \citenamefont {{Grichine}}, \citenamefont {{Grossheim}}, \citenamefont
  {{Guatelli}}, \citenamefont {{Gumplinger}}, \citenamefont {{Hamatsu}},
  \citenamefont {{Hashimoto}}, \citenamefont {{Hasui}}, \citenamefont
  {{Heikkinen}}, \citenamefont {{Howard}}, \citenamefont {{Ivanchenko}},
  \citenamefont {{Johnson}}, \citenamefont {{Jones}}, \citenamefont
  {{Kallenbach}}, \citenamefont {{Kanaya}}, \citenamefont {{Kawabata}},
  \citenamefont {{Kawabata}}, \citenamefont {{Kawaguti}}, \citenamefont
  {{Kelner}}, \citenamefont {{Kent}}, \citenamefont {{Kimura}}, \citenamefont
  {{Kodama}}, \citenamefont {{Kokoulin}}, \citenamefont {{Kossov}},
  \citenamefont {{Kurashige}}, \citenamefont {{Lamanna}}, \citenamefont
  {{Lamp{\'e}n}}, \citenamefont {{Lara}}, \citenamefont {{Lefebure}},
  \citenamefont {{Lei}}, \citenamefont {{Liendl}}, \citenamefont {{Lockman}},
  \citenamefont {{Longo}}, \citenamefont {{Magni}}, \citenamefont {{Maire}},
  \citenamefont {{Medernach}}, \citenamefont {{Minamimoto}}, \citenamefont
  {{Mora de Freitas}}, \citenamefont {{Morita}}, \citenamefont {{Murakami}},
  \citenamefont {{Nagamatu}}, \citenamefont {{Nartallo}}, \citenamefont
  {{Nieminen}}, \citenamefont {{Nishimura}}, \citenamefont {{Ohtsubo}},
  \citenamefont {{Okamura}}, \citenamefont {{O'Neale}}, \citenamefont
  {{Oohata}}, \citenamefont {{Paech}}, \citenamefont {{Perl}}, \citenamefont
  {{Pfeiffer}}, \citenamefont {{Pia}}, \citenamefont {{Ranjard}}, \citenamefont
  {{Rybin}}, \citenamefont {{Sadilov}}, \citenamefont {{di Salvo}},
  \citenamefont {{Santin}}, \citenamefont {{Sasaki}}, \citenamefont {{Savvas}},
  \citenamefont {{Sawada}}, \citenamefont {{Scherer}}, \citenamefont {{Sei}},
  \citenamefont {{Sirotenko}}, \citenamefont {{Smith}}, \citenamefont
  {{Starkov}}, \citenamefont {{Stoecker}}, \citenamefont {{Sulkimo}},
  \citenamefont {{Takahata}}, \citenamefont {{Tanaka}}, \citenamefont
  {{Tcherniaev}}, \citenamefont {{Safai Tehrani}}, \citenamefont {{Tropeano}},
  \citenamefont {{Truscott}}, \citenamefont {{Uno}}, \citenamefont {{Urban}},
  \citenamefont {{Urban}}, \citenamefont {{Verderi}}, \citenamefont
  {{Walkden}}, \citenamefont {{Wander}}, \citenamefont {{Weber}}, \citenamefont
  {{Wellisch}}, \citenamefont {{Wenaus}}, \citenamefont {{Williams}},
  \citenamefont {{Wright}}, \citenamefont {{Yamada}}, \citenamefont
  {{Yoshida}},\ and\ \citenamefont {{Zschiesche}}}]{Agostinelli2003}%
  \BibitemOpen
  \bibfield  {author} {\bibinfo {author} {\bibfnamefont {S.}~\bibnamefont
  {{Agostinelli}}}, \bibinfo {author} {\bibfnamefont {J.}~\bibnamefont
  {{Allison}}}, \bibinfo {author} {\bibfnamefont {K.}~\bibnamefont {{Amako}}},
  \bibinfo {author} {\bibfnamefont {J.}~\bibnamefont {{Apostolakis}}}, \bibinfo
  {author} {\bibfnamefont {H.}~\bibnamefont {{Araujo}}}, \bibinfo {author}
  {\bibfnamefont {P.}~\bibnamefont {{Arce}}}, \bibinfo {author} {\bibfnamefont
  {M.}~\bibnamefont {{Asai}}}, \bibinfo {author} {\bibfnamefont
  {D.}~\bibnamefont {{Axen}}}, \bibinfo {author} {\bibfnamefont
  {S.}~\bibnamefont {{Banerjee}}}, \bibinfo {author} {\bibfnamefont
  {G.}~\bibnamefont {{Barrand}}}, {\it{et al.}},\ }\href@noop {} {\bibfield  {journal}
  {\bibinfo  {journal} {Nucl. Instrum. Methods Phys. Res. A}\ }\textbf
  {\bibinfo {volume} {506}},\ \bibinfo {pages} {250} (\bibinfo {year}
  {2003})}\BibitemShut {NoStop}%
\bibitem [{\citenamefont {Lattuada}\ \emph {et~al.}(2017)\citenamefont
  {Lattuada}, \citenamefont {Balabanski}, \citenamefont {Chesnevskaya},
  \citenamefont {Costa}, \citenamefont {Crucill\`{a}}, \citenamefont {Guardo},
  \citenamefont {Cognata}, \citenamefont {Matei}, \citenamefont {Pizzone},
  \citenamefont {Romano}, \citenamefont {Spitaleri}, \citenamefont {Tumino},\
  and\ \citenamefont {Xu}}]{Lattuada2017}%
  \BibitemOpen
  \bibfield  {author} {\bibinfo {author} {\bibfnamefont {D.}~\bibnamefont
  {Lattuada}}, \bibinfo {author} {\bibfnamefont {D.~L.}\ \bibnamefont
  {Balabanski}}, \bibinfo {author} {\bibfnamefont {S.}~\bibnamefont
  {Chesnevskaya}}, \bibinfo {author} {\bibfnamefont {M.}~\bibnamefont {Costa}},
  \bibinfo {author} {\bibfnamefont {V.}~\bibnamefont {Crucill\`{a}}}, \bibinfo
  {author} {\bibfnamefont {G.~L.}\ \bibnamefont {Guardo}}, \bibinfo {author}
  {\bibfnamefont {M.~L.}\ \bibnamefont {Cognata}}, \bibinfo {author}
  {\bibfnamefont {C.}~\bibnamefont {Matei}}, \bibinfo {author} {\bibfnamefont
  {R.~G.}\ \bibnamefont {Pizzone}}, \bibinfo {author} {\bibfnamefont
  {S.}~\bibnamefont {Romano}}, {\it{et al.}},\ }\href@noop {} {\bibfield  {journal} {\bibinfo
  {journal} {EPJ Web Conf.}\ }\textbf {\bibinfo {volume} {165}},\ \bibinfo
  {pages} {01034} (\bibinfo {year} {2017})}\BibitemShut {NoStop}%
\bibitem [{\citenamefont {S\"{o}derstr\"{o}m}\ \emph
  {et~al.}(2019)\citenamefont {S\"{o}derstr\"{o}m}, \citenamefont {Capponi},
  \citenamefont {Iancu}, \citenamefont {Lattuada}, \citenamefont {Pappalardo},
  \citenamefont {Turturic\u{a}}, \citenamefont {A\c{c}{\i}ks\"{o}z},
  \citenamefont {Balabanski}, \citenamefont {Constantin}, \citenamefont
  {Guardo}, \citenamefont {Ilie}, \citenamefont {Ilie}, \citenamefont {Matei},
  \citenamefont {Nichita}, \citenamefont {Petruse},\ and\ \citenamefont
  {Spataru}}]{Soderstrom2019b}%
  \BibitemOpen
  \bibfield  {author} {\bibinfo {author} {\bibfnamefont {P.-A.}\ \bibnamefont
  {S\"{o}derstr\"{o}m}}, \bibinfo {author} {\bibfnamefont {L.}~\bibnamefont
  {Capponi}}, \bibinfo {author} {\bibfnamefont {V.}~\bibnamefont {Iancu}},
  \bibinfo {author} {\bibfnamefont {D.}~\bibnamefont {Lattuada}}, \bibinfo
  {author} {\bibfnamefont {A.}~\bibnamefont {Pappalardo}}, \bibinfo {author}
  {\bibfnamefont {G.~V.}\ \bibnamefont {Turturic\u{a}}}, \bibinfo {author}
  {\bibfnamefont {E.}~\bibnamefont {A\c{c}{\i}ks\"{o}z}}, \bibinfo {author}
  {\bibfnamefont {D.~L.}\ \bibnamefont {Balabanski}}, \bibinfo {author}
  {\bibfnamefont {P.}~\bibnamefont {Constantin}}, \bibinfo {author}
  {\bibfnamefont {G.~L.}\ \bibnamefont {Guardo}}, {\it{et al.}},\ }\href@noop
  {} {\bibfield  {journal} {\bibinfo  {journal} {J. Instrum.}\ }\textbf
  {\bibinfo {volume} {14}},\ \bibinfo {pages} {T11007} (\bibinfo {year}
  {2019})}\BibitemShut {NoStop}%
\bibitem [{\citenamefont {Brink}(1955)}]{Brink1955}%
  \BibitemOpen
  \bibfield  {author} {\bibinfo {author} {\bibfnamefont {D.~M.}\ \bibnamefont
  {Brink}},\ }\emph {\bibinfo {title} {{Some aspects of the interaction of
  fields with matter}}},\ \href@noop {} {Ph.D. thesis},\ \bibinfo  {school}
  {University of Oxford}, \bibinfo {address} {Oxford, United Kingdom} (\bibinfo
  {year} {1955})\BibitemShut {NoStop}%
\bibitem [{\citenamefont {Axel}(1962)}]{Axel1962}%
  \BibitemOpen
  \bibfield  {author} {\bibinfo {author} {\bibfnamefont {P.}~\bibnamefont
  {Axel}},\ }\href@noop {} {\bibfield  {journal} {\bibinfo  {journal} {Phys.
  Rev.}\ }\textbf {\bibinfo {volume} {126}},\ \bibinfo {pages} {671} (\bibinfo
  {year} {1962})}\BibitemShut {NoStop}%
\bibitem [{\citenamefont {Gilbert}\ and\ \citenamefont
  {Cameron}(1965)}]{Gilbert1965}%
  \BibitemOpen
  \bibfield  {author} {\bibinfo {author} {\bibfnamefont {A.}~\bibnamefont
  {Gilbert}}\ and\ \bibinfo {author} {\bibfnamefont {A.~G.~W.}\ \bibnamefont
  {Cameron}},\ }\href@noop {} {\bibfield  {journal} {\bibinfo  {journal} {Can.
  J. Phys.}\ }\textbf {\bibinfo {volume} {43}},\ \bibinfo {pages} {1446}
  (\bibinfo {year} {1965})}\BibitemShut {NoStop}%
\bibitem [{\citenamefont {Mughabghab}(2018)}]{Mughabghab2018}%
  \BibitemOpen
  \bibfield  {author} {\bibinfo {author} {\bibfnamefont {S.}~\bibnamefont
  {Mughabghab}},\ }\href@noop {} {\emph {\bibinfo {title} {Atlas of Neutron
  Resonances}}}\ (\bibinfo  {publisher} {Elsevier, Amsterdam},\ \bibinfo {year}
  {2018})\BibitemShut {NoStop}%
\bibitem [{\citenamefont {Kopecky}\ and\ \citenamefont
  {Uhl}(1990)}]{Kopecky1990}%
  \BibitemOpen
  \bibfield  {author} {\bibinfo {author} {\bibfnamefont {J.}~\bibnamefont
  {Kopecky}}\ and\ \bibinfo {author} {\bibfnamefont {M.}~\bibnamefont {Uhl}},\
  }\href@noop {} {\bibfield  {journal} {\bibinfo  {journal} {Phys. Rev. C}\
  }\textbf {\bibinfo {volume} {41}},\ \bibinfo {pages} {1941} (\bibinfo {year}
  {1990})}\BibitemShut {NoStop}%
\bibitem [{\citenamefont {{von Egidy}}\ and\ \citenamefont
  {Bucurescu}(2005)}]{vonEgidy2005}%
  \BibitemOpen
  \bibfield  {author} {\bibinfo {author} {\bibfnamefont {T.}~\bibnamefont {{von
  Egidy}}}\ and\ \bibinfo {author} {\bibfnamefont {D.}~\bibnamefont
  {Bucurescu}},\ }\href@noop {} {\bibfield  {journal} {\bibinfo  {journal}
  {Prog. Part. Nucl. Phys.}\ }\textbf {\bibinfo {volume} {72}},\ \bibinfo
  {pages} {044311} (\bibinfo {year} {2005})}\BibitemShut {NoStop}%
\bibitem [{\citenamefont {Bassauer}\ \emph {et~al.}(2020)\citenamefont
  {Bassauer}, \citenamefont {von Neumann-Cosel}, \citenamefont {Reinhard},
  \citenamefont {Tamii}, \citenamefont {Adachi}, \citenamefont {Bertulani},
  \citenamefont {Chan}, \citenamefont {D'Alessio}, \citenamefont {Fujioka},
  \citenamefont {Fujita}, \citenamefont {Fujita}, \citenamefont {Gey},
  \citenamefont {Hilcker}, \citenamefont {Hoang}, \citenamefont {Inoue},
  \citenamefont {Isaak}, \citenamefont {Iwamoto}, \citenamefont {Klaus},
  \citenamefont {Kobayashi}, \citenamefont {Maeda}, \citenamefont {Matsuda},
  \citenamefont {Nakatsuka}, \citenamefont {Noji}, \citenamefont {Ong},
  \citenamefont {Ou}, \citenamefont {Pietralla}, \citenamefont {Ponomarev},
  \citenamefont {Reen}, \citenamefont {Richter}, \citenamefont {Singer},
  \citenamefont {Steinhilber}, \citenamefont {Sudo}, \citenamefont {Togano},
  \citenamefont {Tsumura}, \citenamefont {Watanabe},\ and\ \citenamefont
  {Werner}}]{Bassauer2020}%
  \BibitemOpen
  \bibfield  {author} {\bibinfo {author} {\bibfnamefont {S.}~\bibnamefont
  {Bassauer}}, \bibinfo {author} {\bibfnamefont {P.}~\bibnamefont {von
  Neumann-Cosel}}, \bibinfo {author} {\bibfnamefont {P.-G.}\ \bibnamefont
  {Reinhard}}, \bibinfo {author} {\bibfnamefont {A.}~\bibnamefont {Tamii}},
  \bibinfo {author} {\bibfnamefont {S.}~\bibnamefont {Adachi}}, \bibinfo
  {author} {\bibfnamefont {C.~A.}\ \bibnamefont {Bertulani}}, \bibinfo {author}
  {\bibfnamefont {P.~Y.}\ \bibnamefont {Chan}}, \bibinfo {author}
  {\bibfnamefont {A.}~\bibnamefont {D'Alessio}}, \bibinfo {author}
  {\bibfnamefont {H.}~\bibnamefont {Fujioka}}, \bibinfo {author} {\bibfnamefont
  {H.}~\bibnamefont {Fujita}}, {\it{et al.}},\ }\href@noop {}
  {\bibfield  {journal} {\bibinfo  {journal} {Phys. Rev. C}\ }\textbf {\bibinfo
  {volume} {102}},\ \bibinfo {pages} {034327} (\bibinfo {year}
  {2020})}\BibitemShut {NoStop}%
\bibitem [{\citenamefont {von Neumann-Cosel}\ and\ \citenamefont
  {Tamii}(2019)}]{vonNeumann-Cosel2019a}%
  \BibitemOpen
  \bibfield  {author} {\bibinfo {author} {\bibfnamefont {P.}~\bibnamefont {von
  Neumann-Cosel}}\ and\ \bibinfo {author} {\bibfnamefont {A.}~\bibnamefont
  {Tamii}},\ }\href@noop {} {\bibfield  {journal} {\bibinfo  {journal} {Eur.
  Phys. J. A}\ }\textbf {\bibinfo {volume} {55}},\ \bibinfo {pages} {110}
  (\bibinfo {year} {2019})}\BibitemShut {NoStop}%
\bibitem [{\citenamefont {Koning}\ \emph {et~al.}(2008)\citenamefont {Koning},
  \citenamefont {Hilaire},\ and\ \citenamefont {Duijvestijn}}]{Koning2008}%
  \BibitemOpen
  \bibfield  {author} {\bibinfo {author} {\bibfnamefont {A.~J.}\ \bibnamefont
  {Koning}}, \bibinfo {author} {\bibfnamefont {S.}~\bibnamefont {Hilaire}}, \
  and\ \bibinfo {author} {\bibfnamefont {M.~C.}\ \bibnamefont {Duijvestijn}},\
  }in\ \href@noop {} {\emph {\bibinfo {booktitle} {Proceedings of the
  International Conference on Nuclear Data for Science and Technology}}},\
  Vol.\ \bibinfo {volume} {211},\ \bibinfo {editor} {edited by\ \bibinfo
  {editor} {\bibfnamefont {O.}~\bibnamefont {Bersillon}}, \bibinfo {editor}
  {\bibfnamefont {F.}~\bibnamefont {Gunsing}}, \bibinfo {editor} {\bibfnamefont
  {E.}~\bibnamefont {Bauge}}, \bibinfo {editor} {\bibfnamefont
  {R.}~\bibnamefont {Jacqmin}}, \ and\ \bibinfo {editor} {\bibfnamefont
  {S.}~\bibnamefont {Leray}}}\ (\bibinfo  {publisher} {EDP Sciences},\ \bibinfo
  {year} {2008})\ p.\ \bibinfo {pages} {058}\BibitemShut {NoStop}%
\bibitem [{\citenamefont {Koning}\ and\ \citenamefont
  {Rochman}(2012)}]{Koning2012}%
  \BibitemOpen
  \bibfield  {author} {\bibinfo {author} {\bibfnamefont {A.~J.}\ \bibnamefont
  {Koning}}\ and\ \bibinfo {author} {\bibfnamefont {D.}~\bibnamefont
  {Rochman}},\ }\href@noop {} {\bibfield  {journal} {\bibinfo  {journal} {Nucl.
  Data Sheets}\ }\textbf {\bibinfo {volume} {113}},\ \bibinfo {pages} {2841}
  (\bibinfo {year} {2012})}\BibitemShut {NoStop}%
\bibitem [{\citenamefont {Kadmenskii}\ \emph {et~al.}(1983)\citenamefont
  {Kadmenskii}, \citenamefont {Markushev},\ and\ \citenamefont
  {Furman}}]{Kadmenskii1983}%
  \BibitemOpen
  \bibfield  {author} {\bibinfo {author} {\bibfnamefont {S.~G.}\ \bibnamefont
  {Kadmenskii}}, \bibinfo {author} {\bibfnamefont {V.~P.}\ \bibnamefont
  {Markushev}}, \ and\ \bibinfo {author} {\bibfnamefont {V.~I.}\ \bibnamefont
  {Furman}},\ }\href@noop {} {\bibfield  {journal} {\bibinfo  {journal} {Sov.
  J. Nucl. Phys.}\ }\textbf {\bibinfo {volume} {37}},\ \bibinfo {pages} {345}
  (\bibinfo {year} {1983})}\BibitemShut {NoStop}%
\bibitem [{\citenamefont {Varlamov}\ \emph {et~al.}(2010)\citenamefont
  {Varlamov}, \citenamefont {Ishkhanov}, \citenamefont {Orlin},\ and\
  \citenamefont {Chetvertkova}}]{Varlamov2010}%
  \BibitemOpen
  \bibfield  {author} {\bibinfo {author} {\bibfnamefont {V.~V.}\ \bibnamefont
  {Varlamov}}, \bibinfo {author} {\bibfnamefont {B.~S.}\ \bibnamefont
  {Ishkhanov}}, \bibinfo {author} {\bibfnamefont {V.~N.}\ \bibnamefont
  {Orlin}}, \ and\ \bibinfo {author} {\bibfnamefont {V.~A.}\ \bibnamefont
  {Chetvertkova}},\ }\href@noop {} {\bibfield  {journal} {\bibinfo  {journal}
  {Bull. Russ. Acad. Sci.: Phys.}\ }\textbf {\bibinfo {volume} {74}},\ \bibinfo
  {pages} {833} (\bibinfo {year} {2010})}\BibitemShut {NoStop}%
\bibitem [{\citenamefont {Sorokin}\ \emph {et~al.}(1972)\citenamefont
  {Sorokin}, \citenamefont {Krushchev},\ and\ \citenamefont
  {Yurev}}]{Sorokin1972}%
  \BibitemOpen
  \bibfield  {author} {\bibinfo {author} {\bibfnamefont {Y.}~\bibnamefont
  {Sorokin}}, \bibinfo {author} {\bibfnamefont {V.}~\bibnamefont {Krushchev}},
  \ and\ \bibinfo {author} {\bibfnamefont {B.}~\bibnamefont {Yurev}},\
  }\href@noop {} {\bibfield  {journal} {\bibinfo  {journal} {Bull. Russ. Acad.
  Sci.: Phys.}\ }\textbf {\bibinfo {volume} {36}},\ \bibinfo {pages} {170}
  (\bibinfo {year} {1972})}\BibitemShut {NoStop}%
\bibitem [{\citenamefont {Sorokin}\ and\ \citenamefont
  {Yurev}(1975)}]{Sorokin1975}%
  \BibitemOpen
  \bibfield  {author} {\bibinfo {author} {\bibfnamefont {Y.}~\bibnamefont
  {Sorokin}}\ and\ \bibinfo {author} {\bibfnamefont {B.}~\bibnamefont
  {Yurev}},\ }\href@noop {} {\bibfield  {journal} {\bibinfo  {journal} {Sov. J.
  Nucl. Phys.}\ }\textbf {\bibinfo {volume} {20}},\ \bibinfo {pages} {123}
  (\bibinfo {year} {1975})}\BibitemShut {NoStop}%
\bibitem [{\citenamefont {Loens}\ \emph {et~al.}(2012)\citenamefont {Loens},
  \citenamefont {Langanke}, \citenamefont {Martínez-Pinedo},\ and\
  \citenamefont {Sieja}}]{Loens2012}%
  \BibitemOpen
  \bibfield  {author} {\bibinfo {author} {\bibfnamefont {H.~P.}\ \bibnamefont
  {Loens}}, \bibinfo {author} {\bibfnamefont {K.}~\bibnamefont {Langanke}},
  \bibinfo {author} {\bibfnamefont {G.}~\bibnamefont {Martínez-Pinedo}}, \
  and\ \bibinfo {author} {\bibfnamefont {K.}~\bibnamefont {Sieja}},\
  }\href@noop {} {\bibfield  {journal} {\bibinfo  {journal} {Eur. Phys. J. A}\
  }\textbf {\bibinfo {volume} {48}},\ \bibinfo {pages} {34} (\bibinfo {year}
  {2012})}\BibitemShut {NoStop}%
\bibitem [{\citenamefont {Brun}\ and\ \citenamefont
  {Rademakers}(1997)}]{Brun1997}%
  \BibitemOpen
  \bibfield  {author} {\bibinfo {author} {\bibfnamefont {R.}~\bibnamefont
  {Brun}}\ and\ \bibinfo {author} {\bibfnamefont {F.}~\bibnamefont
  {Rademakers}},\ }\href@noop {} {\bibfield  {journal} {\bibinfo  {journal}
  {Nucl. Instrum. Methods Phys. Res. A}\ }\textbf {\bibinfo {volume} {389}},\
  \bibinfo {pages} {81} (\bibinfo {year} {1997})}\BibitemShut {NoStop}%
\bibitem [{\citenamefont {Antcheva}\ \emph {et~al.}(2009)\citenamefont
  {Antcheva}, \citenamefont {Ballintijn}, \citenamefont {Bellenot},
  \citenamefont {Biskup}, \citenamefont {Brun}, \citenamefont {Buncic},
  \citenamefont {Canal}, \citenamefont {Casadei}, \citenamefont {Couet},
  \citenamefont {Fine}, \citenamefont {Franco}, \citenamefont {Ganis},
  \citenamefont {Gheata}, \citenamefont {Maline}, \citenamefont {Goto},
  \citenamefont {Iwaszkiewicz}, \citenamefont {Kreshuk}, \citenamefont
  {Segura}, \citenamefont {Maunder}, \citenamefont {Moneta}, \citenamefont
  {Naumann}, \citenamefont {Offermann}, \citenamefont {Onuchin}, \citenamefont
  {Panacek}, \citenamefont {Rademakers}, \citenamefont {Russo},\ and\
  \citenamefont {Tadel}}]{Antcheva2009}%
  \BibitemOpen
  \bibfield  {author} {\bibinfo {author} {\bibfnamefont {I.}~\bibnamefont
  {Antcheva}}, \bibinfo {author} {\bibfnamefont {M.}~\bibnamefont
  {Ballintijn}}, \bibinfo {author} {\bibfnamefont {B.}~\bibnamefont
  {Bellenot}}, \bibinfo {author} {\bibfnamefont {M.}~\bibnamefont {Biskup}},
  \bibinfo {author} {\bibfnamefont {R.}~\bibnamefont {Brun}}, \bibinfo {author}
  {\bibfnamefont {N.}~\bibnamefont {Buncic}}, \bibinfo {author} {\bibfnamefont
  {P.}~\bibnamefont {Canal}}, \bibinfo {author} {\bibfnamefont
  {D.}~\bibnamefont {Casadei}}, \bibinfo {author} {\bibfnamefont
  {O.}~\bibnamefont {Couet}}, \bibinfo {author} {\bibfnamefont
  {V.}~\bibnamefont {Fine}}, {\it{et al.}},\ }\href@noop {} {\bibfield  {journal} {\bibinfo
  {journal} {Comp. Phys. Commun.}\ }\textbf {\bibinfo {volume} {180}},\
  \bibinfo {pages} {2499} (\bibinfo {year} {2009})}\BibitemShut {NoStop}%
\bibitem [{\citenamefont {Tsoneva}\ and\ \citenamefont
  {Lenske}(2008)}]{Tsoneva2008}%
  \BibitemOpen
  \bibfield  {author} {\bibinfo {author} {\bibfnamefont {N.}~\bibnamefont
  {Tsoneva}}\ and\ \bibinfo {author} {\bibfnamefont {H.}~\bibnamefont
  {Lenske}},\ }\href@noop {} {\bibfield  {journal} {\bibinfo  {journal} {Phys.
  Rev. C}\ }\textbf {\bibinfo {volume} {77}},\ \bibinfo {pages} {024321}
  (\bibinfo {year} {2008})}\BibitemShut {NoStop}%
\bibitem [{\citenamefont {Markova}\ \emph {et~al.}(2025)\citenamefont
  {Markova}, \citenamefont {{von Neumann-Cosel}},\ and\ \citenamefont
  {Litvinova}}]{Markova2025}%
  \BibitemOpen
  \bibfield  {author} {\bibinfo {author} {\bibfnamefont {M.}~\bibnamefont
  {Markova}}, \bibinfo {author} {\bibfnamefont {P.}~\bibnamefont {{von
  Neumann-Cosel}}}, \ and\ \bibinfo {author} {\bibfnamefont {E.}~\bibnamefont
  {Litvinova}},\ }\href@noop {} {\bibfield  {journal} {\bibinfo  {journal}
  {Phys. Lett. B}\ }\textbf {\bibinfo {volume} {860}},\ \bibinfo {pages}
  {139216} (\bibinfo {year} {2025})}\BibitemShut {NoStop}%
\bibitem [{\citenamefont {Thomas}(1925)}]{Thomas1925}%
  \BibitemOpen
  \bibfield  {author} {\bibinfo {author} {\bibfnamefont {W.}~\bibnamefont
  {Thomas}},\ }\href@noop {} {\bibfield  {journal} {\bibinfo  {journal} {Sci.
  Nat.}\ }\textbf {\bibinfo {volume} {13}},\ \bibinfo {pages} {627} (\bibinfo
  {year} {1925})}\BibitemShut {NoStop}%
\bibitem [{\citenamefont {Reiche}\ and\ \citenamefont
  {Thomas}(1925)}]{Reiche1925}%
  \BibitemOpen
  \bibfield  {author} {\bibinfo {author} {\bibfnamefont {F.}~\bibnamefont
  {Reiche}}\ and\ \bibinfo {author} {\bibfnamefont {W.}~\bibnamefont
  {Thomas}},\ }\href@noop {} {\bibfield  {journal} {\bibinfo  {journal} {Z.
  Phys.}\ }\textbf {\bibinfo {volume} {34}},\ \bibinfo {pages} {510} (\bibinfo
  {year} {1925})}\BibitemShut {NoStop}%
\bibitem [{\citenamefont {Kuhn}(1925)}]{Kuhn1925}%
  \BibitemOpen
  \bibfield  {author} {\bibinfo {author} {\bibfnamefont {W.}~\bibnamefont
  {Kuhn}},\ }\href@noop {} {\bibfield  {journal} {\bibinfo  {journal} {Z.
  Phys.}\ }\textbf {\bibinfo {volume} {33}},\ \bibinfo {pages} {408} (\bibinfo
  {year} {1925})}\BibitemShut {NoStop}%
\bibitem [{\citenamefont {Guttormsen}\ \emph {et~al.}(2001)\citenamefont
  {Guttormsen}, \citenamefont {Hjorth-Jensen}, \citenamefont {Melby},
  \citenamefont {Rekstad}, \citenamefont {Schiller},\ and\ \citenamefont
  {Siem}}]{Guttormsen2001}%
  \BibitemOpen
  \bibfield  {author} {\bibinfo {author} {\bibfnamefont {M.}~\bibnamefont
  {Guttormsen}}, \bibinfo {author} {\bibfnamefont {M.}~\bibnamefont
  {Hjorth-Jensen}}, \bibinfo {author} {\bibfnamefont {E.}~\bibnamefont
  {Melby}}, \bibinfo {author} {\bibfnamefont {J.}~\bibnamefont {Rekstad}},
  \bibinfo {author} {\bibfnamefont {A.}~\bibnamefont {Schiller}}, \ and\
  \bibinfo {author} {\bibfnamefont {S.}~\bibnamefont {Siem}},\ }\href@noop {}
  {\bibfield  {journal} {\bibinfo  {journal} {Phys. Rev. C}\ }\textbf {\bibinfo
  {volume} {63}},\ \bibinfo {pages} {044301} (\bibinfo {year}
  {2001})}\BibitemShut {NoStop}%
\bibitem [{\citenamefont {Nyhus}\ \emph {et~al.}(2012)\citenamefont {Nyhus},
  \citenamefont {Siem}, \citenamefont {Guttormsen}, \citenamefont {Larsen},
  \citenamefont {B\"urger}, \citenamefont {Syed}, \citenamefont {Toft},
  \citenamefont {Tveten},\ and\ \citenamefont {Voinov}}]{Nyhus2012}%
  \BibitemOpen
  \bibfield  {author} {\bibinfo {author} {\bibfnamefont {H.~T.}\ \bibnamefont
  {Nyhus}}, \bibinfo {author} {\bibfnamefont {S.}~\bibnamefont {Siem}},
  \bibinfo {author} {\bibfnamefont {M.}~\bibnamefont {Guttormsen}}, \bibinfo
  {author} {\bibfnamefont {A.~C.}\ \bibnamefont {Larsen}}, \bibinfo {author}
  {\bibfnamefont {A.}~\bibnamefont {B\"urger}}, \bibinfo {author}
  {\bibfnamefont {N.~U.~H.}\ \bibnamefont {Syed}}, \bibinfo {author}
  {\bibfnamefont {H.~K.}\ \bibnamefont {Toft}}, \bibinfo {author}
  {\bibfnamefont {G.~M.}\ \bibnamefont {Tveten}}, \ and\ \bibinfo {author}
  {\bibfnamefont {A.}~\bibnamefont {Voinov}},\ }\href@noop {} {\bibfield
  {journal} {\bibinfo  {journal} {Phys. Rev. C}\ }\textbf {\bibinfo {volume}
  {85}},\ \bibinfo {pages} {014323} (\bibinfo {year} {2012})}\BibitemShut
  {NoStop}%
\bibitem [{\citenamefont {Roy}\ \emph {et~al.}(2021)\citenamefont {Roy},
  \citenamefont {Banerjee}, \citenamefont {Rana}, \citenamefont {Kundu},
  \citenamefont {Pandit}, \citenamefont {{Quang Hung}}, \citenamefont {Ghosh},
  \citenamefont {Mukhopadhyay}, \citenamefont {Mondal}, \citenamefont
  {Mukherjee}, \citenamefont {Manna}, \citenamefont {Sen}, \citenamefont {Pal},
  \citenamefont {Pandey}, \citenamefont {Paul}, \citenamefont {Atreya},\ and\
  \citenamefont {Bhattacharya}}]{Roy2021}%
  \BibitemOpen
  \bibfield  {author} {\bibinfo {author} {\bibfnamefont {P.}~\bibnamefont
  {Roy}}, \bibinfo {author} {\bibfnamefont {K.}~\bibnamefont {Banerjee}},
  \bibinfo {author} {\bibfnamefont {T.~K.}\ \bibnamefont {Rana}}, \bibinfo
  {author} {\bibfnamefont {S.}~\bibnamefont {Kundu}}, \bibinfo {author}
  {\bibfnamefont {D.}~\bibnamefont {Pandit}}, \bibinfo {author} {\bibfnamefont
  {N.}~\bibnamefont {{Quang Hung}}}, \bibinfo {author} {\bibfnamefont {T.~K.}\
  \bibnamefont {Ghosh}}, \bibinfo {author} {\bibfnamefont {S.}~\bibnamefont
  {Mukhopadhyay}}, \bibinfo {author} {\bibfnamefont {D.}~\bibnamefont
  {Mondal}}, \bibinfo {author} {\bibfnamefont {G.}~\bibnamefont {Mukherjee}},
  {\it{et al.}},\ }\href@noop {} {\bibfield  {journal} {\bibinfo  {journal}
  {Eur. Phys. J. A}\ }\textbf {\bibinfo {volume} {57}},\ \bibinfo {pages} {48}
  (\bibinfo {year} {2021})}\BibitemShut {NoStop}%
\bibitem [{\citenamefont {Guttormsen}\ \emph {et~al.}(2003)\citenamefont
  {Guttormsen}, \citenamefont {Chankova}, \citenamefont {Hjorth-Jensen},
  \citenamefont {Rekstad}, \citenamefont {Siem}, \citenamefont {Schiller},\
  and\ \citenamefont {Dean}}]{Guttormsen2003}%
  \BibitemOpen
  \bibfield  {author} {\bibinfo {author} {\bibfnamefont {M.}~\bibnamefont
  {Guttormsen}}, \bibinfo {author} {\bibfnamefont {R.}~\bibnamefont
  {Chankova}}, \bibinfo {author} {\bibfnamefont {M.}~\bibnamefont
  {Hjorth-Jensen}}, \bibinfo {author} {\bibfnamefont {J.}~\bibnamefont
  {Rekstad}}, \bibinfo {author} {\bibfnamefont {S.}~\bibnamefont {Siem}},
  \bibinfo {author} {\bibfnamefont {A.}~\bibnamefont {Schiller}}, \ and\
  \bibinfo {author} {\bibfnamefont {D.~J.}\ \bibnamefont {Dean}},\ }\href@noop
  {} {\bibfield  {journal} {\bibinfo  {journal} {Phys. Rev. C}\ }\textbf
  {\bibinfo {volume} {68}},\ \bibinfo {pages} {034311} (\bibinfo {year}
  {2003})}\BibitemShut {NoStop}%
\bibitem [{\citenamefont {von Helmholtz}(1889)}]{vonHelmholtz1888}%
  \BibitemOpen
  \bibfield  {author} {\bibinfo {author} {\bibfnamefont {H.}~\bibnamefont {von
  Helmholtz}},\ }in\ \href@noop {} {\emph {\bibinfo {booktitle} {Physical
  Memoirs: Selected and Translated from Foregin Sources}}},\ Vol.~\bibinfo
  {volume} {I}\ (\bibinfo  {publisher} {Taylor \& Francis},\ \bibinfo {address}
  {London},\ \bibinfo {year} {1889})\ Chap.\ \bibinfo {chapter} {On the
  Thermodynamics of Chemical Processes}, p.~\bibinfo {pages} {43}\BibitemShut
  {NoStop}%
\bibitem [{\citenamefont {Lee}\ and\ \citenamefont
  {Kosterlitz}(1990)}]{Lee1990}%
  \BibitemOpen
  \bibfield  {author} {\bibinfo {author} {\bibfnamefont {J.}~\bibnamefont
  {Lee}}\ and\ \bibinfo {author} {\bibfnamefont {J.~M.}\ \bibnamefont
  {Kosterlitz}},\ }\href@noop {} {\bibfield  {journal} {\bibinfo  {journal}
  {Phys. Rev. Lett.}\ }\textbf {\bibinfo {volume} {65}},\ \bibinfo {pages}
  {137} (\bibinfo {year} {1990})}\BibitemShut {NoStop}%
\bibitem [{\citenamefont {Lee}\ and\ \citenamefont
  {Kosterlitz}(1991)}]{Lee1991}%
  \BibitemOpen
  \bibfield  {author} {\bibinfo {author} {\bibfnamefont {J.}~\bibnamefont
  {Lee}}\ and\ \bibinfo {author} {\bibfnamefont {J.~M.}\ \bibnamefont
  {Kosterlitz}},\ }\href@noop {} {\bibfield  {journal} {\bibinfo  {journal}
  {Phys. Rev. B}\ }\textbf {\bibinfo {volume} {43}},\ \bibinfo {pages} {3265}
  (\bibinfo {year} {1991})}\BibitemShut {NoStop}%
\bibitem [{\citenamefont {Tsoneva}\ \emph {et~al.}(2004)\citenamefont
  {Tsoneva}, \citenamefont {Lenske},\ and\ \citenamefont
  {Stoyanov}}]{Tsoneva2004}%
  \BibitemOpen
  \bibfield  {author} {\bibinfo {author} {\bibfnamefont {N.}~\bibnamefont
  {Tsoneva}}, \bibinfo {author} {\bibfnamefont {H.}~\bibnamefont {Lenske}}, \
  and\ \bibinfo {author} {\bibfnamefont {C.}~\bibnamefont {Stoyanov}},\
  }\href@noop {} {\bibfield  {journal} {\bibinfo  {journal} {Phys. Lett. B}\
  }\textbf {\bibinfo {volume} {586}},\ \bibinfo {pages} {213} (\bibinfo {year}
  {2004})}\BibitemShut {NoStop}%
\bibitem [{\citenamefont {Tsoneva}\ and\ \citenamefont
  {Lenske}(2016)}]{Tsoneva2016}%
  \BibitemOpen
  \bibfield  {author} {\bibinfo {author} {\bibfnamefont {N.}~\bibnamefont
  {Tsoneva}}\ and\ \bibinfo {author} {\bibfnamefont {H.}~\bibnamefont
  {Lenske}},\ }\href@noop {} {\bibfield  {journal} {\bibinfo  {journal} {Phys.
  At. Nucl.}\ }\textbf {\bibinfo {volume} {79}},\ \bibinfo {pages} {885}
  (\bibinfo {year} {2016})}\BibitemShut {NoStop}%
\bibitem [{\citenamefont {Tonchev}\ \emph {et~al.}(2017)\citenamefont
  {Tonchev}, \citenamefont {Tsoneva}, \citenamefont {Bhatia}, \citenamefont
  {Arnold}, \citenamefont {S.~Goriely}, \citenamefont {Kelley}, \citenamefont
  {E.~Kwan}, \citenamefont {Piekarewicz}, \citenamefont {R.~Raut},
  \citenamefont {Shizuma},\ and\ \citenamefont {Tornow}}]{Tonchev2017}%
  \BibitemOpen
  \bibfield  {author} {\bibinfo {author} {\bibfnamefont {A.}~\bibnamefont
  {Tonchev}}, \bibinfo {author} {\bibfnamefont {N.}~\bibnamefont {Tsoneva}},
  \bibinfo {author} {\bibfnamefont {C.}~\bibnamefont {Bhatia}}, \bibinfo
  {author} {\bibfnamefont {C.}~\bibnamefont {Arnold}}, \bibinfo {author}
  {\bibfnamefont {S.~H.}\ \bibnamefont {S.~Goriely}}, \bibinfo {author}
  {\bibfnamefont {J.}~\bibnamefont {Kelley}}, \bibinfo {author} {\bibfnamefont
  {H.~L.}\ \bibnamefont {E.~Kwan}}, \bibinfo {author} {\bibfnamefont
  {J.}~\bibnamefont {Piekarewicz}}, \bibinfo {author} {\bibfnamefont {G.~R.}\
  \bibnamefont {R.~Raut}}, \bibinfo {author} {\bibfnamefont {T.}~\bibnamefont
  {Shizuma}}, \ and\ \bibinfo {author} {\bibfnamefont {W.}~\bibnamefont
  {Tornow}},\ }\href@noop {} {\bibfield  {journal} {\bibinfo  {journal} {Phys.
  Lett. B}\ }\textbf {\bibinfo {volume} {773}},\ \bibinfo {pages} {20}
  (\bibinfo {year} {2017})}\BibitemShut {NoStop}%
\bibitem [{\citenamefont {Soloviev}(1976)}]{Soloviev1976}%
  \BibitemOpen
  \bibfield  {author} {\bibinfo {author} {\bibfnamefont {V.~G.}\ \bibnamefont
  {Soloviev}},\ }\href@noop {} {\emph {\bibinfo {title} {Theory of complex
  nuclei}}}\ (\bibinfo  {publisher} {Oxford: Pergamon Press},\ \bibinfo {year}
  {1976})\BibitemShut {NoStop}%
\bibitem [{\citenamefont {M.~Grinberg}(1998)}]{Grinberg1998}%
  \BibitemOpen
  \bibfield  {author} {\bibinfo {author} {\bibfnamefont {N.~T.}\ \bibnamefont
  {M.~Grinberg}, \bibfnamefont {Ch.~Stoyanov}},\ }\href@noop {} {\bibfield
  {journal} {\bibinfo  {journal} {Phys. Part. Nucl.}\ }\textbf {\bibinfo
  {volume} {29}},\ \bibinfo {pages} {606} (\bibinfo {year} {1998})}\BibitemShut
  {NoStop}%
\bibitem [{\citenamefont {Shizuma}\ \emph {et~al.}(2022)\citenamefont
  {Shizuma}, \citenamefont {Endo}, \citenamefont {Kimura}, \citenamefont
  {Massarczyk}, \citenamefont {Schwengner}, \citenamefont {Beyer},
  \citenamefont {Hensel}, \citenamefont {Hoffmann}, \citenamefont {Junghans},
  \citenamefont {Römer}, \citenamefont {Turkat}, \citenamefont {Wagner},\ and\
  \citenamefont {Tsoneva}}]{Shizuma2022}%
  \BibitemOpen
  \bibfield  {author} {\bibinfo {author} {\bibfnamefont {T.}~\bibnamefont
  {Shizuma}}, \bibinfo {author} {\bibfnamefont {S.}~\bibnamefont {Endo}},
  \bibinfo {author} {\bibfnamefont {A.}~\bibnamefont {Kimura}}, \bibinfo
  {author} {\bibfnamefont {R.}~\bibnamefont {Massarczyk}}, \bibinfo {author}
  {\bibfnamefont {R.}~\bibnamefont {Schwengner}}, \bibinfo {author}
  {\bibfnamefont {R.}~\bibnamefont {Beyer}}, \bibinfo {author} {\bibfnamefont
  {T.}~\bibnamefont {Hensel}}, \bibinfo {author} {\bibfnamefont
  {H.}~\bibnamefont {Hoffmann}}, \bibinfo {author} {\bibfnamefont
  {A.}~\bibnamefont {Junghans}}, \bibinfo {author} {\bibfnamefont
  {K.}~\bibnamefont {Römer}}, {\it{et al.}},\ }\href@noop {} {\bibfield  {journal} {\bibinfo
   {journal} {Phys. Rev. C}\ }\textbf {\bibinfo {volume} {106}},\ \bibinfo
  {pages} {044326} (\bibinfo {year} {2022})}\BibitemShut {NoStop}%
\bibitem [{\citenamefont {Hauser}\ and\ \citenamefont
  {Feshbach}(1952)}]{Hauser1952}%
  \BibitemOpen
  \bibfield  {author} {\bibinfo {author} {\bibfnamefont {W.}~\bibnamefont
  {Hauser}}\ and\ \bibinfo {author} {\bibfnamefont {H.}~\bibnamefont
  {Feshbach}},\ }\href@noop {} {\bibfield  {journal} {\bibinfo  {journal}
  {Phys. Rev.}\ }\textbf {\bibinfo {volume} {87}},\ \bibinfo {pages} {366}
  (\bibinfo {year} {1952})}\BibitemShut {NoStop}%
\bibitem [{\citenamefont {Goriely}\ \emph {et~al.}(2007)\citenamefont
  {Goriely}, \citenamefont {Hilaire},\ and\ \citenamefont
  {Koning}}]{Goriely2007}%
  \BibitemOpen
  \bibfield  {author} {\bibinfo {author} {\bibfnamefont {S.}~\bibnamefont
  {Goriely}}, \bibinfo {author} {\bibfnamefont {S.}~\bibnamefont {Hilaire}}, \
  and\ \bibinfo {author} {\bibfnamefont {A.~J.}\ \bibnamefont {Koning}},\
  }\href@noop {} {\bibfield  {journal} {\bibinfo  {journal} {Astron.
  Astrophys.}\ }\textbf {\bibinfo {volume} {487}},\ \bibinfo {pages} {767}
  (\bibinfo {year} {2007})}\BibitemShut {NoStop}%
%
\bibitem [{\citenamefont {Koning}\ \emph {et~al.}(2019)\citenamefont
  {Koning}, \citenamefont {Rochman}, \citenamefont {Sublet}, \citenamefont
  {Dzysiuk}, \citenamefont {Fleming},\ and\ \citenamefont
  {van der Marck}}]{Koning2019}%
  \BibitemOpen
  \bibfield  {author} {\bibinfo {author} {\bibfnamefont {A.~J.}~\bibnamefont
  {Koning}}, \bibinfo {author} {\bibfnamefont {D.}~\bibnamefont {Rochman}},
  \bibinfo {author} {\bibfnamefont {J.-Ch.}~\bibnamefont {Sublet}}, \bibinfo
  {author} {\bibfnamefont {N.}~\bibnamefont {Dzysiuk}}, \bibinfo {author}
  {\bibfnamefont {M.}~\bibnamefont {Fleming}}, \ and\ \bibinfo {author}
  {\bibfnamefont {S.}~\bibnamefont {van der Marck}},\ }\href@noop {} {\bibfield
  {journal} {\bibinfo  {journal} {Nucl. Data Sheets}\ }\textbf {\bibinfo
  {volume} {155}},\ \bibinfo {pages} {1} (\bibinfo {year}
  {2019})}\BibitemShut {NoStop}%
\bibitem [{\citenamefont {Rochman}\ \emph {et~al.}(2025)\citenamefont
  {Rochman}, \citenamefont {Koning}, \citenamefont {Goriely},\ and\ \citenamefont
  {Hilarie}}]{Rochman2025}%
  \BibitemOpen
  \bibfield  {author} {\bibinfo {author} {\bibfnamefont {D.}~\bibnamefont
  {Rochman}}, \bibinfo {author} {\bibfnamefont {A.}~\bibnamefont {Koning}},
  \bibinfo {author} {\bibfnamefont {S.}~\bibnamefont {Goriely}}, \ and\ \bibinfo {author}
  {\bibfnamefont {S.}~\bibnamefont {Hilarie}},\ }\href@noop {} {\bibfield
  {journal} {\bibinfo  {journal} {Nucl. Phys. A}\ }\textbf {\bibinfo
  {volume} {1053}},\ \bibinfo {pages} {122951} (\bibinfo {year}
  {2025})}\BibitemShut {NoStop}%
%
\bibitem [{\citenamefont {Goriely}\ \emph {et~al.}(2008)\citenamefont
  {Goriely}, \citenamefont {Hilaire},\ and\ \citenamefont
  {Koning}}]{Goriely2008}%
  \BibitemOpen
  \bibfield  {author} {\bibinfo {author} {\bibfnamefont {S.}~\bibnamefont
  {Goriely}}, \bibinfo {author} {\bibfnamefont {S.}~\bibnamefont {Hilaire}}, \ and\ \bibinfo {author}
  {\bibfnamefont {A.~J.}~\bibnamefont {Koning}},\ }\href@noop {} {\bibfield
  {journal} {\bibinfo  {journal} {Astron. Astrophys.}\ }\textbf {\bibinfo
  {volume} {487}},\ \bibinfo {pages} {767} (\bibinfo {year}
  {2008})}\BibitemShut {NoStop}%
\bibitem [{\citenamefont {Goriely}(2004)}]{Goriely2004}%
  \BibitemOpen
  \bibfield  {author} {\bibinfo {author} {\bibfnamefont {S.}~\bibnamefont
  {Goriely}},\ }\href@noop {} {\bibfield  {journal} {\bibinfo  {journal} {AIP
  Conf. Proc.}\ }\textbf {\bibinfo {volume} {704}},\ \bibinfo {pages} {375}
  (\bibinfo {year} {2004})}\BibitemShut {NoStop}%
\bibitem [{\citenamefont {Xu}\ \emph {et~al.}(2013)\citenamefont {Xu},
  \citenamefont {Goriely}, \citenamefont {Jorissen}, \citenamefont {Chen},\
  and\ \citenamefont {Arnould}}]{Xu2013}%
  \BibitemOpen
  \bibfield  {author} {\bibinfo {author} {\bibfnamefont {Y.}~\bibnamefont
  {Xu}}, \bibinfo {author} {\bibfnamefont {S.}~\bibnamefont {Goriely}},
  \bibinfo {author} {\bibfnamefont {A.}~\bibnamefont {Jorissen}}, \bibinfo
  {author} {\bibfnamefont {G.~L.}\ \bibnamefont {Chen}}, \ and\ \bibinfo
  {author} {\bibfnamefont {M.}~\bibnamefont {Arnould}},\ }\href@noop {}
  {\bibfield  {journal} {\bibinfo  {journal} {Astron. Astrophys.}\ }\textbf
  {\bibinfo {volume} {549}},\ \bibinfo {pages} {A106} (\bibinfo {year}
  {2013})}\BibitemShut {NoStop}%
\end{thebibliography}
\end{document}